\documentclass{aa}  

\usepackage{graphicx}
\usepackage{txfonts}
\usepackage{booktabs}
\usepackage{xcolor}
\usepackage{threeparttable}
\usepackage{multirow}
\usepackage{bm}
\usepackage{placeins}
\usepackage{caption}
\usepackage{comment}
\usepackage{hyperref}
\hypersetup{
    colorlinks=true,
    linkcolor=blue,
    citecolor=blue,
    filecolor=magenta,      
    urlcolor=cyan,
    }
\begin{document}
\newcommand{\hi}{\textsc{H$\,$i}}
\newcommand{\htwo}{\ensuremath{\mathrm{H_2}}}

   \title{Radio-continuum spectra of ram pressure stripped galaxies in the Coma Cluster}


\author{I.D. Roberts\inst{\ref{leiden}}
        \and
        R.J. van Weeren\inst{\ref{leiden}}
        \and
        D.V. Lal\inst{\ref{tata_inst}}
        \and
        M. Sun\inst{\ref{uah}}
        \and
        H. Chen\inst{\ref{hangzhou}}
        \and
        A. Ignesti\inst{\ref{inaf_padova}}
        \and
        M. Br{\"u}ggen\inst{\ref{hamburg}}
        \and
        N. Lyskova\inst{\ref{space_research_moscow},\ref{astro_moscow}}
        \and
        T. Venturi\inst{\ref{inaf_bologna}}
        \and
        M. Yagi\inst{\ref{naoj}}
       }

\institute{
Leiden Observatory, Leiden University, PO Box 9513, 2300 RA Leiden, The Netherlands \label{leiden}
\\
\email{iroberts@strw.leidenuniv.nl}
\and
National Centre for Radio Astrophysics, Tata Institute of Fundamental Research, Post Box 3, Ganeshkhind P.O., Pune 411007, India \label{tata_inst}
\and
Department of Physics \& Astronomy, University of Alabama in Huntsville, 301 Sparkman Drive, Huntsville, AL 35899, USA \label{uah}
\and
Research Center for Intelligent Computing Platforms, Zhejiang Laboratory, Hangzhou 311100, China \label{hangzhou}
\and
INAF--Padova Astronomical Observatory, Vicolo dell'Osservatorio 5, I-35122 Padova, Italy \label{inaf_padova}
\and
Hamburger Sternwarte, Universit{\"a}t Hamburg, Gojenbergsweg 112, 21029, Hamburg, Germany \label{hamburg}
\and
Space Research Institute (IKI), Profsoyuznaya 84/32, Moscow 117997, Russia \label{space_research_moscow}
\and
Astro Space Center, P. N. Lebedev Physical Institute of RAS, Profsojuznaya 84/32, Moscow 117997, Russia \label{astro_moscow}
\and
INAF--Istituto di Radioastronomia, via Gobetti 101, I-40129 Bologna, Italy \label{inaf_bologna}
\and
Astronomy Data Center, National Astronomical Observatory of Japan, Mitaka, Tokyo 181-8588, Japan \label{naoj}
}
 
\abstract
{The population of galaxies in the local Universe is bi-modal in terms of specific star formation rate.  This fact has led to a broad distinction between star-forming galaxies (typically cold-gas rich and late type) and quenched galaxies (typically cold-gas poor and early type). The ratio between quenched and star-forming galaxies is much larger in clusters than the field, and pinpointing which are the physical processes driving this excess quenching in clusters is an open question.}
{We used the nearby Coma Cluster as a laboratory in order to probe the impact of ram pressure on star formation as well as to constrain the characteristic timescales and velocities for the stripping of the non-thermal ISM.}
{We used high-resolution ($6.5\arcsec \approx 3\,\mathrm{kpc}$), multi-frequency ($144\,\mathrm{MHz} - 1.5\,\mathrm{GHz}$) radio continuum imaging of the Coma Cluster to resolve the low-frequency radio spectrum across the discs and tails of 25 ram pressure stripped galaxies. With resolved spectral index maps across these galaxy discs, we constrained the impact of ram pressure perturbations on galaxy star formation. We measured multi-frequency flux-density profiles along each of the ram pressure stripped tails in our sample. We then fit the resulting radio continuum spectra with a simple synchrotron aging model.}
{We showed that ram pressure stripped tails in Coma have steep ($-2 \lesssim \alpha \lesssim -1$) spectral indices. The discs of galaxies undergoing ram pressure stripping have integrated spectral indices within the expected range for shock acceleration from supernovae ($-0.8 \lesssim \alpha \lesssim -0.5$), though there is a tail towards flatter values. In a resolved sense, there are gradients in spectral index across the discs of ram pressure stripped galaxies in Coma. These gradients are aligned with the direction of the observed radio tails, with the flattest spectral indices being found on the `leading half'. From best-fit break frequencies we estimated the projected plasma velocities along the tail to be on the order of hundreds of kilometers per second, with the precise magnitude depending on the assumed magnetic field strength.}
{}


\maketitle
%

\section{Introduction} \label{sec:intro}

The Coma Cluster (Abell 1656) is the nearest, massive ($M_{200c} \gtrsim 10^{15}\,\mathrm{M_\odot}$, \citealt{ho2022}) galaxy cluster to the Milky Way, and is thus an invaluable laboratory for studying the effects of environmentally-driven galaxy evolution.  Galaxies in Coma are primarily passive, early-type galaxies but with a minority population of star-forming disc galaxies that are $\textsc{H\,i}$-deficient relative to non-cluster galaxies \citep[e.g.][]{shapley1934,godwin1983,gavazzi1987_coma,gavazzi1989,bravo-alfaro2000,terlevich2001,poggianti2004,mahajan2010,smith2012,molnar2022}. Among the star-forming galaxy population in Coma, multiple studies have found evidence for anomalously large star formation rates (SFRs), even compared to non-cluster galaxies \citep[e.g.][]{bothun1986,donas1995,miller2009_galaxy_pop,roberts2020,roberts2021_LOFARclust,boselli2022_review,roberts2022_lofar_manga}. At a distance of $\sim\!100\,\mathrm{Mpc}$, Coma is both small enough on the sky ($\sim$a few degrees across) to be efficiently mapped by survey telescopes with $\sim$degree-scale fields of view, but still near enough that $\sim$kiloparsec physical scales can be probed with $\sim$arcsecond angular resolution. The combination of these two facts have led to a number of resolved studies of environmental quenching in Coma, covering a relatively large number of member galaxies. These works suggest that perturbations from ram pressure stripping are likely a key factor driving the environmental quenching observed in Coma today \citep[e.g.][]{poggianti2004,yagi2007,smith2010,yagi2010,gavazzi2018,cramer2019,chen2020,roberts2020,cramer2021,roberts2021_LOFARclust,lal2022}.
\par
Ram pressure stripping \citep[e.g.][]{gunn1972} is a product of the relative motion between satellite galaxies and the intracluster medium (ICM), which drives an external pressure incident on the galaxy. The efficiency of this stripping will dictate the relevant timescale for quenching star formation. In the limit of a `weak' ram pressure, where only the circumgalactic medium (CGM) and diffuse atomic gas are removed, then the quenching timescale will be set by the depletion time of the surviving gas in the disc ($\sim\!1\,\mathrm{Gyr}$ timescales, e.g.\ \citealt{bigiel2011,saintonge2017}). This scenario is analogous to `galaxy starvation', a mechanism often cited to explain the quenching of galaxies in dense environments \citep[e.g.][]{larson1980,balogh2000,peng2015}. If ram pressure is strong enough to overcome not only the gravitational potential on the atomic gas, but also the denser, centrally concentrated molecular gas component, then the star-forming gas can be stripped directly and quenching will proceed on a much shorter timescale.
\par
Ram pressure may also alter the physical conditions of gas in star-forming regions, even if it is not directly removed from the disc. One possibility is that ram pressure compresses gas in the interstellar medium (ISM), both promoting the conversion between atomic and molecular gas as well as increasing the densities of existing molecular gas. Should this occur, then it would also predict enhanced SFRs in cluster galaxies, assuming a normal Kennicutt-Schmidt relation \citep{schmidt1959,kennicutt1998}. This framework has been used to explain large SFRs amongst Coma Cluster galaxies dating back four decades \citep{bothun1986}, and with the onset of recent large CO surveys of cluster galaxies, evidence for ram-pressure enhanced molecular gas densities and star formation in such environments has become more widespread \citep[e.g.][]{gavazzi2001,merluzzi2013,vulcani2018_sf,moretti2020,moretti2020b,vulcani2020,boselli2021_ic3476,cramer2021,durret2021,hess2022,lee2022_enhanced_sfr,roberts2022_lofar_manga,roberts2023_asym,brown2023,moretti2023}. We also note that enhanced SFRs may be a result of gas flows toward the galaxy centre, driven by ram pressure \citep{zhu2023}. Thus, in principle, ram pressure is sufficient to explain both the aforementioned large population of quenched galaxies in Coma and the tendency for many of the remaining star-forming members to show high SFRs.
\par
Cluster galaxies currently experiencing ram pressure stripping are generally identified in one of two ways: either by observing a one-sided tail of stripped material \citep[e.g.][]{gavazzi1978,shostak1982,gavazzi1987_a1367,gavazzi2001,kenney2004,chung2007,yagi2010,merluzzi2013,merluzzi2016,poggianti2017,boselli2018,roberts2021_LOFARclust,roberts2021_LOFARgrp,hess2022,venturi2022,edler2023}, or (often less reliably) by identifying morphological features in the stellar light of a galaxy that are consistent with perturbations from ram pressure \citep{mcpartland2016,poggianti2016,roberts2020,durret2021,roberts2022_UNIONS,vulcani2022}. Most commonly, ram pressure tails are observationally identified through the $\mathrm{H\alpha}$ Balmer line tracing ionized gas, the $21\,\mathrm{cm}$ hydrogen line tracing atomic gas, or the radio continuum tracing synchrotron emission from cosmic ray electrons (CREs), with the earliest examples of late-type galaxies undergoing ram pressure stripping coming from observations in the radio continuum \citep{gavazzi1978,gavazzi1987_a1367}.
\par
A primary asset of observing in the radio continuum is the large primary beam sizes, particularly at low ($\lesssim\!1\,\mathrm{GHz}$) frequencies, that allow for entire clusters at low redshift to be efficiently imaged and thus galaxies undergoing ram pressure stripping to be identified across the full cluster volume. In particular, the high sensitivity ($\sim\!100\,\mathrm{\mu Jy\,bm^{-1}}$) and angular resolution ($\sim\!6\arcsec$) of the Low Frequency Array (LOFAR) Two-metre Sky Survey (LoTSS, \citealt{vanhaarlem2013,shimwell2017,shimwell2019,shimwell2022}) has led to a roughly ten-fold increase in the known number of star-forming galaxies in low-redshift groups and clusters with stripped radio tails \citep{roberts2021_LOFARclust,roberts2021_LOFARgrp,roberts2022_perseus,edler2023,ignesti2023_a2255}. In addition to being clear signs of environmental galaxy evolution, ram pressure stripped radio tails also deposit CREs and magnetic fields into the ICM, which contribute to the seed particle population for diffuse radio halos and relics \citep{ge2019}.
\par
Radio continuum emission in star-forming galaxies is a mix of thermal free-free emission from \textsc{Hii} regions and non-thermal synchrotron emission originating from CREs accelerated by shocks from supernovae. At frequencies below $\sim\!1\,\mathrm{GHz}$, the radio spectrum is dominated by the non-thermal synchrotron component. For galaxies undergoing ram pressure stripping, this includes emission from stripped tails that is believed to be tracing CREs accelerated by supernovae in the disc that are advectively transported out of the galaxy under ram pressure \citep{murphy2009_virgo,vollmer2013,muller2021,roberts2021_LOFARclust,ignesti2022_gasp}. For different CRE acceleration mechanisms and efficiencies, the energy spectral index, $\gamma$, for supernovae is estimated to lie in the range from $-2$ to $-2.6$, which corresponds to a synchrotron spectral index, $\alpha$ between $-0.5$ and $-0.8$ ($\gamma = 2\alpha - 1$, e.g. \citealt{bell1978,bogdan1983,biermann1993}). After CREs are injected into the ISM they are then subject to energy loss mechanisms which alter $\alpha$ in a frequency-dependent fashion. Examples of energy loss mechanisms include ionization losses, synchrotron, and inverse-Compton losses. Synchrotron and inverse-Compton losses most strongly impact high-energy CREs.  This acts to steepen $\alpha$ overall, but in particular, introduces a high-frequency exponential cut-off to the (previously) power-law spectrum, occurring at the so-called `break frequency' ($\nu_b$). If synchrotron and inverse-Compton losses are dominant, then over time the break frequency will evolve to lower frequency and the radiative age of a given plasma can be estimated from the location of this break \citep[e.g.][]{miley1980}. In the context of star formation, this means that young star-forming regions will have spectral indices which are close the injection values ($-0.8 \lesssim \alpha \lesssim -0.5$), whereas older plasma corresponding to a prior episode of star formation will be characterized by a steeper $\alpha$. In the context of ram pressure stripping, radiative age estimates have been used to constrain the bulk velocity of the stripped plasma for some cluster galaxies, generally finding speeds on the order of hundreds of kilometers per second \citep{vollmer2021,ignesti2023_a2255}.
\par
Conversely, ionization losses flatten the spectral index by preferentially affecting low-energy CREs and thus the low-frequency portion of the spectrum. The strength of ionization losses increases with ISM density \citep[e.g.][]{longair2011}, a fact which has been used to explain the flatter-than-injection spectral indices (i.e. $\alpha > -0.5$) that are sometimes observed near strongly star-forming regions in galaxies \citep{basu2015}. There is also tentative evidence for unusually flat spectral indices in the discs of galaxies undergoing ram pressure stripping \citep{ignesti2022_meerkat,roberts2022_perseus}, for which ionization losses provide a potential explanation given the that compression from ram pressure is capable of increasing local ISM densities \citep[e.g.][]{cramer2020,troncoso-iribarren2020,moretti2020,moretti2020b,roberts2022_lofar_manga,roberts2023_asym,brown2023}.
\par
A number of wide-area radio continuum surveys of the Coma Cluster have been published, spanning a decade across the low-frequency spectrum, including: LOFAR at $144\,\mathrm{MHz}$ (\citealt{shimwell2022}, also see \citealt{bonafede2022}), the upgraded Giant Metrewave Telescope (uGMRT) at $400\,\mathrm{MHz}$ and $700\,\mathrm{MHz}$ \citep{lal2020,lal2022}, and the Very Large Array (VLA) at $1.5\,\mathrm{GHz}$ \citep{miller2009, chen2020}. In this work we synergize data from these surveys in order to complete an in-depth study of the low-frequency radio spectrum of ram pressure stripped galaxies in Coma and their stripped tails.  With a working physical resolution of $\sim\!3\,\mathrm{kpc}$, and excellent sensitivity to extended, diffuse emission, we are able to probe spectral properties both across the star-forming discs and along the ram pressure stripped tails. From these datasets we derive constraints on ram-pressure induced star formation, cosmic-ray losses in the ISM, and the characteristic timescales for gas stripping.
\par
The structure of this paper is as follows: In Sect.~\ref{sec:data} we describe the multi-frequency radio continuum imaging used in this work, as well as our sample of ram pressure stripped galaxies. In Sect.~\ref{sec:int_spec_index} we report integrated spectral index measurements over both galaxy discs and stripped tails. In Sect.~\ref{sec:galaxy_specmaps} we show resolved maps of spectral index covering the galaxy discs, primarily between $144\,\mathrm{MHz}$ and $1.5\,\mathrm{GHz}$. In Sect.~\ref{sec:spec_aging} we measure multi-frequency flux-density profiles along observed stripped tails and apply a simple spectral aging model to these observed radio spectra. From this model we extract estimates for synchrotron break frequencies and (projected) bulk stripping velocities along the tails. Finally, in Sects.~\ref{sec:discussion} and \ref{sec:summary} we provide a brief discussion and high-level summary of this work.
\par
Throughout we assume a standard $\Lambda$CDM cosmology with $\Omega_M = 0.3$, $\Omega_\Lambda = 0.7$, and $H_0 = 70\,\mathrm{km\,s^{-1}\,Mpc^{-1}}$. We take the redshift of the Coma Cluster to be $z_\mathrm{Coma} = 0.0234$ \citep{rines2016}, which corresponds to a luminosity distance of $102\,\mathrm{Mpc}$ and a scale of $0.47\,\mathrm{kpc} / \arcsec$. The radio synchrotron spectrum is described as $S_\nu \propto \nu^\alpha$, where $S_\nu$ is the flux density, $\nu$ is the frequency, and $\alpha$ is the spectral index.

\section{Data and galaxy sample} \label{sec:data}

\subsection{Radio imaging}

\subsubsection{LOFAR $144\,\mathrm{MHz}$} \label{sec:data_lofar}

The LOFAR images in this work are from the LOFAR Two-metre Sky Survey (LoTSS, \citealt{shimwell2017,shimwell2022}), a wide-area continuum survey of the northern sky between $120\,\mathrm{MHz}$ and $168\,\mathrm{MHz}$ using the LOFAR high-band antenna (HBA).  We use three $8\,\mathrm{h}$ LoTSS pointings (P192+27, P195+27, P195+30) that combine to map the Coma cluster out to the virial radius.  Each pointing is processed, imaged, and mosiaced following the procedure outlined in \citet{shimwell2022}. This includes initial processing to correct for direction-independent errors, followed by direction-dependent\footnote{\url{https://github.com/mhardcastle/ddf-pipeline}} calibration to account for the variation in ionospheric effects across the large (primary-beam FWHM: $\sim\!4\deg$) field-of-view.  Direction-dependent solutions are applied during imaging with DDFacet \citep{tasse2018} and images for each pointing are mosaicked with neighbouring pointings in order to produce the final, mosaicked image for each field \citep{shimwell2022}.  For each field we checked the LoTSS flux-density scale against both the TGSS \citep{intema2009} and 7C \citep{hales2007} surveys. For all three fields we measured a $\sim\!20\%$ flux-scale offset in LoTSS, both when comparing to TGSS and to 7C.  When directly comparing TGSS and 7C, no flux-scale differences were measured. Thus we applied a $\sim\!20\%$ flux-scale correction (varying slightly from field to field) to each of the three LoTSS mosaics. The final LoTSS images for P192+27, P195+27, and P195+30 reach rms levels at the field centres of 70, 65, 60$\,\mathrm{\mu Jy}$ per $6\arcsec$ beam, respectively. For all frequency bands used in this analysis we work at a common angular resolution of $6.5\arcsec$, corresponding to a physical resolution of $\sim\!3\,\mathrm{kpc}$ at the distance of the Coma Cluster. We convolve the LoTSS images to this working resolution using the \texttt{imsmooth} function in \texttt{CASA} \citep{casa2022}.

\subsubsection{uGMRT $400$ and $700\,\mathrm{MHz}$}

At 400 and $700\,\mathrm{MHz}$ we use imaging from uGMRT of the Coma Cluster from \citet{lal2020} and \citet{lal2022}. This imaging covers an area of $\sim 4\deg$ focused on the central and south-west regions of Coma, which is made up of three pointings in Band 3 ($250 - 550\,\mathrm{MHz}$) and eight pointings in Band 4 ($550 - 850\,\mathrm{MHz}$). The on-source times range between $1.8\,\mathrm{h}$ and $4.7\,\mathrm{h}$ for pointings in Band 3 and $1.3\,\mathrm{h}$ and $3.4\,\mathrm{h}$ for pointings in Band 4. We refer to \citet{lal2020} and \citet{lal2022} for all further details on the observing set-up.
\par
To process the raw data we used the Source Peeling and Atmospheric Modeling pipeline (\texttt{SPAM}, \citealt{intema2009}). We followed the recommended workflow for wide-band uGMRT data\footnote{\url{http://www.intema.nl/doku.php?id=huibintemaspampipeline}}, which first splits the dataset into $\sim\!50\,\mathrm{MHz}$ sub-bands that are each processed individually by the main \texttt{SPAM} pipeline. After processing, sub-bands were imaged together with \texttt{WSClean} \citep{offringa2014} to form a final wide-band image. For final imaging we use Briggs weighting \citep{briggs1995} with a robust parameter of $-0.5$. This results in rms levels between 26 and $30\,\mathrm{\mu Jy}$ (depending on the pointing) per $\sim\!5.5\arcsec$ beam in Band 3, and rms levels between 13 and $26\,\mathrm{\mu Jy}$ per $\sim\!3.5\arcsec$ beam in Band 4. All Band 3 and Band 4 pointings are convolved to our working resolution of $6.5\arcsec$, reprojected onto a common pixel grid, and then mosaicked together to produce a single mosaic image for each frequency band.  The largest angular scale for the uGMRT is $\sim\!33\arcmin$ in Band 3 and $\sim\!20\arcmin$ in Band 4, both of which are well larger than the angular size of $2\arcmin$ at $144\,\mathrm{MHz}$ for the largest source (NGC4848) in our sample.  

\subsubsection{VLA $1.5\,\mathrm{GHz}$} \label{sec:data_vla}

For our highest frequency data we used two different VLA $1.5\,\mathrm{GHz}$ data-sets, from \citet{chen2020} and \citet{miller2009}, which together cover more than a square degree of the Coma Cluster. Throughout this work we assumed that the observed flux density at $1.5\,\mathrm{GHz}$ is solely due to nonthermal, synchrotron emission -- in other words, a `thermal fraction' ($f_\mathrm{th}$) of zero. While this is unlikely to be strictly true, previous works have found typical thermal fractions of $f_\mathrm{th} \lesssim 10\%$ at $1.5\,\mathrm{GHz}$ for nearby star-forming galaxies \citep[e.g.][]{condon1992,niklas1997,tabatabaei2017,ignesti2022_meerkat}.
\par
The observations from \citet{chen2020} include one pointing centred on D100 and one pointing centred on NGC4848, each well-known examples of ram pressure stripped galaxies in the Coma Cluster. Here we briefly highlight some main features of the data and processing and we refer to \citet{chen2020} for a more comprehensive discussion.
\par
The NGC4848 and D100 fields were observed for an on-source time of $12.1\,\mathrm{h}$ and $6.3\,\mathrm{h}$, respectively, in the VLA B-configuration. Data were processed following standard procedures in \texttt{CASA}.  After calibration the continuum data was imaged using \texttt{tclean} with Briggs weighting and a robust parameter of 0.5, giving a synthesized beam size of $\sim\! 4\arcsec$ for both fields. At the pointing centre the resulting rms noise is $5.9\,\mathrm{\mu Jy\,bm^{-1}}$ for the NGC4848 field and $8.0\,\mathrm{\mu Jy\,bm^{-1}}$ for the D100 field.  Both pointings were then convolved to our working resolution of $6.5\arcsec$.
\par
The observations from \citet{miller2009} also include two fields, one covering the centre of the Coma Cluster (Coma 1, which has substantial overlap with the D100 field from \citealt{chen2020}) and one to the southwest of the cluster centre (Coma 3).  Again, here we only provide a brief description of the data processing, but a detailed outline is available in \citet{miller2009}.
\par
For each field, 11 separate pointings were obtained in order to achieve roughly uniform sensitivity across a $30\arcmin \times 50\arcmin$ area. Each pointing was observed for $\sim\!1\,\mathrm{h}$ in the VLA B-configuration. The data were reduced and imaged with the Astronomical Image Processing Software (AIPS). Final images for each field have a restoring beam size of $4.4\arcsec$ and reach rms levels of $20 - 25\,\mathrm{\mu Jy}$ at the image centre. We obtained the Coma 1 and Coma 3 images from the CDS \footnote{\url{https://cdsarc.cds.unistra.fr/viz-bin/cat/J/AJ/137/4436}} and both were then smoothed to our working resolution of $6.5\arcsec$.
\par
Lastly, we re-projected all four VLA images from \citet{miller2009} and \citet{chen2020} (after smoothing to a $6.5\arcsec$ beam) onto a common pixel grid and mosaicked them together in the image plane.  The image combination was weighted by the inverse of the local rms level in each image. Where the \citet{miller2009} and \citet{chen2020} images overlap, the majority of the weight in the combined mosaic is thus given to the deeper \citet{chen2020} data. In most cases the \citet{miller2009} images are not deep enough to detect the stripped tails, however these data still provide important insight into the spectral properties over the discs of galaxies not covered by the \citet{chen2020} observations. We note that the largest angular scale for the VLA in B-configuration is $120\arcsec$ for full synthesis observations\footnote{\url{https://science.nrao.edu/facilities/vla/docs/manuals/propvla}}. This is nearly identical to the angular size of the largest source in our sample at $144\,\mathrm{MHz}$, therefore for the longest tails in our sample (e.g.\ NGC4848, MRK0057) it is possible that there is a small amount of missing flux density at $1.5\,\mathrm{GHz}$. That said, for an aging synchrotron population, it is expected that the observed tail length will scale inversely with the square root of the frequency \citep{ignesti2022_meerkat} and therefore we expect that the tails lengths at $1.5\,\mathrm{GHz}$ will be intrinsically shorter than at $144\,\mathrm{MHz}$ (by roughly a factor of three). This would bring the expected tail lengths at $1.5\,\mathrm{GHz}$ within the B-configuration largest angular scale.

\begin{figure}
    \centering
    \includegraphics[width = \columnwidth]{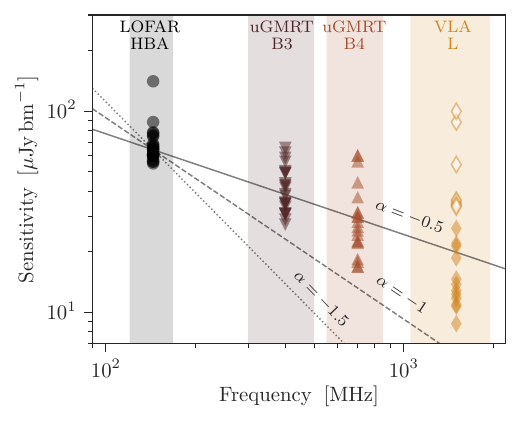}
    \caption{One-sigma sensitivity (rms) as a function of frequency for each of the observing bands in this work. All values are measured from images at our working resolution of $6.5\arcsec$. Data points correspond to the local rms measured around each galaxy. For the L-band, the filled markers correspond to galaxies that are covered by the \citet{chen2020} observations and the open markers correspond to galaxies only covered by the \citet{miller2009} observations. Shaded bands correspond to the width of each observing band. For reference we plot line corresponding to spectral indices of $-0.5$, $-1$, and $-1.5$ anchored to the lowest frequency band.}
    \label{fig:rms}
\end{figure}

\subsection{Ram pressure stripped galaxy sample} \label{sec:galaxy_sample}

\begin{table*}[!ht]
\scriptsize
\centering
\begin{threeparttable}
\caption{Galaxy sample}
\label{tab:galaxy_table}
\begin{tabular}{l c c c c c c c c c c c}
\toprule
\toprule
Galaxy & GMP & R.A. & Decl. & $z$ & $\log\,M_\bigstar$ & $\log\,\mathrm{SFR}$ & $r_{90}$ & $144\,\mathrm{MHz}$ & $400\,\mathrm{MHz}$ & $700\,\mathrm{MHz}$ & $1.4\,\mathrm{GHz}$ \\
&& $\deg$ & $\deg$ && $\log(\mathrm{M_\odot})$ & $\log(\mathrm{M_\odot\,yr^{-1}})$ & $\arcsec$ &&&& \\
(1) & (2) & (3) & (4) & (5) & (6) & (7) & (8) & (9) & (10) & (11) \\
\midrule
NGC4848 & 4471 & 194.5235 & 28.2427 & 0.0240 & $10.77 \pm 0.03$ & $0.99 \pm 0.02$ & 24.7 & \checkmark & \checkmark & \checkmark & \checkmark \\
NGC4853 & 4156 & 194.6466 & 27.5964 & 0.0256 & $10.81 \pm 0.03$ & $0.41 \pm 0.09$ & 10.2 & \checkmark & \checkmark & \checkmark & \checkmark \\
NGC4858 & 3816 & 194.7585 & 28.1158 & 0.0314 & $10.16 \pm 0.05$ & $0.56 \pm 0.06$ & 10.1 & \checkmark & \checkmark & \checkmark & \checkmark \\
NGC4911 & 2374 & 195.2337 & 27.7908 & 0.0266 & $11.21 \pm 0.04$ & $0.05 \pm 0.51$ & 34.1 & \checkmark & \checkmark & \checkmark & \checkmark \\
IC3913 & 5422 & 194.1191 & 27.2913 & 0.0251 & $9.97 \pm 0.04$ & $-0.12 \pm 0.07$ & 22.4 & \checkmark & \checkmark & \checkmark & \checkmark \\
IC3949 & 3896 & 194.7334 & 27.8334 & 0.0251 & $10.60 \pm 0.02$ & $-0.16 \pm 0.06$ & 20.3 & \checkmark & \checkmark & \checkmark & \checkmark \\
IC4040 & 2559 & 195.1580 & 28.0574 & 0.0255 & $10.28 \pm 0.07$ & $0.64 \pm 0.07$ & 15.1 & \checkmark & \checkmark & \checkmark & \checkmark \\
KUG1250+276 & --- & 193.2038 & 27.4018 & 0.0258 & $9.75 \pm 0.06$ & $0.14 \pm 0.07$ & 10.6 & \checkmark & \checkmark & \checkmark & $\times$ \\
KUG1255+275 & 4351 & 194.5776 & 27.3108 & 0.0247 & $9.70 \pm 0.07$ & $-0.09 \pm 0.05$ & 9.1 & \checkmark & \checkmark & \checkmark & \checkmark \\
KUG1255+283 & 4555 & 194.4905 & 28.0618 & 0.0271 & $9.92 \pm 0.06$ & $0.07 \pm 0.11$ & 17.8 & \checkmark & \checkmark & \checkmark & \checkmark \\
KUG1256+287B & 3509 & 194.8465 & 28.4887 & 0.0234 & $9.57 \pm 0.04$ & $-0.50 \pm 0.06$ & 9.0 & \checkmark & \checkmark & \checkmark & \checkmark \\
KUG1257+278 & 3271 & 194.9159 & 27.5765 & 0.0167 & $9.07 \pm 0.06$ & $-0.66 \pm 0.07$ & 11.6 & \checkmark & \checkmark & \checkmark & \checkmark \\
KUG1257+288B & 3253 & 194.9172 & 28.6308 & 0.0178 & $9.37 \pm 0.07$ & $-0.32 \pm 0.03$ & 16.7 & \checkmark & \checkmark & $\times$ & $\times$ \\
KUG1258+287 & 2536 & 195.1691 & 28.5190 & 0.0298 & --- & --- & 13.1 & \checkmark & \checkmark & $\times$ & \checkmark \\
KUG1258+279A & 2599 & 195.1406 & 27.6377 & 0.0250 & $9.78 \pm 0.03$ & $0.06 \pm 0.01$ & 17.8 & \checkmark & \checkmark & \checkmark & \checkmark \\
KUG1259+279 & 1616 & 195.5329 & 27.6483 & 0.0230 & $10.23 \pm 0.06$ & $0.42 \pm 0.15$ & 15.8 & \checkmark & \checkmark & $\times$ & $\times$ \\
KUG1259+284 & 1576 & 195.5532 & 28.2149 & 0.0273 & $9.96 \pm 0.07$ & $0.18 \pm 0.09$ & 11.3 & \checkmark & \checkmark & $\times$ & $\times$ \\
MRK0053 & 5643 & 194.0254 & 27.6781 & 0.0164 & $9.20 \pm 0.08$ & $0.04 \pm 0.08$ & 9.1 & \checkmark & \checkmark & \checkmark & \checkmark \\
MRK0056 & 4159 & 194.6472 & 27.2649 & 0.0245 & $9.77 \pm 0.05$ & $0.24 \pm 0.05$ & 14.1 & \checkmark & \checkmark & \checkmark & \checkmark \\
MRK0057 & 4135 & 194.6554 & 27.1763 & 0.0256 & $9.84 \pm 0.07$ & $0.25 \pm 0.07$ & 10.2 & \checkmark & \checkmark & \checkmark & \checkmark \\
MRK0058 & 3779 & 194.7720 & 27.6446 & 0.0181 & $9.75 \pm 0.05$ & $-0.04 \pm 0.07$ & 18.5 & \checkmark & \checkmark & \checkmark & \checkmark \\
D100 & 2910 & 195.0382 & 27.8666 & 0.0177 & $9.26 \pm 0.06$ & $-0.08 \pm 0.04$ & 7.8 & \checkmark & \checkmark & \checkmark & \checkmark \\
GMP2601 & 2601 & 195.1398 & 27.5042 & 0.0186 & $8.96 \pm 0.05$ & $-0.542 \pm 0.03$ & 12.2 & \checkmark & \checkmark & \checkmark & \checkmark \\
GMP3618 & 3618 & 194.8195 & 27.1061 & 0.0280 & $10.14 \pm 0.02$ & $0.09 \pm 0.09$ & 6.3 & \checkmark & \checkmark & $\times$ & \checkmark \\
GMP5226 & 5226 & 194.2132 & 26.8989 & 0.0208 & $10.37 \pm 0.02$ & $0.01 \pm 0.07$ & 16.1 & \checkmark & \checkmark & \checkmark & \checkmark \\
\bottomrule
\end{tabular}
\begin{tablenotes}[flushleft]
    \item \textbf{Notes:} Table columns are: (1) galaxy name, (2) galaxy number from the Coma GMP catalog \citep{godwin1983}, (3) right ascension (J2000), (4) declination (J2000), (5) galaxy redshift, (6) galaxy stellar mass from \citet{salim2016,salim2018}, (7) galaxy star formation rate from \citet{salim2016,salim2018}, (8) $r$-band Petrosian 90\% light radius, (9) -- (12) frequency coverage.  We do not list a stellar mass or star formation rate for KUG1258+287 as it does not have an SDSS spectrum and thus is not in the \citet{salim2016,salim2018} catalogue.
\end{tablenotes}
\end{threeparttable}
\end{table*}

\citet{roberts2021_LOFARclust} and \citet{roberts2021_LOFARgrp} published a large sample of $\sim\!150$ galaxies with one-sided radio tails at $144\,\mathrm{MHz}$ that are likely undergoing ram pressure stripping in low-$z$ groups and clusters. This sample includes 29 galaxies in the Coma Cluster.  The galaxy sample used in this work includes all ram pressure stripped Coma galaxies from \citet{roberts2021_LOFARclust} that are also within the observing area of at least one of the $400\,\mathrm{MHz}$, $700\,\mathrm{MHz}$, and $1.5\,\mathrm{GHz}$ datasets described above. This restriction results in 25 galaxies with observations at at least two frequencies.
\par
We also included two supplemental ram pressure stripped galaxies from Coma in our sample. These two galaxies, NGC4911 and KUG1258+287, were not part of the \citet{roberts2021_LOFARclust} sample for technical reasons, but do show one-sided stripped tails at $144\,\mathrm{MHz}$ (and other frequencies).  NGC4911 is known to have an $\mathrm{H\alpha}$ stripped tail identified by \citet{yagi2010}, and also shows a clear one-sided tail in LoTSS at $144\,\mathrm{MHz}$ (see Fig.~\ref{fig:example_imgs_NGC4911}) in the same direction as the $\mathrm{H\alpha}$ tail.  The reason that NGC4911 was not included in the \citet{roberts2021_LOFARclust} sample, despite the radio continuum tail, is that NGC4911 has a specific star formation rate ($\mathrm{sSFR} = \mathrm{SFR} / M_\bigstar = 10^{-11.2}\,\mathrm{yr^{-1}}$) that falls just below the star-forming galaxy selection of $\mathrm{sSFR} > 10^{-11}\,\mathrm{yr^{-1}}$ imposed by \citet{roberts2021_LOFARclust}. However, NGC4911 clearly shows signatures of ongoing star formation \citep{gregg2003,yagi2010} and we opt to include it in our sample for completeness.
\par
KUG1258+287 has a one-sided, $\sim\!40\,\mathrm{kpc}$ tail at $144\,\mathrm{MHz}$ in LoTSS (see Fig.~\ref{fig:example_imgs_KUG1258+287}) but lacks an SDSS redshift, presumably due to its close proximity to a bright foreground star. The \citet{roberts2021_LOFARclust} sample relies on SDSS spectroscopic redshifts and therefore does not include KUG1258+287.  However, KUG1258+287 does have a spectroscopic redshift of $z = 0.0298$ from \citet{edwards2011}, and thus has a projected velocity offset of $\sim\!1700\,\mathrm{km\,s^{-1}}$ from the systemic velocity of Coma. The velocity offset, alongside a projected cluster-centric radius of $0.34\,\mathrm{R_{180}}$, places KUG1258+287 as a Coma member galaxy according to the criteria of \citet{roberts2021_LOFARclust}, and therefore we include it in our ram pressure galaxy sample.
\par
This results in a total ram pressure stripping galaxy sample of 25 galaxies.  We list these galaxies along with their general properties and frequency coverage in Table~\ref{tab:galaxy_table}.  In Fig.~\ref{fig:rms} we show the local rms measured from $\sim\!3\arcmin$ cutout images centred on each galaxy, for each of our frequency bands.  Generally speaking, we reach comparable sensitivity to freshly injected electrons in all frequency bands ($\alpha \approx -0.7$), but in most cases do not reach equivalent sensitivity in our high-frequency bands to steep-spectrum emission ($\alpha \lesssim -1$) detected by the LOFAR HBA.  The points skewed to high rms at $1.5\,\mathrm{GHz}$ in Fig.~\ref{fig:rms} correspond to those galaxies that are covered by the L-band observations from \citet{miller2009} but not from \citet{chen2020}.

\section{Integrated spectral properties: galaxy and tail} \label{sec:int_spec_index}

\begin{figure}
    \centering
    \includegraphics[width = \columnwidth]{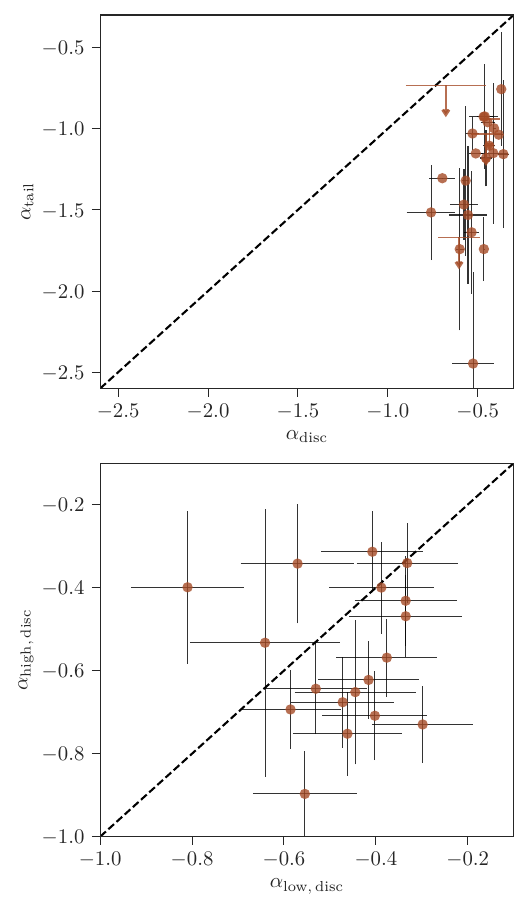}
    \caption{\textit{Top:} Tail spectral index versus galaxy disc spectral index. When emission is detected in three or more frequency bands, the spectral index and its error are determined from orthogonal distance regression fits.  When emission is only detected in two frequency bands, the spectral index and its error are determined via the direct method (Eq.~\ref{eq:spec_index_direct}). For galaxies where emission in the tail is only detected at one frequency (see Table.~\ref{tab:flux_table}), $\alpha_\mathrm{tail}$ is shown as a $3\sigma$ upper limit.  \textit{Bottom:} Radio colour-colour diagram for the galaxy discs.  The high-frequency spectral index ($y$-axis) is measured between $700\,\mathrm{MHz}$ and $1.5\,\mathrm{GHz}$ and the low-frequency spectral index ($x$-axis) is measured between $144\,\mathrm{MHz}$ and $400\,\mathrm{MHz}$.  In both the top and bottom panels the dashed lines correspond to the one-to-one relation.}
    \label{fig:int_specindex}
\end{figure}

\begin{table*}[!ht]
\footnotesize
\centering
\begin{threeparttable}
\caption{Flux density and spectral index measurements}
\label{tab:flux_table}
\begin{tabular}{l c c c c c c c}
\toprule
\toprule
Galaxy & Component & \multicolumn{4}{c}{$S_\nu \;\; \mathrm{[mJy]}$} & $\alpha_\mathrm{disc}$ & $\alpha_\mathrm{tail}$ \\[0.25em]
&& $\mathrm{144\,MHz}$ & $\mathrm{400\,MHz}$ & $\mathrm{700\,MHz}$ & $\mathrm{1.4\,GHz}$ && \\
\midrule
\multirow{2}{*}{NGC4848} & Disc & $69.4 \pm 7.0$ & $47.3 \pm 4.7$ & $36.5 \pm 3.7$ & $23.6 \pm 2.4$ & \multirow{2}{*}{$-0.45 \pm 0.03$} & \multirow{2}{*}{$-1.2 \pm 0.2$} \\
& Tail & $28.0 \pm 2.9$ & $5.1 \pm 0.7$ & $3.3 \pm 0.7$ & $2.2 \pm 0.2$ &&\\[2pt]
\multirow{2}{*}{NGC4853} & Disc & $10.3 \pm 1.0$ & $6.0 \pm 0.6$ & $4.4 \pm 0.4$ & $2.7 \pm 0.3$ & \multirow{2}{*}{$-0.57 \pm 0.02$} & \multirow{2}{*}{$-1.3 \pm 0.5$} \\
& Tail & $2.2 \pm 0.3$ & $0.3 \pm 0.1$ & $0.4 \pm 0.1$ & $<0.3$ &&\\[2pt]
\multirow{2}{*}{NGC4858} & Disc & $22.7 \pm 2.3$ & $16.7 \pm 1.7$ & $13.0 \pm 1.3$ & $7.5 \pm 0.7$ & \multirow{2}{*}{$-0.45 \pm 0.07$} & \multirow{2}{*}{$-0.9 \pm 0.1$} \\
& Tail & $16.9 \pm 1.7$ & $5.7 \pm 0.6$ & $3.7 \pm 0.4$ & $2.1 \pm 0.1$ &&\\[2pt]
\multirow{2}{*}{NGC4911} & Disc & $93.0 \pm 9.4$ & $51.1 \pm 5.1$ & $37.7 \pm 3.8$ & $22.2 \pm 2.2$ & \multirow{2}{*}{$-0.60 \pm 0.02$} & \multirow{2}{*}{$-1.7 \pm 0.5$} \\
& Tail & $9.5 \pm 1.1$ & $0.6 \pm 0.3$ & $<0.6$ & $0.3 \pm 0.1$ &&\\[2pt]
\multirow{2}{*}{IC3913} & Disc & $11.9 \pm 1.3$ & $5.2 \pm 0.6$ & $3.6 \pm 0.4$ & $2.6 \pm 0.4$ & \multirow{2}{*}{$-0.70 \pm 0.07$} & \multirow{2}{*}{$-1.3 \pm 0.0$} \\
& Tail & $3.0 \pm 0.4$ & $0.8 \pm 0.1$ & $0.4 \pm 0.1$ & $<0.3$ &&\\[2pt]
\multirow{2}{*}{IC3949} & Disc & $5.4 \pm 0.6$ & $3.4 \pm 0.3$ & $3.2 \pm 0.3$ & $1.8 \pm 0.2$ & \multirow{2}{*}{$-0.44 \pm 0.07$} & \multirow{2}{*}{$<-1.0$} \\
& Tail & $1.4 \pm 0.3$ & $<0.3$ & $<0.3$ & $<0.1$ &&\\[2pt]
\multirow{2}{*}{IC4040} & Disc & $56.3 \pm 5.7$ & $36.8 \pm 3.7$ & $27.0 \pm 2.7$ & $16.8 \pm 1.7$ & \multirow{2}{*}{$-0.51 \pm 0.04$} & \multirow{2}{*}{$-1.2 \pm 0.1$} \\
& Tail & $14.3 \pm 1.6$ & $6.0 \pm 0.7$ & $2.2 \pm 0.3$ & $1.0 \pm 0.1$ &&\\[2pt]
\multirow{2}{*}{KUG1250+276} & Disc & $5.2 \pm 0.6$ & $3.5 \pm 0.4$ & $2.2 \pm 0.2$ & --- & \multirow{2}{*}{$-0.54 \pm 0.10$} & \multirow{2}{*}{$-1.5 \pm 0.4$} \\
& Tail & $1.7 \pm 0.3$ & $0.4 \pm 0.1$ & $<0.3$ & --- &&\\[2pt]
\multirow{2}{*}{KUG1255+275} & Disc & $8.3 \pm 0.8$ & $5.9 \pm 0.6$ & $4.5 \pm 0.4$ & $3.4 \pm 0.4$ & \multirow{2}{*}{$-0.38 \pm 0.02$} & \multirow{2}{*}{$-1.0 \pm 0.1$} \\
& Tail & $11.4 \pm 1.2$ & $3.6 \pm 0.4$ & $2.4 \pm 0.3$ & $1.0 \pm 0.3$ &&\\[2pt]
\multirow{2}{*}{KUG1255+283} & Disc & $27.6 \pm 2.8$ & $17.0 \pm 1.7$ & $13.0 \pm 1.4$ & $7.8 \pm 0.8$ & \multirow{2}{*}{$-0.53 \pm 0.03$} & \multirow{2}{*}{$-1.0 \pm 0.1$} \\
& Tail & $5.8 \pm 0.7$ & $1.7 \pm 0.3$ & $<1.2$ & $0.6 \pm 0.1$ &&\\[2pt]
\multirow{2}{*}{KUG1256+287B} & Disc & $1.6 \pm 0.2$ & $0.5 \pm 0.1$ & $0.2 \pm 0.1$ & $0.4 \pm 0.1$ & \multirow{2}{*}{$-0.69 \pm 0.24$} & \multirow{2}{*}{$<-0.7$} \\
& Tail & $0.8 \pm 0.1$ & $<0.3$ & $<0.3$ & $<0.2$ &&\\[2pt]
\multirow{2}{*}{KUG1257+278} & Disc & $3.4 \pm 0.4$ & $2.2 \pm 0.2$ & $1.8 \pm 0.2$ & $1.1 \pm 0.1$ & \multirow{2}{*}{$-0.46 \pm 0.04$} & \multirow{2}{*}{$-0.9 \pm 0.3$} \\
& Tail & $1.0 \pm 0.2$ & $0.2 \pm 0.1$ & $<0.3$ & $0.1 \pm 0.1$ &&\\[2pt]
\multirow{2}{*}{KUG1257+288B} & Disc & $8.9 \pm 0.9$ & $4.1 \pm 0.5$ & --- & --- & \multirow{2}{*}{$-0.76 \pm 0.16$} & \multirow{2}{*}{$-1.9 \pm 0.4$} \\
& Tail & $7.4 \pm 0.8$ & $1.1 \pm 0.4$ & --- & --- &&\\[2pt]
\multirow{2}{*}{KUG1258+287} & Disc & $9.8 \pm 1.0$ & $6.1 \pm 0.6$ & --- & $3.3 \pm 0.4$ & \multirow{2}{*}{$-0.47 \pm 0.00$} & \multirow{2}{*}{$-1.7 \pm 0.2$} \\
& Tail & $16.9 \pm 1.7$ & $2.9 \pm 0.5$ & --- & $<2.1$ &&\\[2pt]
\multirow{2}{*}{KUG1258+279A} & Disc & $12.2 \pm 1.3$ & $8.1 \pm 0.8$ & $7.7 \pm 0.8$ & $4.5 \pm 0.5$ & \multirow{2}{*}{$-0.40 \pm 0.07$} & \multirow{2}{*}{$-1.1 \pm 0.3$} \\
& Tail & $10.9 \pm 1.2$ & $3.3 \pm 0.4$ & $0.7 \pm 0.3$ & $1.2 \pm 0.1$ &&\\[2pt]
\multirow{2}{*}{KUG1259+279} & Disc & $12.0 \pm 1.2$ & $6.5 \pm 0.7$ & --- & --- & \multirow{2}{*}{$-0.60 \pm 0.14$} & \multirow{2}{*}{$<-1.7$} \\
& Tail & $4.8 \pm 0.6$ & $<0.9$ & --- & --- &&\\[2pt]
\multirow{2}{*}{KUG1259+284} & Disc & $10.0 \pm 1.0$ & $5.8 \pm 0.6$ & --- & --- & \multirow{2}{*}{$-0.53 \pm 0.14$} & \multirow{2}{*}{$-2.4 \pm 0.6$} \\
& Tail & $6.1 \pm 0.7$ & $0.5 \pm 0.3$ & --- & --- &&\\[2pt]
\multirow{2}{*}{MRK53} & Disc & $12.3 \pm 1.2$ & $8.7 \pm 0.9$ & $6.2 \pm 0.6$ & $4.5 \pm 0.5$ & \multirow{2}{*}{$-0.44 \pm 0.03$} & \multirow{2}{*}{$-1.0 \pm 0.1$} \\
& Tail & $13.2 \pm 1.4$ & $5.7 \pm 0.6$ & $2.8 \pm 0.3$ & $<1.5$ &&\\[2pt]
\multirow{2}{*}{MRK56} & Disc & $10.8 \pm 1.1$ & $7.3 \pm 0.7$ & $5.7 \pm 0.6$ & $4.2 \pm 0.4$ & \multirow{2}{*}{$-0.41 \pm 0.01$} & \multirow{2}{*}{$-1.0 \pm 0.1$} \\
& Tail & $6.6 \pm 0.7$ & $2.7 \pm 0.3$ & $1.3 \pm 0.2$ & $<0.6$ &&\\[2pt]
\multirow{2}{*}{MRK57} & Disc & $12.2 \pm 1.2$ & $8.0 \pm 0.8$ & $5.8 \pm 0.6$ & $4.6 \pm 0.5$ & \multirow{2}{*}{$-0.43 \pm 0.03$} & \multirow{2}{*}{$-1.2 \pm 0.1$} \\
& Tail & $20.4 \pm 2.1$ & $7.2 \pm 0.8$ & $2.5 \pm 0.3$ & $1.5 \pm 0.3$ &&\\[2pt]
\multirow{2}{*}{MRK58} & Disc & $17.7 \pm 1.8$ & $10.1 \pm 1.0$ & $8.5 \pm 0.9$ & $4.3 \pm 0.4$ & \multirow{2}{*}{$-0.57 \pm 0.07$} & \multirow{2}{*}{$-1.5 \pm 0.2$} \\
& Tail & $5.4 \pm 0.6$ & $1.0 \pm 0.2$ & $0.3 \pm 0.1$ & $0.3 \pm 0.1$ &&\\[2pt]
\multirow{2}{*}{MRK60} & Disc & $2.7 \pm 0.3$ & $1.9 \pm 0.2$ & $1.6 \pm 0.2$ & $1.2 \pm 0.1$ & \multirow{2}{*}{$-0.35 \pm 0.03$} & \multirow{2}{*}{$-1.2 \pm 0.5$} \\
& Tail & $2.5 \pm 0.4$ & $0.4 \pm 0.1$ & $0.2 \pm 0.1$ & $0.4 \pm 0.0$ &&\\[2pt]
\multirow{2}{*}{GMP2601} & Disc & $2.6 \pm 0.3$ & $1.3 \pm 0.2$ & $1.1 \pm 0.2$ & $0.7 \pm 0.1$ & \multirow{2}{*}{$-0.54 \pm 0.04$} & \multirow{2}{*}{$-1.6 \pm 0.4$} \\
& Tail & $2.6 \pm 0.3$ & $0.5 \pm 0.2$ & $<0.6$ & $<0.3$ &&\\[2pt]
\multirow{2}{*}{GMP3618} & Disc & $3.6 \pm 0.4$ & $2.5 \pm 0.3$ & --- & $1.5 \pm 0.2$ & \multirow{2}{*}{$-0.37 \pm 0.00$} & \multirow{2}{*}{$-0.7 \pm 0.3$} \\
& Tail & $1.9 \pm 0.2$ & $0.5 \pm 0.1$ & --- & $0.5 \pm 0.1$ &&\\[2pt]
\multirow{2}{*}{GMP5226} & Disc & $7.6 \pm 0.8$ & $4.3 \pm 0.5$ & $3.7 \pm 0.4$ & $2.8 \pm 0.4$ & \multirow{2}{*}{$-0.44 \pm 0.05$} & \multirow{2}{*}{$<-0.9$} \\
& Tail & $0.6 \pm 0.1$ & $<0.2$ & $<0.1$ & $<0.2$ &&\\[2pt]
\bottomrule
\end{tabular}
\end{threeparttable}
\end{table*}

We measured integrated spectral indices for each galaxy disc and corresponding stripped tail in our sample. Flux densities for the galaxy disc were measured over an elliptical aperture centred on the right ascension and declination of each galaxy (see Table~\ref{tab:galaxy_table}). The semi-major axis of the ellipse was taken to be the Petrosian 90\% light radius (listed in Table~\ref{tab:galaxy_table}) and the axis ratio and position angle were determined through 2D S{\'e}rsic fits. All structural parameters were measured in the $r$-band and obtained from the NASA-Sloan Atlas\footnote{\url{http://nsatlas.org/}}.
\par
Flux densities for the stripped tails were measured over rectangular apertures that extend from the outer edge of the galaxy disc (as defined above) along the direction of the tail. The length and width of these apertures were set by hand in order to enclose tail emission within the $2\sigma$ contour at $144\,\mathrm{MHz}$. We measured tail flux densities over the same aperture for each frequency band. The galaxy disc and tail apertures for each galaxy are shown in Appendix~\ref{sec:galaxy_appendix}.
\par
A minority of galaxies in the sample overlap, in projection, with the diffuse radio halo in Coma that is well detected by LOFAR at $144\,\mathrm{MHz}$ \citep{bonafede2022}. Thus direct $144\,\mathrm{MHz}$ flux density measurements for these galaxies will include both a contribution from CREs originating from the galaxy and from CREs originating from the radio halo.  To account for this, we measure the median `background' level around (but not including) each galaxy and its stripped tail.  This level is then taken to be the local level of the radio halo and is subtracted off of each galaxy cutout image. The flux densities quoted in Table~\ref{tab:flux_table}, and used throughout the rest of the paper, are then measured from images after this subtraction. While the radio halo is not detected in our $400\,\mathrm{MHz}$, $700\,\mathrm{MHz}$, or $1.5\,\mathrm{GHz}$ images, we still follow the same procedure when measuring flux densities at these higher frequencies in case there is any low-level contribution from the halo that is not immediately obvious from the images. We are implicitly assuming here that the brightness of the radio halo is constant over the length scale of the galaxy plus stripped tail ($\sim\!10-50\,\mathrm{kpc}$ for our sample).
\par
We used a threshold of $3\sigma$ to determine whether or not significant emission was detected in each aperture. If the measured flux density was below this threshold then we instead quote a $3\sigma$ upper limit. We consider two sources of error on the measured flux densities. First a random error set by the noise in the image, and second an error on the flux calibration which is specific to each observing band. To measure the random error we measure flux densities within 500 randomly-positioned apertures in source-free regions of the image surrounding the target galaxy. We then take the random error to be the sigma-clipped standard deviation of the resulting 500 aperture flux densities.  For flux calibration we assume a relative error of $10\%$ for LOFAR HBA, $5\%$ for uGMRT B3 and B4, and $5\%$ for VLA L-band. The total error on the flux density ($\delta S_\nu$) in each aperture is then given by the sum in quadrature of the random error and the calibration error. Generally speaking, the flux calibration uncertainties are the dominant sources of error. Measured flux densities are listed in Table~\ref{tab:flux_table}. The `disc + tail' flux densities from Table~\ref{tab:flux_table} agree well with measured flux densities for overlapping sources from \citet{lal2022} (see Table 3 in \citealt{lal2022}). Any small differences are due to the fact that our disc apertures, while internally consistent for this paper, do not always enclose all of the radio emission around the galaxies in our sample (see images in Appendix~\ref{sec:galaxy_appendix}).
\par
Spectral indices were calculated using two different methods, depending on the number of frequency bands where significant flux density was detected.  When three or more frequency bands are detected, we calculated the spectral index as the slope of the best-fit linear relationship between $\ln S_\nu$ and $\ln \nu$.  Fits are done using orthogonal distance regression, specifically \texttt{scipy.odr}, and the uncertainty on the spectral index is taken to be the statistical uncertainty on the best-fit slope. When only two frequency bands are detected, we calculated spectral indices via the direct method, namely
\begin{equation} \label{eq:spec_index_direct}
    \alpha = \frac{\ln(S_{\nu,1} / S_{\nu,2})}{\ln(\nu_1 / \nu_2)} \pm \delta \alpha,
\end{equation}
\noindent
with
\begin{equation}
    \delta \alpha = \frac{1}{\ln(\nu_1 / \nu_2)} \sqrt{\left(\frac{\delta_1}{S_{\nu,1}}\right)^2 + \left(\frac{\delta_2}{S_{\nu,2}}\right)^2}.
\end{equation}
\noindent
Spectral indices for both disc and tail regions are listed in Table~\ref{tab:flux_table}.
\par
In Fig.~\ref{fig:int_specindex} (top) we show the spectral index integrated over the tail versus the spectral index integrated over the disc for each galaxy in our sample.  For the galaxy discs, spectral indices are broadly consistent with the typical injection spectral indices expected from star formation ($\sim-0.8$ to $-0.5$, e.g. \citealt{bell1978,bogdan1983,condon1992}). The median spectral index over galaxy discs is $-0.47 \pm 0.1$. This agrees with the mean $144\,\mathrm{MHz} - 1.5\,\mathrm{GHz}$ spectral index of $-0.55 \pm 0.14$ for the sample of 76 nearby star-forming galaxies from \citet{heesen2022}. Thus we do not find evidence for different broadband spectral indices between `normal' star-forming discs and galaxy discs undergoing ram pressure stripping. This is consistent with previous work in the Virgo Cluster that makes the same conclusion based off of spectral index measurements at gigahertz frequencies \citep{vollmer2010}.
\par
There is a collection of galaxies with spectral indices that are marginally flatter than typical injection values (i.e.\ $>-0.5$). We discuss potential origins of this spectral flattening in more detail in Sect.~\ref{sec:discussion_galaxy}, but it is likely connected to increased ionization losses (due to ISM compression from ram pressure) and/or shock acceleration due to the galaxy--ISM interaction.
\par
Spectral indices integrated over the tail region are clearly steeper than the disc spectral indices.  This is true for the sample as a whole but also true for each galaxy individually.  The contrast between disc and tail spectral indices is consistent with a framework where the synchrotron emission from the tail is due to aged cosmic ray electrons removed from the galaxy disc by ram pressure (see also: \citealt{chen2020,muller2021,ignesti2022_meerkat,venturi2022}). We extend our analysis of spectral aging along the tail in Sect.~\ref{sec:aging_model}, where we fit flux density profiles along the tail with a synchrotron aging model in order to derive characteristic stripping timescales for this sample of galaxies.
\par
There has been some evidence for flatter-than-injection spectral indices, specifically at low frequencies, in ram pressure stripped galaxies \citep{ignesti2022_meerkat,roberts2022_perseus}. To probe whether this is the case for our sample we measured low- and high-frequency spectral indices for each galaxy disc.  The low-frequency spectral index ($\alpha_\mathrm{low,\,disc}$) is measured between $144\,\mathrm{MHz}$ and $400\,\mathrm{MHz}$ and the high-frequency spectral index ($\alpha_\mathrm{high,\,disc}$) is measured between $700\,\mathrm{MHz}$ and $1.5\,\mathrm{GHz}$. The results are shown in Fig.~\ref{fig:int_specindex} (bottom) where we plot the corresponding colour-colour diagram. Given the short separations between frequency bands, the error bars on $\alpha_\mathrm{low,\,disc}$ and $\alpha_\mathrm{high,\,disc}$ are large and the galaxy discs are broadly consistent with no curvature, i.e.\ $\alpha_\mathrm{low,\,disc} = \alpha_\mathrm{high,\,disc}$. The median value for $\alpha_\mathrm{low,\,disc}$ is $-0.43 \pm 0.20$ and $-0.59 \pm 0.24$ for $\alpha_\mathrm{high,\,disc}$. This may be a hint of spectral curvature and flat low-frequency spectral indices in these galaxies, but we do not have the statistical power to confirm this.

\section{Disc spectral index maps} \label{sec:galaxy_specmaps}

\begin{figure*}
    \centering
    \includegraphics[width = 0.91\textwidth]{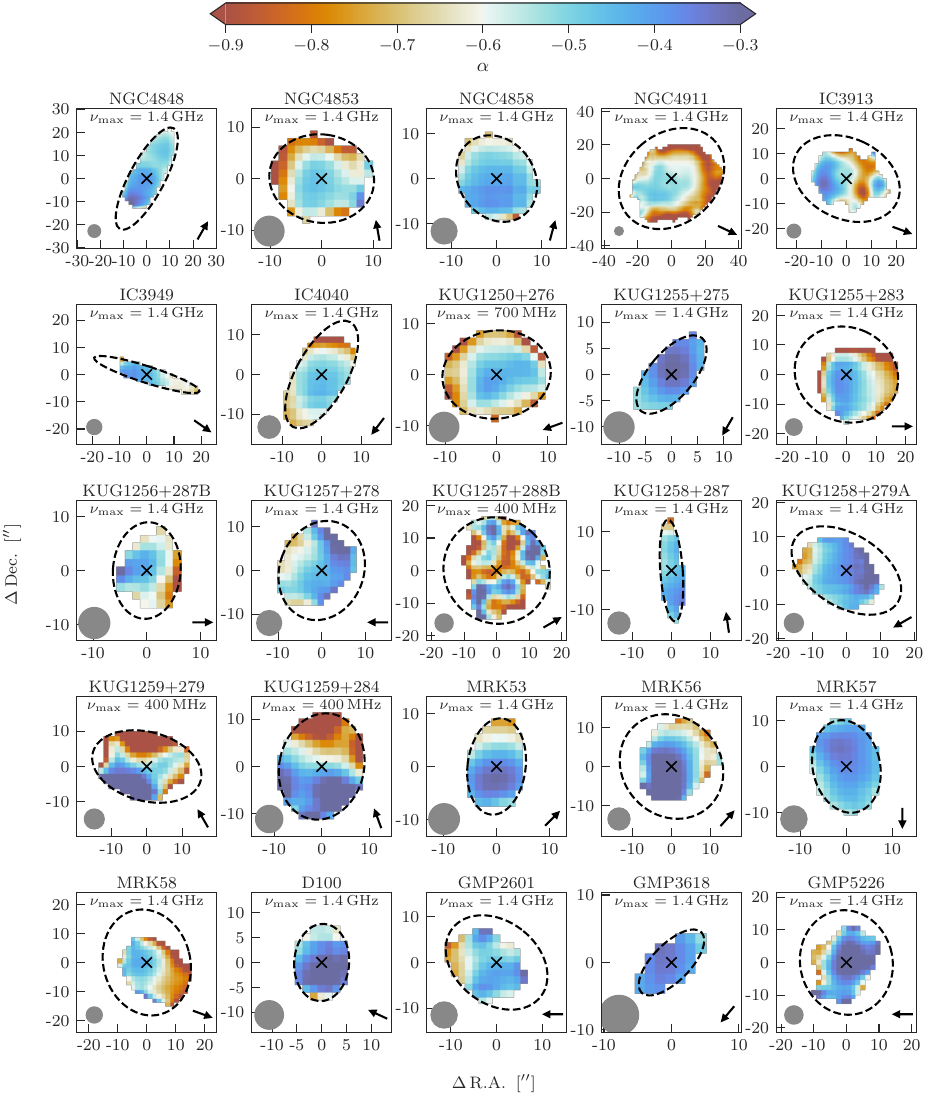}
    \caption{Spectral index maps for each galaxy disc in our sample. Spectral indices are measured between $144\,\mathrm{MHz}$ and $\nu_\mathrm{max}$, as listed in each figure panel. The dashed ellipse in each panel corresponds to the $r$-band Petrosian 90\% radius, the `x' marks the optical galaxy centre, and the filled circle shows the $6.5\arcsec$ beam FWHM. In the lower right of each panel we show an arrow which points in the projected direction of the stripped tail. The typical uncertainties on the spectral index maps are $\sim\!0.05$ between $144\,\mathrm{MHz} - 1.5\,\mathrm{GHz}$, $\sim\!0.07$ between $144\,\mathrm{MHz} - 700\,\mathrm{MHz}$, and $\sim\!0.11$ between $144\,\mathrm{MHz} - 400\,\mathrm{MHz}$.}
    \label{fig:galaxy_specmaps}
\end{figure*}

\begin{figure}
    \centering
    \includegraphics[width = \columnwidth]{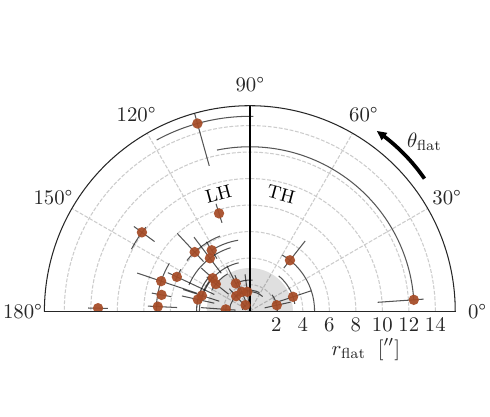}
    \caption{In polar coordinates, the location of the peak (i.e.\ flattest) spectral index within each disc, relative to the direction of the radio tail. The azimuthal axis shows $\theta_\mathrm{flat}$ and the radial axis shows $r_\mathrm{flat}$ (see text for details). The shaded region corresponds to $3.25\arcsec$, the half-width at half-maximum of the beam. Angles greater than $90^\circ$ correspond to points on the leading half (LH) and angles less than $90^\circ$ correspond to points on the trailing half (TH). The peak spectral index is systematically found on the leading half ($\theta_\mathrm{flat} > 0$) of ram pressure stripped galaxies.}
    \label{fig:spec_polar}
\end{figure}

In Fig.~\ref{fig:galaxy_specmaps} we show non-thermal spectral index maps covering the disc of each galaxy, as mentioned in Sect.~\ref{sec:data_vla}, the thermal contribution is largely negligible at these frequencies.  For the majority of galaxies, $\alpha$ is measured using fluxes at $144\,\mathrm{MHz}$ and $1.5\,\mathrm{GHz}$, in order to maximize the lever arm between the two frequencies, which in turn minimizes the uncertainty of the pixel-by-pixel spectral index measurements. For galaxies that are not covered by the $1.5\,\mathrm{GHz}$ imaging, we measure the spectral index between $144\,\mathrm{MHz}$ and the highest available frequency. This corresponds to $700\,\mathrm{MHz}$ for KUG1250+276, and $400\,\mathrm{MHz}$ for KUG1257+288B, KUG1259+279, and KUG1259+284. Each panel in Fig.~\ref{fig:galaxy_specmaps} lists the highest frequency that is used to derive the spectral index map. The disc region is defined as it was in the previous section, the $r$-band Petrosian 90\% light radius, and is shown by the dashed lines in Fig.~\ref{fig:galaxy_specmaps}. We only include pixels in the spectral index map that have $>2\sigma$ detections in both of the frequency bands that are used. For reference, when only including calibration uncertainties (which dominate over the galaxy discs), the typical uncertainty on the spectral index maps between $144\,\mathrm{MHz} - 1.5\,\mathrm{GHz}$ is $\sim\!0.05$.  For spectral index maps between $144\,\mathrm{MHz} - 700\,\mathrm{MHz}$ and $144\,\mathrm{MHz} - 400\,\mathrm{MHz}$, the typical uncertainties are $\sim\!0.07$ and $\sim\!0.11$, respectively.
\par
For 18/25 galaxies, the spectral index at the galaxy centre (marked by the `x' in Fig.~\ref{fig:galaxy_specmaps}), is consistent with values typical of recent injection from star formation (i.e.\ $-0.8$ to $-0.5$).  The remaining seven galaxies have central spectral indices that are flatter than $-0.5$. In the bottom right of each panel in Fig.~\ref{fig:galaxy_specmaps} we also show an arrow pointing in the projected direction of the radio continuum tail for each galaxy. For many galaxies in Fig.~\ref{fig:galaxy_specmaps}, the synchrotron emission is truncated within the stellar disc which reflects the outside-in nature of ram pressure stripping. In particular, galaxies are preferentially `missing' radio emission on their leading-halves (i.e.\ opposite to the direction of the tail), corresponding to CREs being transported from the leading half along the direction of the ram pressure wind to form the stripped radio tail. This truncation is present, to varying extents, for 14/25 galaxies. Those galaxies with radio emission that fills the optical disc may represent galaxies at an earlier stage of stripping, though we note that the majority of galaxies that do not show this truncation are also only marginally resolved at $6.5\arcsec$.  For these galaxies there may still be a truncated CRE distribution that is masked by the beam size. We show spectral index maps in Fig.~\ref{fig:galaxy_specmaps}, which requires detections at both frequencies, but note that this truncation signature is not a product of differences in depth between the two frequencies. When plotting single-frequency flux density maps instead, the same truncation signature is visible regardless of which frequency is used.
\par
Across the disc there is a gradient pattern in the spectral index maps that is present for a majority of galaxies in the sample. For galaxies that show a spectral index gradient, the orientation of this gradient is roughly aligned with the direction of the stripped tail. The steepest spectral indices are generally found in the direction of the tail (relative to the galaxy centre). This pattern is consistent with cosmic rays being transported from `further upstream' in the galaxy and aging as they are transported along the tail direction. For most galaxies in Fig.~\ref{fig:galaxy_specmaps}, the flattest spectral indices are not always found at the galaxy centre, but instead are often on the leading half of the galaxy, opposite to the direction of the stripped tail. We quantitatively demonstrate this tendency for the flattest spectral indices to be found on the leading halves of galaxies by identifying the pixel in the spectral index map with the flattest spectrum, given by pixel coordinates ($i_\mathrm{flat}, j_\mathrm{flat}$). When determining $i_\mathrm{flat}$ and $j_\mathrm{flat}$ we filtered each spectral index map with a $3 \times 3$ uniform filter so that each pixel was averaged with its eight nearest neighbours. We also only considered pixels where all eight of the surrounding pixels have detected spectral indices, in other words we did not consider pixels on the edge of the map. Lastly, given the pixel ($i_\mathrm{flat}, j_\mathrm{flat}$), we calculated the radial offset of this pixel from the optical galaxy centre ($r_\mathrm{flat}$) as well as the angular orientation between this pixel and the direction of the radio tail ($\theta_\mathrm{flat}$). Formally, $\theta_\mathrm{flat}$ is defined such that
\begin{equation}
    \cos \theta_\mathrm{flat} = \bm{u}_\mathrm{flat} \cdot \bm{u}_\mathrm{tail},
\end{equation}
\noindent
where $\bm{u}_\mathrm{flat}$ and $\bm{u}_\mathrm{tail}$ are unit vectors pointing between the optical galaxy centre and ($i_\mathrm{flat}, j_\mathrm{flat}$) and away from the galaxy centre in the direction of the radio tail, respectively. In this definition, $\theta_\mathrm{flat} = 180\deg$ corresponds to ($i_\mathrm{flat}, j_\mathrm{flat}$) being directly opposite to the tail direction and $\theta_\mathrm{flat} > 90\deg$ corresponds to ($i_\mathrm{flat}, j_\mathrm{flat}$) being on the leading half of the galaxy disc.
\par
In Fig.~\ref{fig:spec_polar} we show $\theta_\mathrm{flat}$ (azimuthal axis) and $r_\mathrm{flat}$ (radial axis) for each galaxy disc on a polar plot. It is immediately clear that the distribution of $\theta_\mathrm{flat}$ values does not uniformly cover the parameter space but instead is skewed towards large values, with the majority of galaxies having $120 \ge \theta_\mathrm{flat} \ge 180\deg$ and $r_\mathrm{flat}$ offset from zero. Only for a minority of galaxies in the sample is $r_\mathrm{flat}$ is smaller than the HWHM beam size.  These cases are consistent with the spectral index being flattest at the galaxy centre, and thus $\theta_\mathrm{flat}$ is not particularly meaningful in this scenario.
\par
Such an offset between the galaxy centre and the location of the flattest spectral index for galaxies undergoing ram pressure stripping was first observed for NGC4522 in the Virgo Cluster by \citet{vollmer2004_ngc4522}, and seems to be relatively commonplace for such galaxies in the Coma Cluster. While a majority of galaxies show this pattern, not all do.  For example, IC4040, KUG1255+275, D100, and GMP3618 all have spectral index distributions that are quite symmetric about the galaxy centre, though KUG1255+275, D100, and GMP3618 are also only marginally resolved. KUG1257+288B shows an irregular spectral index map. There is a region of flat emission on the leading half and then a second region of flat emission to the north that appears to follow a spiral arm in the galaxy (see Fig.~\ref{fig:example_imgs_KUG1257+288B}).  We also note that the short frequency spacing ($144\,\mathrm{MHz}$ to $400\,\mathrm{MHz}$), and the relatively low S/N detection of KUG1257+288B at $400\,\mathrm{MHz}$, may be contributing to the irregular map in Fig.~\ref{fig:galaxy_specmaps}.

\section{Spectral properties along the tail} \label{sec:spec_aging}

\subsection{Radio tail flux-density profiles} \label{sec:tail_flux_profiles}

\begin{figure*}
    \centering
    \includegraphics[width = 0.9\textwidth]{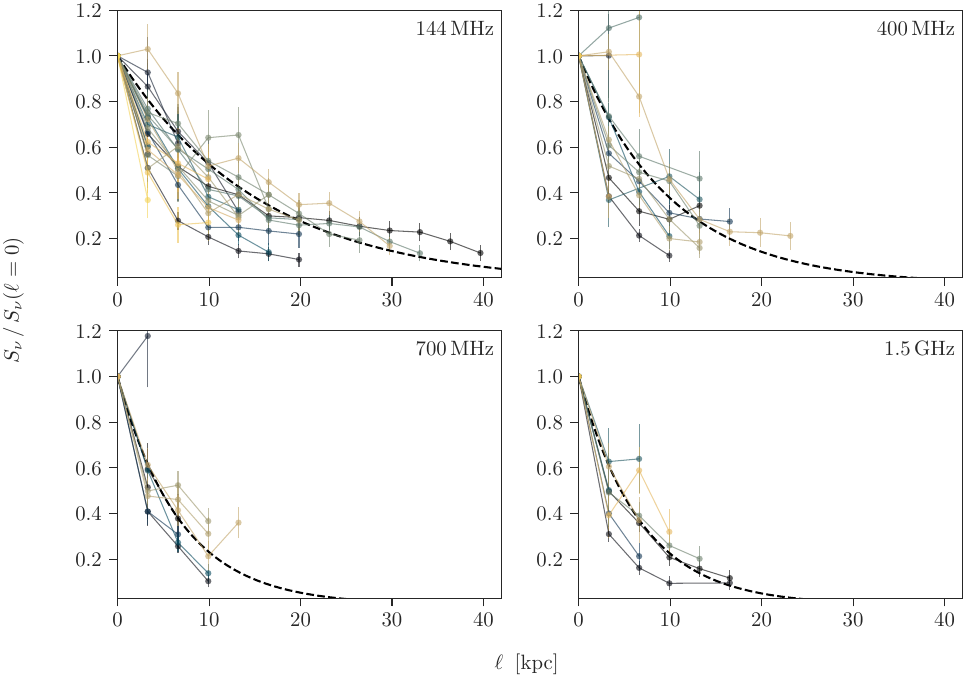}
    \caption{Normalized flux-density radial profiles along the stripped tails. Each solid line and markers correspond to an individual galaxy in the sample. Line colours for individual galaxies are consistent across the four panels. The dashed line in each panel shows the best-fit exponential decline for each frequency band.}
    \label{fig:rad_profiles}
\end{figure*}

\begin{figure}
    \centering
    \includegraphics[width = \columnwidth]{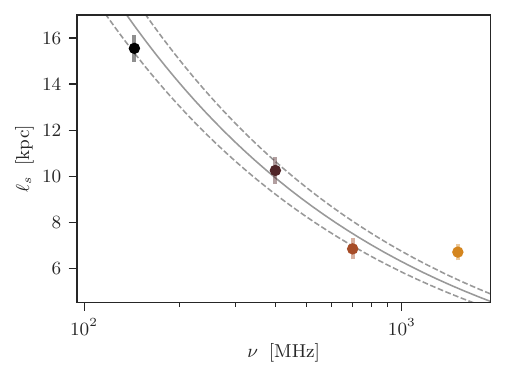}
    \caption{Exponential scale-length as a function of frequency for the radial profiles shown in Fig.~\ref{fig:rad_profiles}.  The solid and dashed lines shows the best-fit model of square-root decline, and its $1\sigma$ confidence interval. Such a decline is the expectation for the very simple scenario of constant stripping velocity in a uniform magnetic field (see text for details).}
    \label{fig:rs_scaling}
\end{figure}

For each galaxy, and for each frequency band, we extracted flux-density profiles along the observed radio tail in rectangular apertures. The height (i.e.\ the length parallel to the tail) of each aperture is $7\arcsec \approx 3.5\,\mathrm{kpc}$, slightly larger than the FWHM common beam size of $6.5\arcsec$, and the width (i.e.\ the length perpendicular to the tail) of each aperture differs from galaxy to galaxy and is set in order to enclose the lowest contour level (i.e.\ $2\sigma$) shown in Appendix~\ref{sec:galaxy_appendix}. The uncertainty on flux density measurements was computed following the procedure described in Sect.~\ref{sec:int_spec_index}.
\par
The extent of the tail flux-density profiles is determined by the length of the observed tail at $144\,\mathrm{MHz}$, as for all galaxies the observed tails are longest in this frequency band. The first tail aperture is set such that the inner edge of the aperture is set against the edge of the optical galaxy disc.  Apertures continue along the tail in $7\arcsec$ steps until significant emission is no longer detected in the $144\,\mathrm{MHz}$ band. We consider significant emission to be apertures where $S_\nu > 3\,\delta S_\nu$. This approach ensures that for all flux-density profiles there is detected emission in at least one frequency band, namely $144\,\mathrm{MHz}$.
\par
In Fig.~\ref{fig:rad_profiles} we show normalized flux-density for our four frequency bands as a function distance along the tail, $\ell$, which is defined such that $\ell = 0$ in the first bin (i.e.\ just off of the galaxy disc). The flux-density profiles are normalized by the value at $\ell = 0$ and therefore Fig.~\ref{fig:rad_profiles} shows the relative declines as a function of distance along the tail. We characterize the rate of this decline with a simple model of exponential decay, where the normalized flux density is given by
\begin{equation}
    \frac{S_\nu}{S_\nu(\ell = 0)} = e^{-\ell / \ell_s},
\end{equation}
\noindent
where $\ell_s$ is the scale length of the flux-density decline. For each panel in Fig.~\ref{fig:rad_profiles} we fit for the best-fit scale length in order to estimate a quantitative length scale associated with the ram pressure tails as a function of frequency. The best-fit exponential model is shown by the dashed lines in Fig.~\ref{fig:rad_profiles} and in Fig.~\ref{fig:rs_scaling} we show $\ell_s$ as a function of frequency for the four bands in this work.
\par
For a very simple model of constant stripping velocity in a constant magnetic field with no re-acceleration, one expects that the observed tail lengths will decline as the square root of the frequency \citep{ignesti2022_meerkat}. Therefore in Fig.~\ref{fig:rs_scaling} we also overplot the best-fit model of square-root decline as a function of frequency. This expectation broadly fits the observed scale lengths, signifying that this simple synchrotron aging model in a relatively constant magnetic field may be a reasonable approximation to reality for our galaxy sample. Though we note that the scale length at $1.5\,\mathrm{GHz}$ is a $2 - 3\sigma$ outlier from this simple model. The scale lengths shown in Fig.~\ref{fig:rs_scaling} are derived from the full galaxy sample together, and thus should be thought of as a description of the average behaviour for ram pressure stripped radio tails in Coma.

\subsection{Synchrotron aging model} \label{sec:aging_model}

\begin{figure*}
    \centering
    \includegraphics[width = \textwidth]{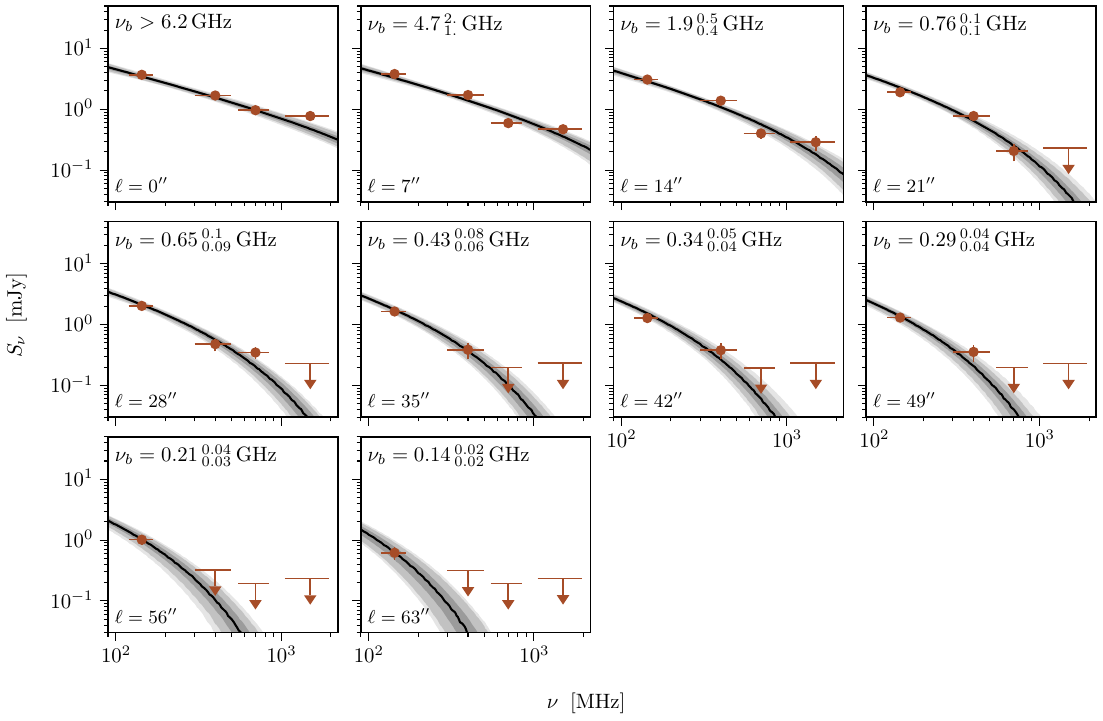}
    \caption{Radio continuum spectra and best-fit aging model along the stripped tail of MRK0057. Data points show observed flux densities and solid black line shows the best-fit (median) aging model along with the 84\%, 95\%, and 99.7\% credible regions (shading). Error bars on flux densities are determined following the procedure outlined in Sect.~\ref{sec:int_spec_index} and error bars on the frequency axis correspond to the bandwidth of the observing band. If significant ($>3\sigma$) flux density is not detected in an aperture than we show $3\sigma$ upper limits. We list the best-fit break frequency ($\nu_b$) as well as the distance along the stripped tail ($\ell$) in each panel.}
    \label{fig:example_fit_result}
\end{figure*}

Given the support for a scenario of spectral aging, both from the observed radial profiles and the difference in spectral index between the galaxy discs and tails, here we explicitly fitted a model of synchrotron aging to the observed spectra along the tail on a galaxy-by-galaxy basis. This model shares many qualitative similarities with the aging model from \citet{ignesti2023_a2255}, though  the two differ in their specific implementations. One main difference is that the model from \citet{ignesti2023_a2255} explicitly assumes that plasma is transported along the tail at a constant velocity, whereas this is not assumed in our aging model.  We only applied our aging model to galaxies with detected flux density in at least one frequency band, for at least three apertures along the tail.  Of the galaxies listed in Table~\ref{tab:galaxy_table}, six do not satisfy this criterion (NGC4853, IC3949, KUG1250+276, KUG1256+287B, GMP3618, and GMP5226) and were therefore excluded from this portion of the analysis.
\par
Our aging model is based on a \citet{jaffe1973} model (`JP model') for a population of electrons subject to synchrotron and inverse-Compton losses.  The JP model has three free parameters: the dimensionless normalization of the spectrum ($S_0$), the injection spectral index ($\alpha_0$), and the break frequency ($\nu_b$). We fit the radio continuum spectrum with a JP model for each distance bin along the stripped tail.  For a given galaxy, we assumed that $S_0$ and $\alpha_0$ are constant across all distance bins of the tail. This is equivalent to assuming that the removal of cosmic rays from the galaxy, and subsequent transport along the tail, is roughly in a steady state over the radiative timescales that we are sensitive to ($\sim\!100\,\mathrm{Myr}$, depending on the magnetic field strength). Thus $S_0$ and $\alpha_0$ were fit simultaneously across all distance bins while a different value for $\nu_b$ was fit for each bin. The result of this approach is that the relative difference in flux density, for a given frequency, across distance bins is determined solely by the location of the break frequency in the best-fit spectrum.  For neighbouring distance bins $d_1$, $d_2$, where $d_2 > d_1$ (i.e. the distance $d_2$ is further offset from the galaxy than $d_1$), we enforced that $\nu_{b,1} > \nu_{b,2}$. In other words, our model requires the break frequency to shift monotonically to lower frequencies as you move along the stripped tail.  This is an aging requirement and reflects the assumption that more distant electrons should have been stripped earlier and thus subject to greater radiative losses.
\par
We emphasize that the aging model described above may or may not provide a good description of the observed continuum spectra along the stripped tails.  For example, if electrons in the stripped tails are subject to fresh injection or substantial re-acceleration (extra-planar star formation could be a potential source of this) then this model will likely be a poor descriptor of the data. Indeed, we intentionally defined our aging model in this way so that we can identify both stripped tails that are consistent with a simple synchrotron aging framework and those that are not.
\par
Furthermore, there are a number of assumptions that are implicitly made in our model construction, including: the assumption of a uniform magnetic field along each tail, that the plasma in each distance bin has the same geometrical properties (i.e.\ volume, filling fraction), that the break frequency is constant within each distance bin, and that adiabatic losses are negligible. The latter appears at least roughly true given the spectral index decline that has been observed along ram pressure stripped tails \citep[e.g.][]{vollmer2004_ngc4522,muller2021,ignesti2022_meerkat}, suggesting that the timescale for synchrotron losses is shorter than that for adiabatic losses.  The remainder of these assumptions are difficult to validate observationally and thus add uncertainties to this model that are important to bear in mind when interpreting the subsequent scientific results.
\par
The aging model was implemented using the \texttt{synchrofit}\footnote{\url{https://github.com/synchrofit/synchrofit}} package \citep{turner2018a,turner2018b} in \texttt{Python} to calculate theoretical spectra, assuming a JP model, given values for $S_0$, $\alpha_0$, and $\nu_b$. We then found the best-fit values for $S_0$, $\alpha_0$ (constant across all distance bins), and $\nu_b$ (varying across all distance bins) using Markov chain Monte Carlo (MCMC) implemented with the \texttt{Python} package \texttt{emcee}\footnote{\url{https://emcee.readthedocs.io/en/stable/}} \citep{foreman-mackey2013}.  For each distance bin we included all four frequencies when fitting, even if emission is not detected at one or more frequencies. Though we did require that there is detected emission in at least one frequency band.  For fitting purposes, when significant emission was not detected (for a given frequency and aperture), we included $3\sigma$ upper limits in the likelihood following \citet{sawicki2012}.  We used the following uniform priors for all fit parameters: 
\begin{align*}
    S_0 & \in (1 \times 10^4, 5 \times 10^8) \\
    \alpha_0 & \in (-0.8, -0.5) \\
    \nu_{\mathrm{b},\,i} & \in
        \begin{cases}
            (0.05, 10),& \text{if } i = 1 \\
            (0.05, \nu_{\mathrm{b},\,i-1}),& \text{otherwise}
        \end{cases}
        ,
\end{align*}
\noindent
where $S_0$ and $\alpha_0$ are dimensionless and $v_{\mathrm{b},\,i}$ is the break frequency for each of the distance bins along the tail in GHz. We assessed convergence using the auto-correlation time, $\tau$, and considered our MCMC chain to be `converged' if the chain length, $N$, satisfied: $N > 100\times \tau$\footnote{\url{https://emcee.readthedocs.io/en/stable/tutorials/autocorr/}}. For each fit we run the MCMC chain until this inequality is satisfied, which typically corresponded to a few hundred thousand steps.
\par
In Fig.~\ref{fig:example_fit_result} we show an example of our spectral modelling for MRK0057, with panels in Fig.~\ref{fig:example_fit_result} showing the observed continuum spectrum and best-fit aging model at different distances, $\ell$, along the tail. In Appendix~\ref{sec:galaxy_appendix} we include analogous plots for all galaxies in our samples, in addition to `corner plots' showing the posterior distributions for all of the model parameters in our fits.
\par
We use Fig.~\ref{fig:example_fit_result} as an example to highlight some general characteristics of the spectral fitting results.  MRK0057 is an example of a galaxy where the radio continuum spectra along the stripped tail are well described by our aging model.  There are no clear offsets between the data and the best-fit model spectra in Fig.~\ref{fig:example_fit_result} for any of the distance bins. MRK0057 is also an example where the spectrum in the first distance bin, i.e. $\ell = 0$, is consistent with a powerlaw. In other words, we can only derive a lower limit on the break frequency. We find this to be the case for the majority (but not all) galaxies in our sample.  Widening our frequency range moving forward will allow us to determine whether these cases are truly represented by powerlaw spectra, or if they are simply well approximated by a powerlaw over the frequency range considered here.

\begin{figure}
    \centering
    \includegraphics[width = 0.95\columnwidth]{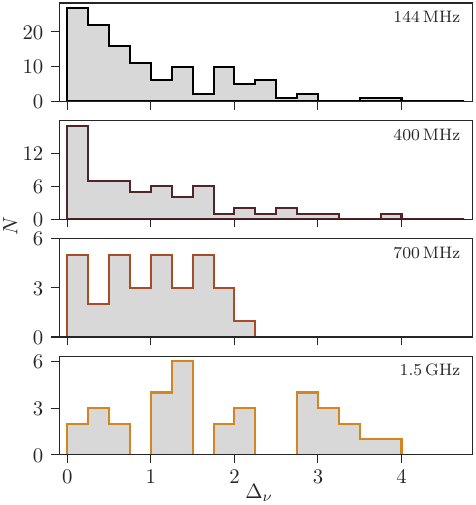}
    \caption{Distributions of offsets ($\Delta_\nu$, see Eq.~\ref{eq:delta}) from best-fit models for each frequency band.}
    \label{fig:model_offsets}
\end{figure}

Generally speaking, all galaxies in our sample have continuum spectra along their observed tails that are broadly consistent with our aging model. Inspection of the spectrum plots for individual galaxies in Appendix~\ref{sec:galaxy_appendix} shows that the vast majority of flux-density measurements are consistent with the 99\% model envelopes. Where there are deviations from the aging model they are most often found at high frequency. We show this more quantitatively with the parameter $\Delta$ which we define for each frequency band as:
\begin{equation} \label{eq:delta}
    \Delta_\nu = \frac{| S_\nu - S_{\nu,\,\mathrm{model}}^{50} |}{\sqrt{(\delta S_\nu)^2 + (\sigma_{\nu,\,\mathrm{model}})^2}},
\end{equation}
\noindent
where $S_\nu$ and $\delta S_\nu$ are the observed flux density and uncertainty, $S_{\nu,\,\mathrm{model}}^{50}$ is the median model flux density, and $\sigma_{\nu,\,\mathrm{model}}$ is the standard deviation of the model realizations at a given frequency, $\nu$. Thus $\Delta_\nu$ is a measure of the agreement between observations and the median model, accounting both for the observational error and the spread in the model posteriors.
\par
Fig.~\ref{fig:model_offsets} shows the distribution of $\Delta_\nu$ for all measured flux densities across all galaxies in our sample, divided by frequency band. The vast majority of flux-density measurements have $\Delta_\nu < 2$, indicating modest deviations between model and data. This quantitatively confirms our previous statement that the majority of flux-density measurements along the tail are reasonably-well reproduced by our aging model. It does seem that deviations between model and data become somewhat larger at high-frequency (i.e.\ $1.5\,\mathrm{GHz}$), though given the relatively small number of $1.5\,\mathrm{GHz}$ detections relative to the lower frequency bands, this statement is subject to small-number statistics. Fig~\ref{fig:model_offsets} only includes formal detections but we note that the vast majority of the upper limits along the tail are also consistent with the best-fit models (see Appendix~\ref{sec:galaxy_appendix}). The bulk of the model outliers ($\Delta_\nu \gtrsim 3$) seen in Fig.~\ref{fig:model_offsets} are from a single galaxy in our sample, NGC4858. In Appendix~\ref{sec:extra_sfr} we provide a detailed discussion of the possible origins of these model deviations at high frequency. In particular, we show that some, though likely not all, of the high-frequency deviations could be attributed to extraplanar star formation in the stripped tail.
\begin{figure}
    \centering
    \includegraphics[width = \columnwidth]{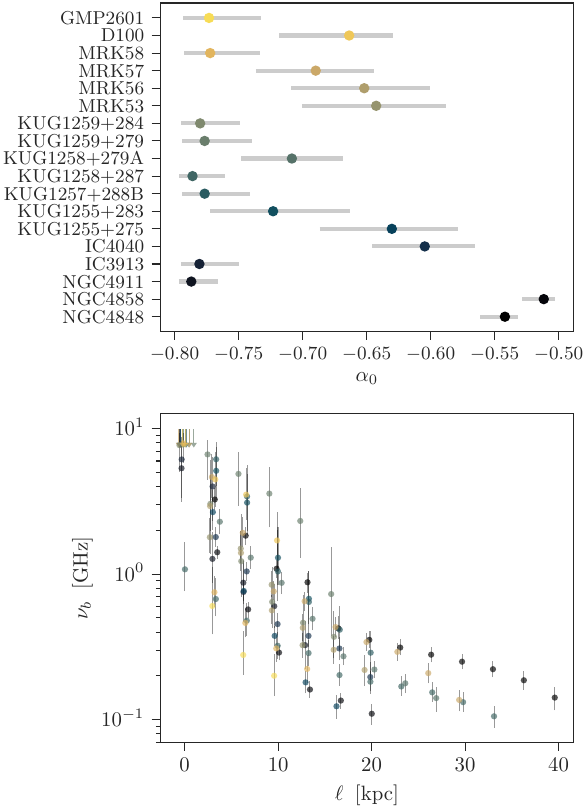}
    \caption{Best-fit parameters from the aging model fits to the stripped tails.  Data points are coloured according to galaxy and are consistent between panels. \textit{Top:} Distribution of best-fit injection spectral indices on a galaxy-by-galaxy basis.  Error bars correspond to 68\% credible regions from the posterior distribution.  \textit{Bottom:} Best-fit break frequency as a function of distance along tail.  Each marker corresponds to a single measurement for an individual galaxy and errorbars span the 68\% credible region. In the $\ell = 0$ bin, upper limits correspond to galaxies with purely powerlaw spectra and are shown at $\nu_b = 10\,\mathrm{GHz}$, which is the upper limit of our prior distribution.  To improve readability, in each distance bin data points are randomly shifted along the $x$-axis according to a Gaussian distribution with $\mu = 0$ and $\sigma = 0.5\,\mathrm{kpc}$.}
    \label{fig:aging_model_results}
\end{figure}
\par
Lastly, now having established that our aging model is a good descriptor of the data in the vast majority of cases, in Fig.~\ref{fig:aging_model_results} we show the best-fit parameter values derived from this fitting.  The top panel of Fig.~\ref{fig:aging_model_results} shows the best fit injection spectral indices for each galaxy. Roughly half of the galaxies have best-fit injection indices between $-0.7$ and $-0.5$ and the other half have steeper spectral indices of $\sim\!-0.8$, pushing up against the boundary of our prior. We note that the majority of galaxies with best-fit injection indices near $-0.8$ have sparse frequency coverage (e.g.\ only covered by $144\,\mathrm{MHz}$ and $400\,\mathrm{MHz}$ imaging) and/or have a small number of flux density detections along the tail at frequencies higher than $144\,\mathrm{MHz}$. Thus it may be that the relatively steep injection indices in Fig.~\ref{fig:aging_model_results} is due to some aging in the spectrum, but that the corresponding curvature is not distinguished as a result of the poor frequency coverage.  The bottom panel of Fig.~\ref{fig:aging_model_results} shows the best-fit break frequencies as a function of distance along the tail.  By construction the break frequency decreases monotonically for a given galaxy with increasing distance along the tail.  At fixed $\ell$ there is substantial scatter in the break frequency across the galaxies in our sample. This may indicate a range of plasma ages across different tails, or equally possible a range of magnetic field strengths (or both). The smallest break frequencies observed are $\sim\!100-200\,\mathrm{MHz}$. This reflects our requirement for detected emission at $144\,\mathrm{MHz}$, which becomes increasingly less likely for $\nu_b < 100\,\mathrm{MHz}$. Probing potential extensions of the stripped tails where $\nu_b < 100\,\mathrm{MHz}$ may be possible moving forward with observations from the LOFAR low-band antenna (LBA).

\subsection{Stripping timescales} \label{sec:stripping_timescales}

\begin{figure*}
    \centering
    \includegraphics[width = \textwidth]{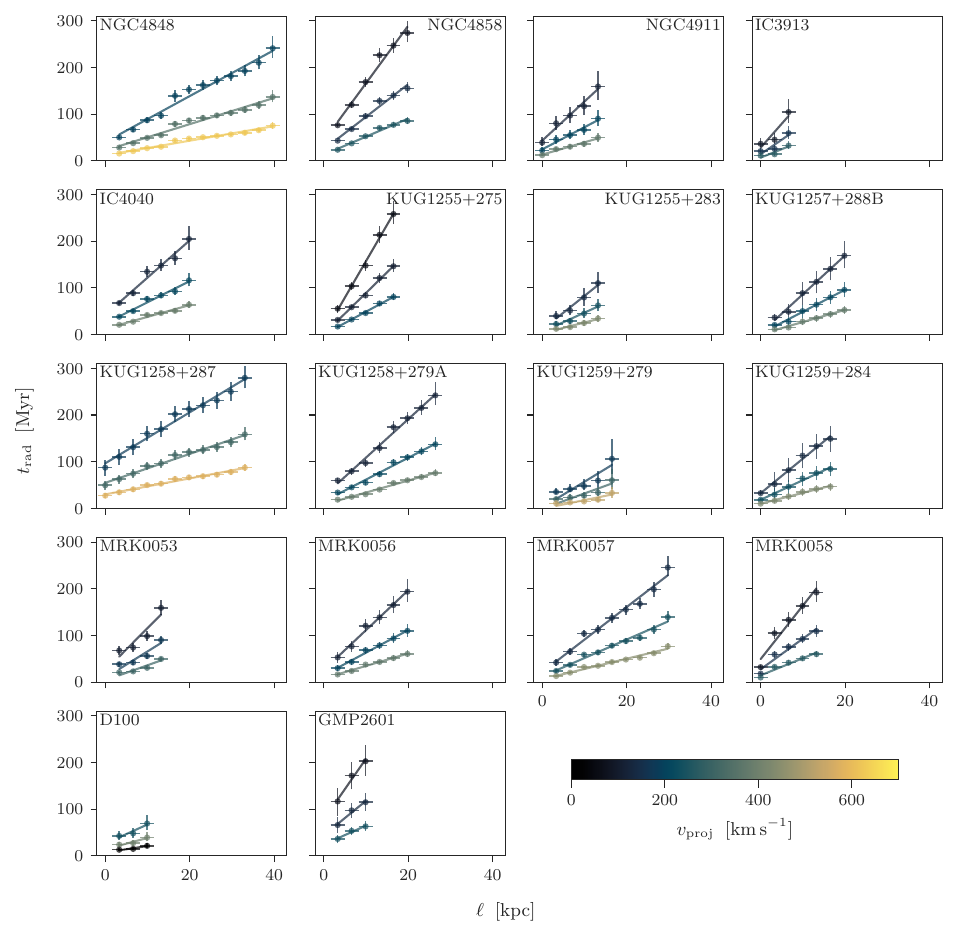}
    \caption{Radiative age versus projected distance along the tail for each of the galaxies in our sample. For each distance bin we plot three estimates of radiative age corresponding to assumed magnetic field strengths of $1.96$ (minimum-loss field), 6, and $10\,\mathrm{\mu G}$. Radiative age increases monotonically with decreasing magnetic field strength. For each assumed magnetic field strength, we show the best-fit linear relationship between radiative age and distance along tail, which gives an estimate for the projected bulk velocity of cosmic rays along the tail. Data markers and best-fit lines are coloured according to the best-fit projected velocity.}
    \label{fig:velocity_profiles}
\end{figure*}

Given that our aging model shows broad agreement with measured tail flux-densities, we can constrain the timescales over which material is removed from these galaxies by estimating the radiative age of the plasma as a function of distance along the radio tail. The radiative age \citep{miley1980} is given by
\begin{equation}
    t_\mathrm{rad} \approx 3.2 \times 10^{4}\,\frac{\sqrt{B}}{B^2 + B_\mathrm{CMB}^2}\frac{1}{\sqrt{\nu_b (1 + z)}}\;\mathrm{Myr},
\end{equation}
\noindent
where $\nu_b$ is the break frequency in MHz, $z$ is the source redshift, $B$ is the magnetic field in $\mathrm{\mu G}$, and $B_\mathrm{CMB}$ is CMB equivalent magnetic field given by $B_\mathrm{CMB} = 3.25 (1 + z)^2\;\mathrm{\mu G}$. Thus given the best-fit break frequency from our aging model, and an assumed magnetic field strength, we can estimate a radiative age for each distance bin along the observed radio tails.
\par
In Fig.~\ref{fig:velocity_profiles} we show $t_\mathrm{rad}$ as a function of $\ell$ for each of the galaxies in our sample.  Since we do not have constraints on the magnetic field strength in the stripped tails, we consider a range of possible values. We set the low end of this range to be the minimum energy loss magnetic field which is given by $B_\mathrm{min} = B_\mathrm{CMB} / \sqrt{3} \approx 1.96\,\mathrm{\mu G}$ for the Coma Cluster, we consider an intermediate field strength of $6\,\mathrm{\mu G}$, and a maximum field strength of $10\,\mathrm{\mu G}$. This maximum of $10\,\mathrm{\mu G}$ is set in order to encapsulate the small number of observational estimates on field strengths in ram pressure stripped tails in other clusters (\citealt{muller2021}: $2-4\,\mathrm{\mu G}$, \citealt{vollmer2021}: $6-7\,\mathrm{\mu G}$, \citealt{ignesti2022_meerkat}: $<\!10\,\mathrm{\mu G}$). Depending on the assumed magnetic field strength, we find radiative ages that increase from tens of megayears near the galaxy disc out to two or three hundred megayears at the edge of the tail.
\begin{table}[!ht]
\centering
\caption{Stripped plasma velocity estimates}
\label{tab:velocity_table}
\begin{tabular}{l c c c}
\toprule
\toprule
Galaxy & \multicolumn{3}{c}{$v\;\;\mathrm{[km\,s^{-1}]}$} \\
\cmidrule(lr){2-4}
& $B \approx 2\,\mathrm{\mu G}$ & $B = 6\,\mathrm{\mu G}$ & $B = 10\,\mathrm{\mu G}$ \\
\midrule
NGC4848 & $198 \pm 10$ & $348 \pm 18$ & $626 \pm 32$ \\
NGC4858 & $79 \pm 6$ & $140 \pm 11$ & $253 \pm 19$ \\
NGC4911 & $117 \pm 12$ & $206 \pm 20$ & $371 \pm 36$ \\
IC3913 & $95 \pm 41$ & $158 \pm 64$ & $269 \pm 108$ \\
IC4040 & $123 \pm 10$ & $216 \pm 18$ & $386 \pm 32$ \\
KUG1255+275 & $63 \pm 2$ & $111 \pm 4$ & $201 \pm 8$ \\
KUG1255+283 & $135 \pm 18$ & $239 \pm 32$ & $426 \pm 56$ \\
KUG1257+288B & $118 \pm 6$ & $207 \pm 10$ & $377 \pm 19$ \\
KUG1258+287 & $179 \pm 8$ & $316 \pm 14$ & $573 \pm 26$ \\
KUG1258+279A & $119 \pm 5$ & $209 \pm 9$ & $380 \pm 16$ \\
KUG1259+279 & $177 \pm 47$ & $298 \pm 79$ & $529 \pm 139$ \\
KUG1259+284 & $132 \pm 7$ & $233 \pm 12$ & $423 \pm 22$ \\
MRK0053 & $108 \pm 31$ & $174 \pm 47$ & $303 \pm 82$ \\
MRK0056 & $115 \pm 6$ & $202 \pm 10$ & $367 \pm 19$ \\
MRK0057 & $140 \pm 8$ & $247 \pm 14$ & $445 \pm 26$ \\
MRK0058 & $85 \pm 11$ & $149 \pm 20$ & $266 \pm 34$ \\
MRK0060 & $238 \pm 79$ & $400 \pm 130$ & $710 \pm 229$ \\
GMP2601 & $75 \pm 12$ & $132 \pm 22$ & $240 \pm 39$ \\
\bottomrule
\end{tabular}
\end{table}

\par
At fixed magnetic field strength, Fig.~\ref{fig:velocity_profiles} shows that a linear relationship between $t_\mathrm{rad}$ and $\ell$ is a good descriptor for all galaxies in our sample. This is consistent with the commonly-made assumption that CRE transport in ram pressure stripped tails is dominated by advection. As a result of this, we determined the slope of best-fit linear trend between $t_\mathrm{rad}$ and $\ell$ in order to estimate a constant bulk velocity for the plasma being stripped along the tail. The data points and best-fit trends in Fig.~\ref{fig:velocity_profiles} are then coloured according to this bulk velocity.  Roughly speaking, the best-fit velocities are on the order of hundreds of kilometers per second. The specific values for each galaxy and magnetic field assumption are listed in Table~\ref{tab:velocity_table}. We stress that these are projected velocities, since $\ell$ is projected on the sky, and therefore must be lower limits on the 3D stripping speed.

\section{Discussion} \label{sec:discussion}

\subsection{Radio continuum properties in the discs of ram pressure stripped galaxies} \label{sec:discussion_galaxy}

The first part of the analysis in this work is focused on probing the spectral properties of radio continuum emission within the discs of the ram pressure stripped galaxies in our sample. While ram pressure is oft associated with tails of stripped debris (e.g.\ jellyfish galaxies), it also can significantly alter the conditions of both the thermal and non-thermal ISM within the disc. For example, through outside-in quenching \citep[e.g.][]{cortese2012,schaefer2017,yoon2017,schaefer2019}, gas compression \citep[e.g.][]{cramer2020,troncoso-iribarren2020,roberts2022_lofar_manga}, localized starbursts \citep[e.g.][]{gavazzi2001,tomicic2018,boselli2021_ic3476,hess2022,roberts2022_perseus,roberts2022_lofar_manga}, magnetic field strength amplification \citep[e.g.][]{gavazzi1999,murphy2009_virgo,vollmer2013,chen2020}, etc.  All of these examples can be constrained, to some extent, with multi-frequency radio continuum observations like those in this work.
\par
Many previous works have found evidence for enhanced star formation in galaxies undergoing ram pressure stripping \citep[e.g.][]{ebeling2014,poggianti2016,vulcani2018_sf,roberts2020,wang2020,roberts2021_LOFARclust}. This is thought to be a product of the external pressure both increasing molecular gas densities, as well as increasing the efficiency with which atomic gas in converted to the molecular form, which in both cases can lead to an increase in star formation prior to gas depletion \citep[e.g.][]{bothun1986,gavazzi2001,merluzzi2013,lee2017,cramer2020,troncoso-iribarren2020,moretti2020,moretti2020b,boselli2021_ic3476,hess2022,roberts2022_lofar_manga,roberts2023_asym}. One can then ask, for this scenario, what is the expected effect on radio continuum emission. In principle, the increased ISM densities should lead to increased ionization losses and thus a flattening of the low-frequency spectral index \citep[e.g.][]{basu2015,chyzy2018}. This has previously been invoked to explain flat spectral indices at low frequencies for galaxies undergoing ram pressure stripping \citep{ignesti2022_meerkat,roberts2022_perseus}. In this work we found tentative evidence for this flattening. In the colour-colour plot in Fig.~\ref{fig:int_specindex}, the median low-frequency spectral index is marginally flatter than the high-frequency spectral index. Furthermore, there are a handful of galaxies with $\alpha_\mathrm{low,\,disc} > \alpha_\mathrm{high,\,disc}$ at $>\!2\sigma$ significance, but no galaxies with $\alpha_\mathrm{low,\,disc} < \alpha_\mathrm{high,\,disc}$ at the same significance level. Ultimately this is still unconvincing in a statistical sense in this work, and a wider range in frequency is necessary for stronger constraints to be possible.  This should include both higher (e.g.\ $5\,\mathrm{GHz}$) and lower ($50\,\mathrm{MHz}$, LOFAR LBA) frequency data in order to extend the lever arm in both directions.
\par
The spectral index maps in Fig~\ref{fig:galaxy_specmaps} also support such a scenario. For a majority of galaxies in the sample, the flattest spectral indices are found on the the leading half, opposite to the tail direction (Fig.~\ref{fig:spec_polar}).  The leading half is the site that should be associated with the strongest gas compression, though it is also subject to shocks are driven into the ISM by ram pressure at the galaxy -- ICM interface \citep{vollmer2004_ngc4522,murphy2009_virgo,pedrini2022}. Such an offset between the galaxy centre and the location of the flattest spectrum for a galaxy undergoing ram pressure stripping, was first noted by \citet{vollmer2004_ngc4522} for NGC4522 in the Virgo Cluster. \citet{vollmer2004_ngc4522} suggest that this is a signature of a ram pressure induced shock, which accelerates CREs leading to a flat spectral index. This is corroborated by a ridge of polarized emission on the leading edge of NGC4522, which would not be expected from the turbulent motions associated with star formation. It is possible, if not likely, that a combination of gas compression/star formation and shocks is contributing to the observed flat spectral indices on the leading halves of the galaxies in this work.  \citet{roberts2022_lofar_manga} have shown that many of the galaxies in this sample do show evidence for enhanced star formation on their leading halves. Moving forward, observations of polarized radio flux for this sample will be insightful in order to determine whether similar ridges of polarized emission are present for ram pressure stripped galaxies in Coma, as have been seen for similar types of galaxies in the Virgo Cluster \citep{vollmer2004_ngc4522,chyzy2007,murphy2009_virgo,vollmer2010,vollmer2013}.

\subsection{Spectral aging along the stripped tails}
For the ram pressure tails considered in this work, the decline in flux density with increasing distance from the galaxy is broadly consistent with a simple model of synchrotron aging. This is generally true for all four frequencies, though there are some hints of deviations at $1.5\,\mathrm{GHz}$ (but this is also the frequency band with the fewest detections in the tails).  This is consistent with previous studies where it is argued that these stripped tails visible in the radio continuum are formed from CREs that are accelerated by star formation in the galaxy disc, and subsequently removed from the galaxy by ram pressure \citep[e.g.][]{chen2020,roberts2021_LOFARclust}. In the absence of newly injected CREs or re-acceleration, this framework will lead to steeper spectral indices in the stripped tails than the galaxy discs (assuming that the disc is actively star forming). This has been previously confirmed for a relatively small number of galaxies \citep{vollmer2004_ngc4522,chen2020,muller2021,ignesti2022_meerkat,roberts2022_perseus,venturi2022,ignesti2023_a2255}, and here we show that it holds for a relatively large set of ram pressure stripped galaxies in Coma (Fig.~\ref{fig:int_specindex}). The steep spectral indices imply that low-frequency imaging is particularly useful for identifying galaxies with ram pressure tails in low-$z$ groups and clusters, as evidenced by the large number of such tailed sources identified by \citet{roberts2021_LOFARclust} compared to surveys at higher frequency \citep[e.g.][]{miller2009,chen2020}.
\par
The dominant mode of CRE transport in forming ram pressure stripped tails is believed to be advection/streaming \citep{murphy2009_virgo,vollmer2013,muller2021,ignesti2022_gasp}. We estimate a constant streaming velocity ($t_\mathrm{rad} \propto \ell$) for each galaxy in Fig.~\ref{fig:velocity_profiles}. While our estimates on the bulk velocity cover a relatively broad range due to the uncertainty in the magnetic field strength, they are consistent with previous estimates in the literature that derive stripping timescales from synchrotron radiative ages.  \citet{vollmer2021} estimate a bulk velocity limit of $\gtrsim\!140\,\mathrm{km\,s^{-1}}$ based on $1.4$ and $5\,\mathrm{GHz}$ observations of the Virgo galaxy NGC4330. \citet{ignesti2023_a2255} have published estimates for plasma velocities in the ram pressure stripped tails of galaxies in the cluster Abell 2255. For five galaxies with monotonically decreasing flux-density profiles, \citet{ignesti2023_a2255} constrain the projected bulk velocity to be between $160$ and $430\,\mathrm{km\,s^{-1}}$, based off of synchrotron cooling timescale arguments. These estimates are derived assuming a minimum-loss magnetic field, and thus formally are lower limits. Constraints from hydrodynamical simulations generally show that gas velocities during ram pressure stripping are $\lesssim\!1000\,\mathrm{km\,s^{-1}}$ \citep[e.g.][]{tonnesen2010,steinhauser2016,choi2022}. These gas velocity constraints can be compared to to bulk plasma velocities constrained in this work under the assumption that magnetic fields remain frozen in to the ISM during stripping, and thus that the stripped cosmic rays also track the stripped gaseous ISM.

\subsection{A broad picture of the impact from ram pressure on the non-thermal ISM in Coma satellites} \label{sec:impact_nt}

To conclude, we discuss a general picture of the impact from ram pressure on the radio continuum spectra of star-forming satellites, that is consistent with the observations of Coma Cluster galaxies presented in this work.
\par
As galaxies traverse the ICM in galaxy clusters, the non-thermal ISM is both perturbed within the galaxy disc as well as stripped off of the galaxy to form a radio tail. The formation of the radio tail can also be aided by the presence of ICM magnetic field accreted onto the galaxy via magnetic draping \citep{dursi2008,pfrommer2010,muller2021}. In clusters, this typically occurs on first infall towards orbital pericenter \citep{roberts2021_LOFARclust,smith2022}. For a snapshot in time, the fraction of star-forming cluster galaxies (at low-$z$) with stripped radio tails is roughly twenty per cent \citep{roberts2021_LOFARgrp}, though this figure does not include those galaxies that may have been previously tailed and are now sufficiently stripped, nor those galaxies in a pre-stripping phase that may develop a tail in the future. It may be that most, or all, star-forming galaxies in clusters form a radio tail, but this remains an open question.
\par
Perturbations within the disc manifest themselves in the synchrotron spectral index -- both in a resolved and integrated sense.  Perturbations are strongest on the leading half of the disc where the spectral index becomes flattest.  In some cases the spectral index becomes flatter than the typical expected range for star formation (Fig.~\ref{fig:galaxy_specmaps}; see also, \citealt{ignesti2022_meerkat, roberts2022_perseus}). This could be a result of increased ionization losses related to compression of the ISM by the external ram pressure \citep[e.g.][]{moretti2020,moretti2020b,troncoso-iribarren2020,ignesti2022_meerkat,roberts2022_perseus,roberts2022_lofar_manga,roberts2023_asym,moretti2023}. Magnetic field lines should also be compressed in the disc by ram pressure, which may contribute to the unusually high radio luminosity to SFR ratios in cluster galaxies -- particularly those with ram pressure stripped tails \citep[e.g.][]{gavazzi1999, murphy2009_virgo, vollmer2013, chen2020}. Ridges of polarized emission have also been observed along the leading edge of ram pressure stripped galaxies \citep[e.g.][]{vollmer2004_ngc4522,vollmer2007,chyzy2007}, a further signature of the interaction at the ISM -- ICM interface, and that may indicate that CRE acceleration from shocks is also playing a role in the observed flat spectral indices.
\par
In many cases the distribution of synchrotron emission becomes truncated, particularly on the leading half, due to the outside-in removal of CREs from the disc. Through this truncation, CREs are transported, likely through advection/streaming \citep[e.g.][]{murphy2009_virgo,vollmer2013,ignesti2022_gasp}, from the disc downstream along the stripped tail. As CREs move further along the tail, their spectra age through synchrotron and inverse-Compton losses, leading to a decrease in radio flux density and curvature in the spectrum. This aging along the tail is consistent with a relatively constant projected stripping velocity, on the order of hundreds of kilometers per second. Observational estimates of this stripping velocity is highly dependent on the assumed magnetic field strength, which is very poorly constrained observationally. Most galaxies show no evidence for fresh injection of CREs or re-acceleration in the stripped tail, suggesting that any radio flux density produced by extra-planar star formation in the tail is subdominant to that from the aging plasma stripped off of the disc.  This is consistent with the low SFRs observed in ram pressure stripped tails \citep[e.g.][]{sun2007,hester2010,fumagalli2011,vulcani2018_sf,poggianti2019_extraplanar_sf,junais2021}.  For example, a typical extra-planar \textsc{Hii} region with a SFR of $10^{-3}\,\mathrm{M_\odot\,yr^{-1}}$ will only produce a $1.5\,\mathrm{GHz}$ flux density of $\sim\!1\,\mathrm{\mu Jy}$ (at the distance of Coma).
\par
The framework outlined above is consistent with the observations of ram pressure stripped galaxies in Coma presented here. Future observations of polarized intensity for these galaxies would provide a further test of its consistency. Constraining the applicability of this picture beyond requires both the identification of a significant number of ram pressure stripped galaxies in another (likely nearby) cluster, and multi-frequency radio continuum observations of said galaxies. The stripped tails of jellyfish galaxies in Abell 2255 are broadly consistent with this picture \citep{ignesti2023_a2255}, and moving forward, the Virgo Cluster is another obvious testing ground where significant work has already been done at high frequencies \citep[e.g.][]{crowl2005,chyzy2007,murphy2009_virgo,vollmer2004_ngc4522,vollmer2010,vollmer2013}.  With recent low-frequency data (see the VICTORIA project, \citealt{edler2023}), direct comparisons between galaxies in Virgo and this work can be made.

\section{Conclusions} \label{sec:summary}

In this work we have presented a comprehensive radio continuum study of 25 ram pressure stripped satellite galaxies in the Coma Cluster. We use nearly continuous frequency coverage between $144\,\mathrm{MHz}$ and $1.5\,\mathrm{GHz}$ to constrain radio spectral properties from galaxy discs all the way across their stripped radio tails. Below we itemize the main scientific conclusions from this work:

\begin{itemize}
    \itemsep0.5em
    \item Spectral indices integrated over the galaxy discs cover a range between $-0.8$ and $-0.3$.  Roughly half of the galaxy disc sample have integrated spectral indices that are $>\!-0.5$, i.e.\ flatter than the typical value for injection by star formation (Fig.~\ref{fig:int_specindex}).

    \item Ram pressure stripped tails have steep integrated spectral indices.  All tails in our sample are consistent with $\alpha_\mathrm{tail} < -1$. The integrated spectral indices over the tails are clearly steeper than those measured over the disc, consistent with spectral aging as the plasma is removed from the galaxy (Fig.~\ref{fig:int_specindex}).

    \item Resolved spectral index maps of the discs show that the leading halves of the disc (i.e.\ the direction opposite to the tail direction) have systematically flatter spectral indices compared to the rest of the disc. This is consistent with a scenario where gas is compressed by ram pressure leading to enhanced local star formation and/or increased CRE ionization losses and/or the presence of ram pressure induced shocks (Figs.~\ref{fig:galaxy_specmaps}, \ref{fig:spec_polar}).

    \item The scale lengths of the stripped radio tails decrease with increasing observing frequency.  This decrease is roughly in proportion to the inverse square root of the frequency, which is the expectation from a simple model of constant stripped plasma velocity in a uniform magnetic field (Figs.~\ref{fig:rad_profiles}, \ref{fig:rs_scaling}).

    \item For the vast majority of galaxies in our sample, radio continuum spectra extracted at various distances along the observed tails are well reproduced by a simple model of spectral aging due to synchrotron and inverse-Compton losses (Figs.~\ref{fig:example_fit_result}, \ref{fig:aging_model_results}, \ref{fig:model_offsets}).

    \item The radio tails in our sample show linear relationships between radiative age and distance along the tail, implying roughly constant (projected) bulk stripping velocities with magnitudes on the order of hundreds of kilometers per second (Figs.~\ref{fig:velocity_profiles}).
\end{itemize}

This work is an example of the value that can be derived from sensitive, high-resolution radio continuum observations in the context of advancing our understanding of ram pressure stripping in galaxy clusters. Moving forward this analysis could be furthered by homogenizing the frequency coverage for all galaxies in the sample.  Specifically, this will require filling in gaps in the $700\,\mathrm{MHz}$ and $1.5\,\mathrm{GHz}$ imaging.  Lower frequency data from the LOFAR LBA will likely also become available moving forward. An ultimate goal of broadband, multi-frequency imaging (e.g.\ from $\sim$tens of megahertz to $\sim$a few gigahertz) covering the full Coma Cluster within the virial radius would allow for a complete study of the environmental impact on the non-thermal ISM in galaxies part of the nearest, massive cluster. This is, in principle, achievable thanks to the large primary beams of radio facilities like LOFAR, the uGMRT, the VLA, and MeerKAT (not to mention next-generation facilities), but will still require a significant observational commitment.

\begin{acknowledgements}
The authors thank William Forman for comments on a draft of the paper and his effort invested into obtaining the uGMRT data used in this work, Neal Miller for comments on a draft of the paper, and Larry Rudnick for helpful discussions on flux-density scale during the preparation of the paper. IDR and RJvW acknowledge support from the ERC Starting Grant Cluster Web 804208. DVL acknowledges the support of the Department of Atomic Energy, Government of India, under project no. 12-R\&D-TFR-5.02-0700. MS acknowledges support from the NASA grant 80NSSC22K1508 and the USRA SOFIA grant 09\_0221 provided by NASA. AI acknowledges funding from the European Research Council (ERC) under the European Union's Horizon 2020 research and innovation programme (grant agreement No. 833824, PI Poggianti), and the INAF founding program `Ricerca Fondamentale 2022' (PI A. Ignesti). This work would not have been possible without the following software packages: \texttt{AstroPy} \citep{astropy2013}, \texttt{CASA} \citep{casa2022}, \texttt{ChainConsumer} \citep{hinton2016}, \texttt{CMasher} \citep{vandervelden2020}, \texttt{DS9} \citep{ds9_2003}, \texttt{Emcee} \citep{foreman-mackey2013}, \texttt{Matplotlib} \citep{hunter2007}, \texttt{NumPy} \citep{harris2020}, \texttt{Photutils} \citep{bradley2022}, \texttt{Regions} \citep{bradley2022_regions}, \texttt{RSMF} (\url{https://rsmf.readthedocs.io/en/latest/source/howto.html}), \texttt{SciPy} \citep{virtanen2020}, \texttt{Synchrofit} \citep{turner2018a,turner2018b}. This work is based in part on data collected at Subaru Telescope, which is operated by the National Astronomical Observatory of Japan. We thank the staff of the GMRT that made these observations possible. The GMRT is run by the National Centre for Radio Astrophysics of the Tata Institute of Fundamental Research. The National Radio Astronomy Observatory is a facility of the National Science Foundation operated under cooperative agreement by Associated Universities, Inc. LOFAR \citep{vanhaarlem2013} is the Low Frequency Array designed and constructed by ASTRON. It has observing, data processing, and data storage facilities in several countries, which are owned by various parties (each with their own funding sources), and which are collectively operated by the ILT foundation under a joint scientific policy. The ILT resources have benefitted from the following recent major
funding sources: CNRS-INSU, Observatoire de Paris and Université d’Orléans, France; BMBF, MIWF-NRW, MPG, Germany; Science Foundation Ireland (SFI), Department of Business, Enterprise and Innovation (DBEI), Ireland; NWO, The Netherlands; The Science and Technology Facilities Council, UK; Ministry of Science and Higher Education, Poland. Funding for the NASA-Sloan Atlas has been provided by the NASA Astrophysics Data Analysis Program (08-ADP08-0072) and the NSF (AST-1211644). Funding for SDSS-III has been provided by the Alfred P. Sloan Foundation, the Participating Institutions, the National Science Foundation, and the U.S. Department of Energy. The SDSS-III web site is \url{http://www.sdss3.org}. SDSS-III is managed by the Astrophysical Research Consortium for the Participating Institutions of the SDSS-III Collaboration including the University of Arizona, the Brazilian Participation Group, Brookhaven National Laboratory, University of Cambridge, University of Florida, the French Participation Group, the German Participation Group, the Instituto de Astrofisica de Canarias, the Michigan State/Notre Dame/JINA Participation Group, Johns Hopkins University, Lawrence Berkeley National Laboratory, Max Planck Institute for Astrophysics, New Mexico State University, New York University, Ohio State University, Pennsylvania State University, University of Portsmouth, Princeton University, the Spanish Participation Group, University of Tokyo, University of Utah, Vanderbilt University, University of Virginia, University of Washington, and Yale University.
\end{acknowledgements}

\FloatBarrier

%
%

\bibliographystyle{aa}
\bibliography{main}

\appendix

\onecolumn

\section{Deviations from the aging model for NGC4858: a contribution from extra-planar star formation?} \label{sec:extra_sfr}

\begin{figure*}
    \centering
    \includegraphics[width = \textwidth]{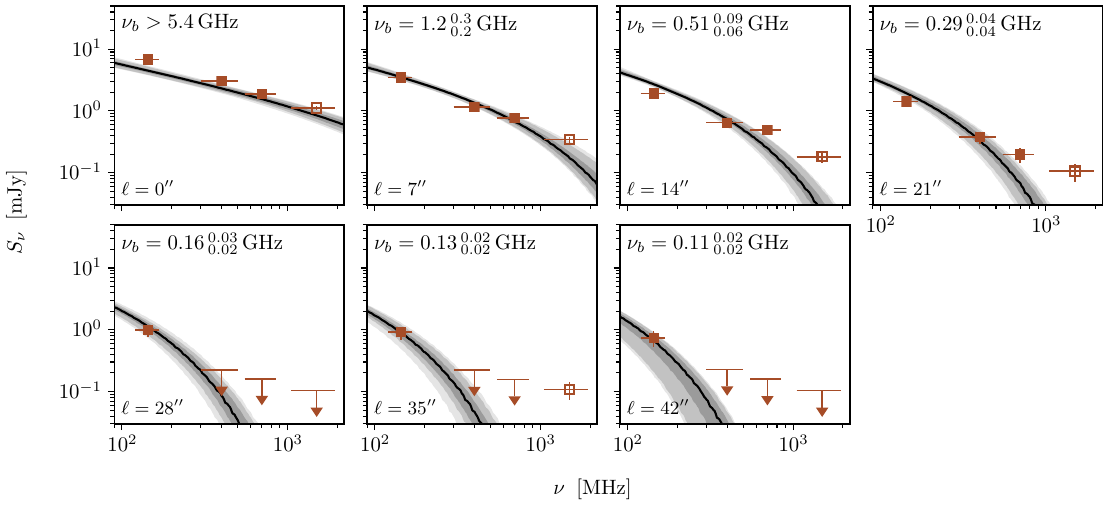}
    \caption{Radio continuum spectra and best-fit aging model along the stripped tail of NGC4858, fit only to the $144$, $400$, and $700\,\mathrm{MHz}$ data. Panels follow the same formatting as Fig.~\ref{fig:example_fit_result}.}
    \label{fig:specs_NGC4858_noL}
\end{figure*}

While the vast majority of flux-density measurements are well reproduced by the aging model (see Fig.~\ref{fig:model_offsets}), the galaxy in the sample with the largest deviations from the aging model predictions is NGC4858. This is visible in the spectrum plots in Fig.~\ref{fig:NGC4858_spec} where two clear offsets between the model and data are visible. First, at $\ell = 0\arcsec$ the model significantly under-predicts the observed $144\,\mathrm{MHz}$ flux density, and second, at $\ell = 14\arcsec$ and $\ell = 21\arcsec$ the model under-predicts the observed $1.5\,\mathrm{GHz}$ flux density.  The $\mathrm{S/N}$ of the deviations at $1.5\,\mathrm{GHz}$ are 5.4, 3.1, respectively. We note that there is also a flux excess in the $\ell = 35\arcsec$ bin in Fig.~\ref{fig:NGC4858_spec}, though we believe that this is most likely a noise spike since it is detached from the rest of the $1.5\,\mathrm{GHz}$ tail and it does not overlap with the observed extra-planar $\mathrm{H\alpha}$ emission for NGC4858 (see below). \citet{muller2021} present evidence for extra-planar star formation in a ram pressure stripped tail in the form of a flat spectral index between $1.4$ and $3\,\mathrm{GHz}$. It is possible that the aforementioned flux excesses at $1.5\,\mathrm{GHz}$ are also connected to ongoing star formation. We stress that this section is only meant to be a broad evaluation of the plausibility of this scenario. We recognize the substantial uncertainties, both in terms of the robustness of the $1.5\,\mathrm{GHz}$ excess outlined above, as well as on estimating extra-planar SFRs from narrowband imaging (see below).  Towards the end of this section we discuss these uncertainties more thoroughly and provide a forecast for how they can be reduced (though certainly not eliminated) moving forward.
\par
NGC4858 is one of the poster children in the Coma Cluster for ongoing star formation in a ram pressure stripped tail \citep{smith2010,yagi2010}. Qualitatively, extra emission from ongoing star formation could potentially explain the offsets between the data and model in Fig.~\ref{fig:NGC4858_spec}. Synchrotron emission from recent star formation will have a flat spectral index, and thus when superimposed with steep-spectrum, aged plasma, the relative contribution from star formation will be most apparent at the high-frequency end. Therefore extra-planar star formation could, at least partially, account for the $1.5\,\mathrm{GHz}$ excess while only making a relatively small contribution at low frequencies. In this scenario, the observed tail spectrum would be a composite of an aged spectrum (e.g.\ a JP model) and a flat, powerlaw spectrum.
\begin{equation} \label{eq:composite_model}
    S_\nu = \mathrm{JP}(S_0^\mathrm{aged}, \alpha_0, \nu_b) + S_{1.5}^\mathrm{PL} \left(\frac{\nu}{1.5\,\mathrm{GHz}}\right)^{-0.5}
\end{equation}
\noindent
where $\mathrm{JP}(S_0^\mathrm{aged}, \alpha_0, \nu_b)$ is an aged JP model as described in Sect.~\ref{sec:aging_model}, $S_{1.5}^\mathrm{PL}$ is the powerlaw flux density at $1.5\,\mathrm{GHz}$, and we opt for a powerlaw spectral index of $-0.5$ which is similar to the best-fit injection index for NGC4858 (see Figs.~\ref{fig:aging_model_results}, \ref{fig:NGC4858_corner}). Assuming a steeper injection index of $-0.6$ or $-0.7$ does not alter the qualitative conclusions.  We now explore whether such a composite model is able to account for the observed discrepancies between the data and a solely aging model for NGC4858.
\par
We first re-fit the data with the aging model described in Sect.~\ref{sec:aging_model}, but only include $144$, $400$, and $700\,\mathrm{MHz}$ flux densities in the fit, since in this scenario the $1.5\,\mathrm{GHz}$ flux density may be contaminated by emission from recent star formation. It is true that emission from star formation will affect flux densities at all frequencies, but again, superimposed on an aged plasma the relative impact will be largest at high frequency (especially for $\nu_b \lesssim 1\,\mathrm{GHz}$) and therefore omitting the $1.5\,\mathrm{GHz}$ data from the fit will give a better estimate for the aged plasma component. Fig.~\ref{fig:specs_NGC4858_noL} shows the spectra along the tail for NGC4858 (same data points as Fig.~\ref{fig:NGC4858_spec}), but now overlaid with an aging model fit only to the $144$, $400$, and $700\,\mathrm{MHz}$ data, not $1.5\,\mathrm{GHz}$. Compared to Fig.~\ref{fig:NGC4858_spec} the break frequencies are shifted to slightly lower frequencies, and the $144\,\mathrm{MHz}$ flux density at $\ell = 0\arcsec$ is no longer a $>3\sigma$ outlier. This is a result of the aging model not being compelled to account for the relatively large observed flux density at $1.5\,\mathrm{GHz}$. For the two previously mentioned distance bins along the tail ($\ell = 14\arcsec, 21\arcsec$) there is now slightly stronger evidence for excess $1.5\,\mathrm{GHz}$ emission with respect to the updated aging model.
\begin{figure}
    \centering
    \includegraphics[width = 0.5\textwidth]{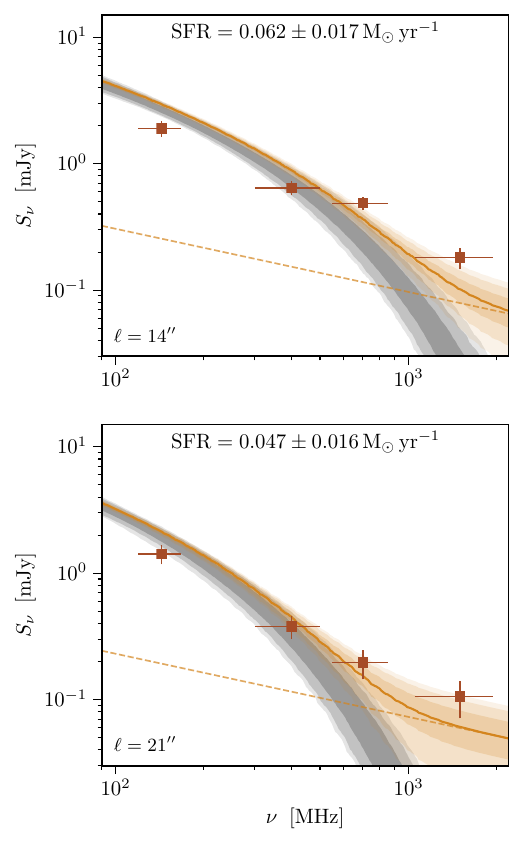}
    \caption{Radio continuum spectra for distance bins that show an excess of $1.5\,\mathrm{GHz}$ emission relative to the aging model. Data points correspond to observed flux densities, grey shading corresponds to the aging model, dashed gold line shows the best-fit powerlaw component, and the gold shading shows the composite aging plus powerlaw model. Each panel lists the best-fit SFR required in order to reproduce the observed flux density excess at $1.5\,\mathrm{GHz}$.}
    \label{fig:NGC4858_composite}
\end{figure}
\par
To determine the level of star formation required in the tail to account for this excess $1.5\,\mathrm{GHz}$ flux density, we add a powerlaw component to the aging model shown in Fig.~\ref{fig:specs_NGC4858_noL} and fit to the observed $144\,\mathrm{MHz}$ to $1.5\,\mathrm{GHz}$ flux densities to determine the best-fit powerlaw normalization at $1.5\,\mathrm{GHz}$ for the composite model described by Eq.~\ref{eq:composite_model}. This gives an estimate for the flux density at $1.5\,\mathrm{GHz}$ due to star formation that is required to account for the observed excess. We use the $\mathrm{SFR} - L_{1.5}$ calibration from \citet{kennicutt2012},
\begin{equation}
    \frac{\mathrm{SFR}}{\mathrm{M_\odot\,yr^{-1}}} = \left(\frac{L_{1.5}}{\mathrm{erg\,s^{-1}\,Hz^{-1}}}\right)\,10^{-28.20},
\end{equation}
\noindent
to convert this value to an expected SFR. In Fig.~\ref{fig:NGC4858_composite} we show the results of this composite model fitting for the two distance bins along the tail where there is an observed $1.5\,\mathrm{GHz}$ excess. Fig.~\ref{fig:NGC4858_composite} shows the aging component (grey) as well as the best-fit powerlaw component and composite model (gold). We also list the required SFR for each distance bin derived from the best-fit normalization for the powerlaw component. The required SFRs are between $10^{-2}$ and $10^{-1}\,\mathrm{M_\odot\,yr^{-1}}$, for a total of $0.11 \pm 0.02\,\mathrm{M_\odot\,yr^{-1}}$. Not unprecedented, but on the high end relative to extra-planar SFRs that have been observed in other galaxies \citep[e.g.][]{sun2007,hester2010,fumagalli2011,vulcani2018_sf,cramer2019,poggianti2019_extraplanar_sf,boselli2021_ic3476,junais2021}. Because there are existing observations showing $\mathrm{H\alpha}$ emission within the stripped tail of NGC4858, we can test more directly if there is sufficient star formation in the tail to match the required SFRs listed in Fig.~\ref{fig:NGC4858_composite}. We note that the distance bins considered above, both overlap spatially with the observed $\mathrm{H\alpha}$ tail for NGC4858 (see \citealt{yagi2010}).
\par
To constrain observed SFRs in the tail of NGC4858 we use narrowband $\mathrm{H\alpha} + \textsc{Nii}$ imaging from \citet{yagi2010} taken with the Subaru Suprime-Cam. The line emission is isolated by using the $R$-band for continuum subtraction, though we note that there is significant uncertainty associated with the continuum subtraction for regions where there is strong underlying continuum emission (see \citealt{yagi2010,yagi2017} for a more detailed discussion), e.g.\ galaxy discs. Here we only consider extra-planar emission in the tail, where strong continuum emission is not expected, and therefore the continuum subtraction in order to isolate the line emission is relatively reliable.
\par
\begin{figure}
    \centering
    \includegraphics[width = 0.5\textwidth]{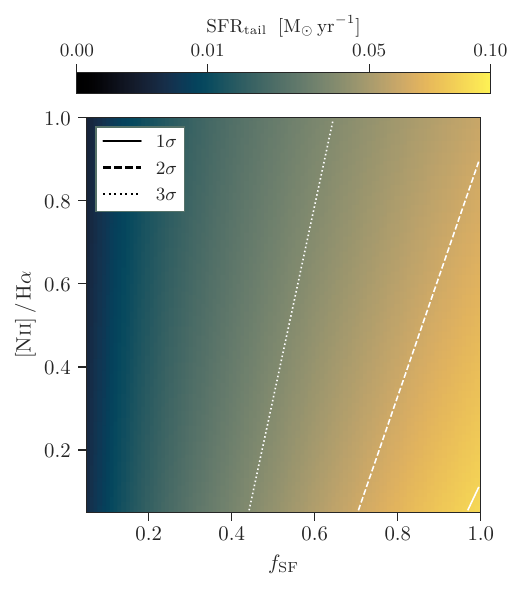}
    \caption{The SFR derived from observed $\mathrm{H\alpha} + \textsc{Nii}$ emission in the tail, for a grid of different assumed values for the $\mathrm{[\textsc{Nii}]}$-to-$\mathrm{H\alpha}$ ratio and $f_\mathrm{SF}$.  We also plot $1\sigma$, $2\sigma$, and $3\sigma$ contours for the `required' SFRs from the composite fits shown in Fig.~\ref{fig:NGC4858_composite}.}
    \label{fig:sfr_ratio}
\end{figure}

We measure the narrowband flux over the tail regions and this flux is converted to a SFR assuming the calibration from \citet{kennicutt2012},
\begin{equation}
    \frac{\mathrm{SFR}}{\mathrm{M_\odot\,yr^{-1}}} = \left(\frac{L_\mathrm{H\alpha}}{\mathrm{erg\,s^{-1}}}\right)\,10^{-41.27}.
\end{equation}
\noindent
We calculate SFRs for a range of assumed $\mathrm{[\textsc{Nii}]}$-to-$\mathrm{H\alpha}$ ratios, as this cannot be directly constrained from the narrowband imaging.  We also consider a range in $f_\mathrm{SF}$, which we define as the fraction of $\mathrm{H\alpha} + \textsc{Nii}$ flux originating from extraplanar star formation, as the ionization source (and therefore, loosely, $f_\mathrm{SF}$) cannot be constrained directly from narrowband imaging.
\par
In Fig.~\ref{fig:sfr_ratio} we plot the `observed' SFR, derived from the $\mathrm{H\alpha} + \textsc{Nii}$ flux in the tail, for a grid of different assumed values for the $\mathrm{[\textsc{Nii}]}$-to-$\mathrm{H\alpha}$ ratio and $f_\mathrm{SF}$.  We also plot $1\sigma$, $2\sigma$, and $3\sigma$ contours corresponding to the best-fit, total SFR in the tail from the composite fits shown in Fig.~\ref{fig:NGC4858_composite}.  It is possible to account for the best-fit SFR from the observed $\mathrm{H\alpha} + \textsc{Nii}$ emission, within the fit uncertainties, but doing so requires both a very large value for $f_\mathrm{SF}$ and a relatively low value for the $\mathrm{[\textsc{Nii}]}$-to-$\mathrm{H\alpha}$ ratio. Not impossible, but in tension with most previous constraints on these parameters \citep[e.g.][]{yagi2007,cramer2019,poggianti2019_extraplanar_sf}. While it is difficult to account for the full excess $1.5\,\mathrm{GHz}$ emission in the tail, for NGC4858, Fig.~\ref{fig:sfr_ratio} shows that a significant portion of this excess could be explained by extraplanar star formation.
\par
Of course it is paramount to acknowledge the uncertainties beyond the $\mathrm{[\textsc{Nii}]}$-to-$\mathrm{H\alpha}$ ratio and $f_\mathrm{SF}$ that go into Fig.~\ref{fig:sfr_ratio}.  These include uncertainties around internal extinction in star-forming regions in the tail, continuum subtraction in the narrowband, and likely most significant, uncertainties on the calibrations from $\mathrm{H\alpha}$ and $1.5\,\mathrm{GHz}$ luminosities to SFR. Not to mention the assumptions that go into the aging model (see Sect.~\ref{sec:aging_model}) that directly feed into $\mathrm{SFR_{fit}}$. Bearing all this in mind, factor of two uncertainties on the SFRs in this section are likely a minimum. Thus while we cannot rule out the scenario where extra-planar star formation is driving the $1.5\,\mathrm{GHz}$ excess, a stronger statement is not possible. Forthcoming integral field unit (IFU) spectroscopy with the WEAVE Large IFU (PI Roberts) of the $\sim\!30$ LoTSS jellyfish galaxies \citep{roberts2021_LOFARclust} in the Coma Cluster (including NGC4858) and their stripped tails, will alleviate some of the aforementioned uncertainties on the measured SFR in the tail. Deep radio continuum imaging at higher frequencies (e.g. $3-5\,\mathrm{GHz}$) will also be important for confirming the existence of excess emission at high frequency.

\section{Plots and images for individual galaxies} \label{sec:galaxy_appendix}

In this appendix we include images of all galaxies from our analysis with the disc and tail apertures overlaid.  For galaxies that are included in the tail spectral modelling analysis, we also show plots of continuum spectra at various distances along the tail as well as corner plots showing the results of the MCMC fits.

\clearpage

\subsection{NGC4848}

\par\smallskip\noindent
\centerline{\begin{minipage}{\textwidth}
\includegraphics[width = \textwidth]{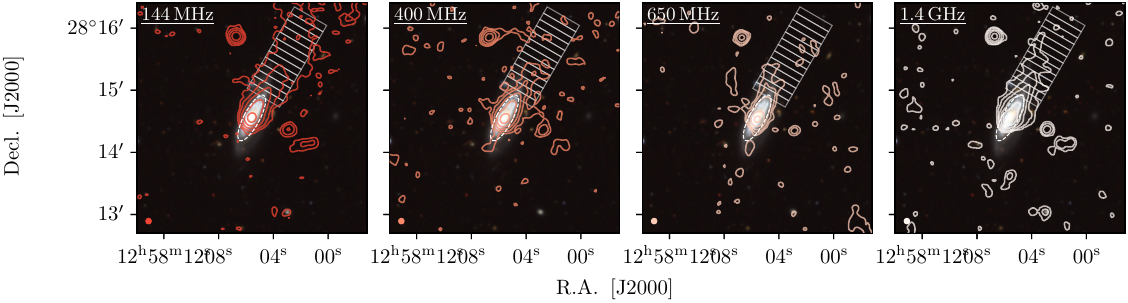}
\captionof{figure}{Radio -- optical overlay images for the galaxy NGC4848. Background image is a three-colour $grz$ image from the Legacy Surveys \citep{dey2019}. Contour levels begin at $2.5\times\mathrm{rms}$ and increase by a factor of three with each step. In the lower left we show the common FWHM beam size of $6.5\arcsec$. In each panel we also overlay the galaxy disc aperture (dashed ellipse) and the apertures used to measure radio continuum flux densities along the tail (solid rectangles).}
\label{fig:example_imgs_NGC4848}
\end{minipage}}
\par\smallskip

\par\smallskip\noindent
\centerline{\begin{minipage}{\textwidth}
\centering
\includegraphics[width = 0.95\textwidth]{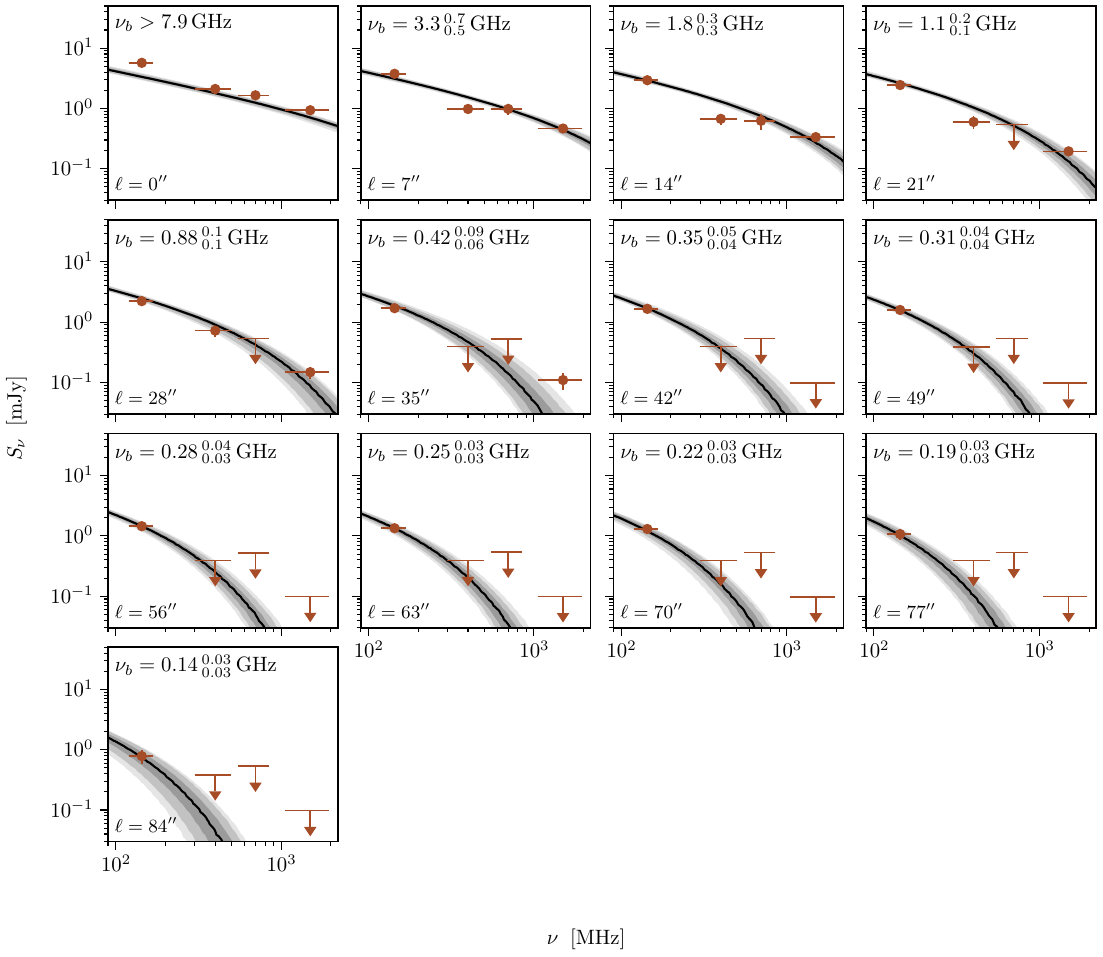}
\captionof{figure}{Tail spectral fitting results for NGC4848.  See Fig.~\ref{fig:example_fit_result} for details.}
\label{fig:NGC4848_spec}
\end{minipage}}
\par\smallskip

\par\smallskip\noindent
\centerline{\begin{minipage}{\textwidth}
\centering
\includegraphics[width = \textwidth]{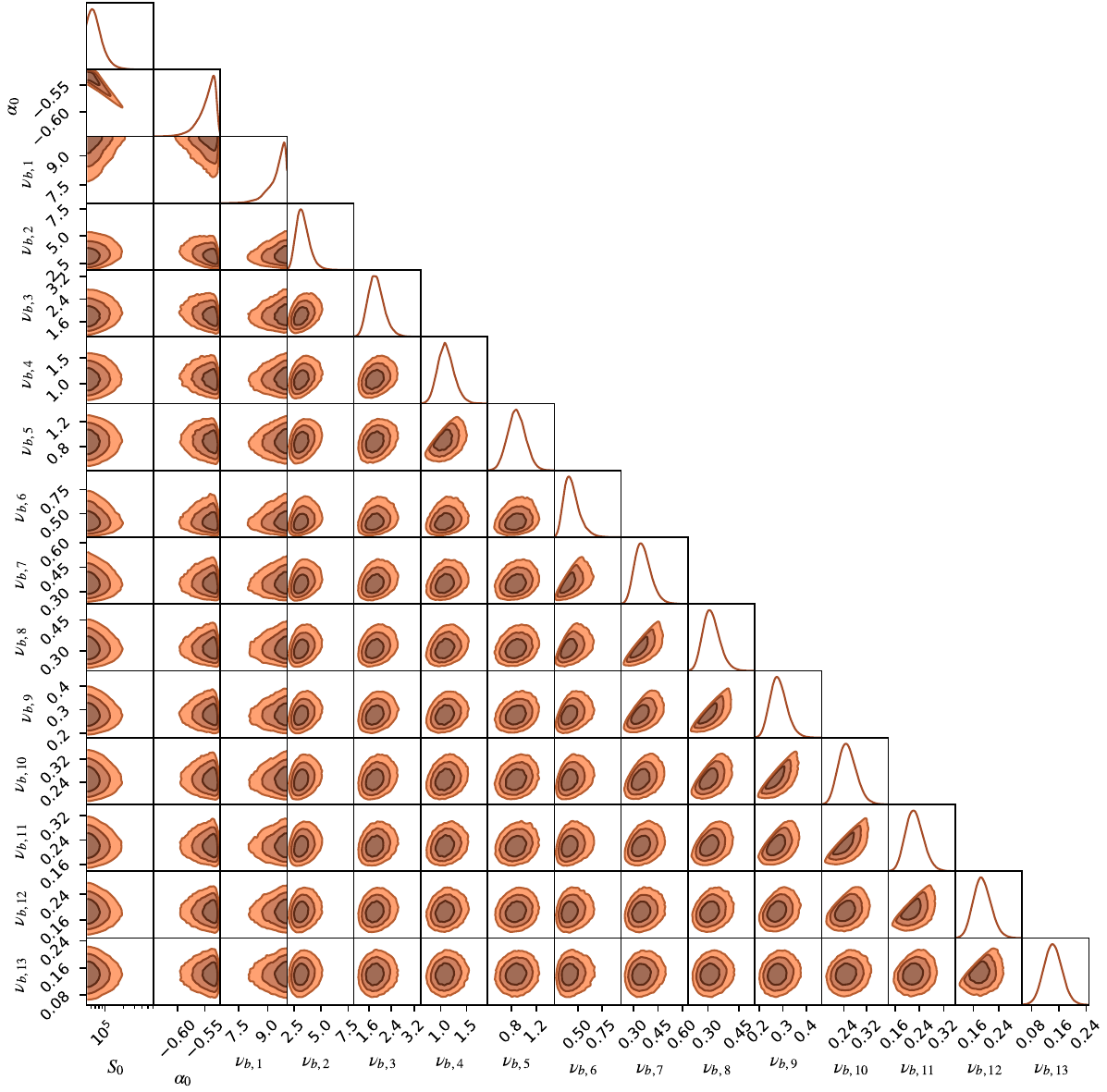}
\captionof{figure}{MCMC fitting results for the aging model along the tail.}
\label{fig:NGC4848_corner}
\end{minipage}}
\par\smallskip

\clearpage

\subsection{NGC4853}

\par\smallskip\noindent
\centerline{\begin{minipage}{\textwidth}
\centering
\includegraphics[width = \textwidth]{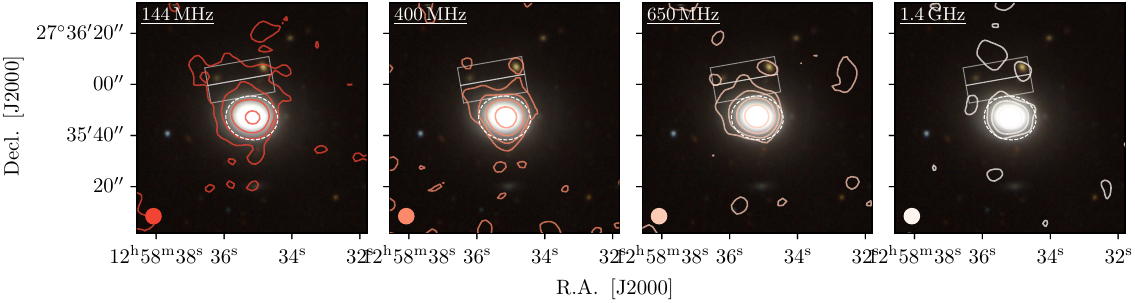}
\captionof{figure}{NGC4853, see Fig.~\ref{fig:example_imgs_NGC4848} for details.}
\label{fig:example_imgs_NGC4858}
\end{minipage}}
\par\smallskip

\clearpage

\subsection{NGC4858}

\par\smallskip\noindent
\centerline{\begin{minipage}{\textwidth}
\centering
\includegraphics[width = \textwidth]{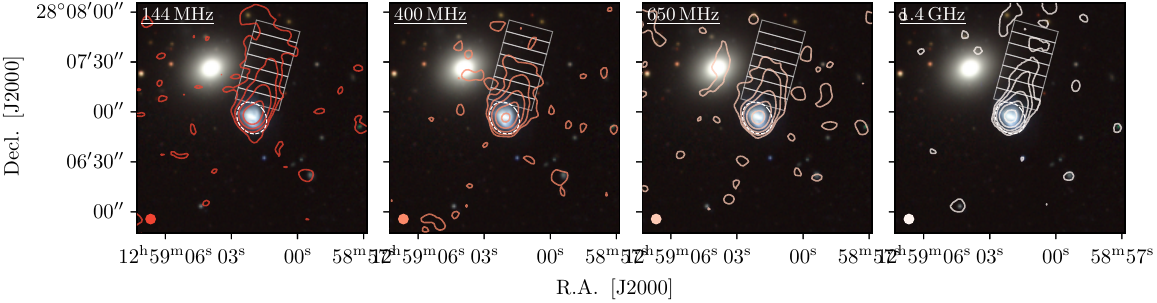}
\captionof{figure}{NGC4858, see Fig.~\ref{fig:example_imgs_NGC4848} for details.}
\label{fig:example_imgs_NGC4858}
\end{minipage}}
\par\smallskip

\par\smallskip\noindent
\centerline{\begin{minipage}{\textwidth}
\centering
\includegraphics[width = \textwidth]{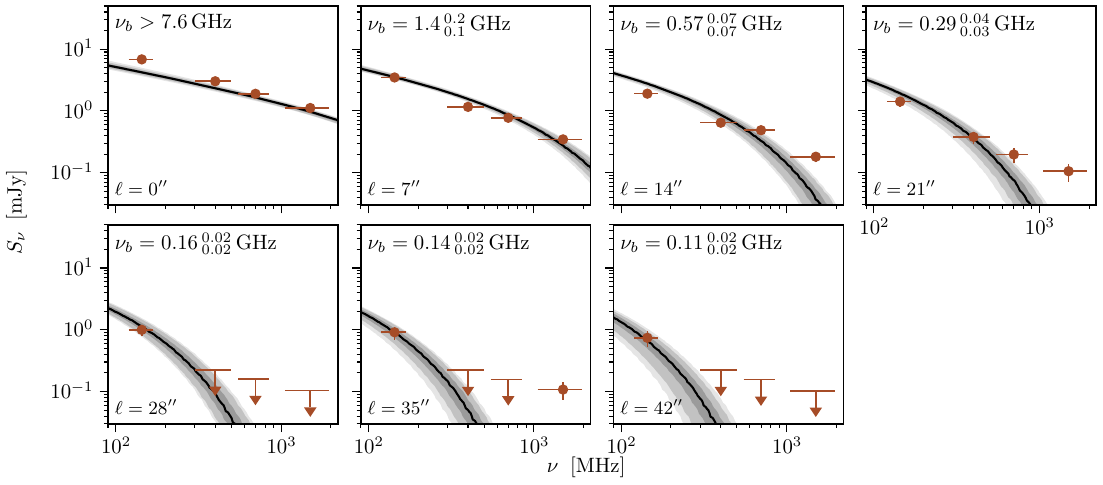}
\captionof{figure}{Tail spectral fitting results for NGC4858.  See Fig.~\ref{fig:example_fit_result} for details.}
\label{fig:NGC4858_spec}
\end{minipage}}
\par\smallskip

\par\smallskip\noindent
\centerline{\begin{minipage}{\textwidth}
\centering
\includegraphics[width = \textwidth]{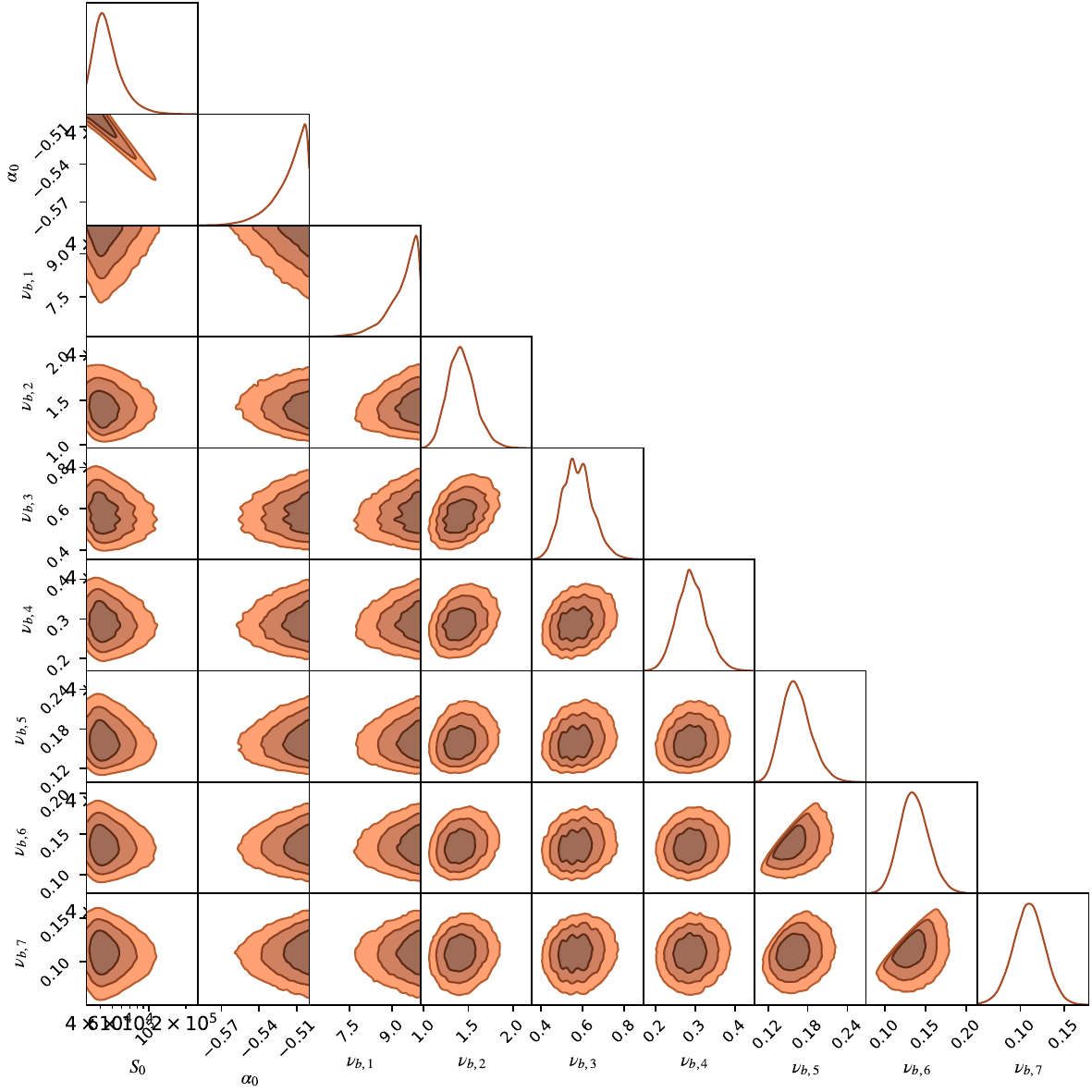}
\captionof{figure}{MCMC fitting results for the aging model along the tail.}
\label{fig:NGC4858_corner}
\end{minipage}}
\par\smallskip

\clearpage

\subsection{NGC4911}

\par\smallskip\noindent
\centerline{\begin{minipage}{\textwidth}
\centering
\includegraphics[width = \textwidth]{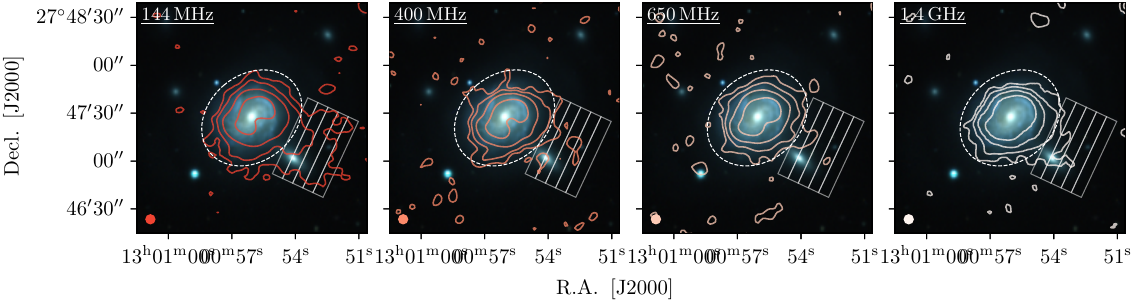}
\captionof{figure}{NGC4911, see Fig.~\ref{fig:example_imgs_NGC4848} for details.}
\label{fig:example_imgs_NGC4911}
\end{minipage}}
\par\smallskip

\par\smallskip\noindent
\centerline{\begin{minipage}{\textwidth}
\centering
\includegraphics[width = \textwidth]{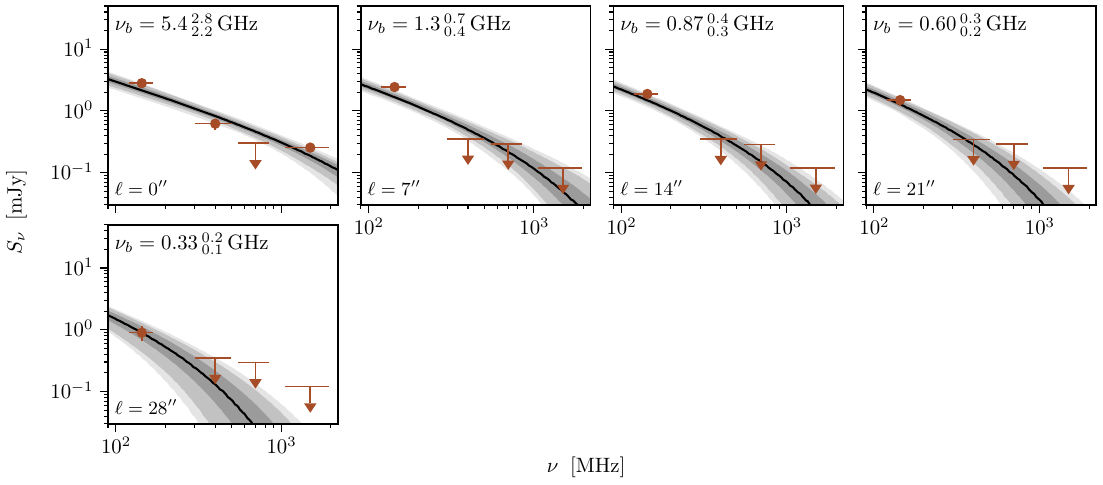}
\captionof{figure}{Tail spectral fitting results for NGC4911.  See Fig.~\ref{fig:example_fit_result} for details.}
\label{fig:NGC4911_spec}
\end{minipage}}
\par\smallskip

\par\smallskip\noindent
\centerline{\begin{minipage}{\textwidth}
\centering
\includegraphics[width = \textwidth]{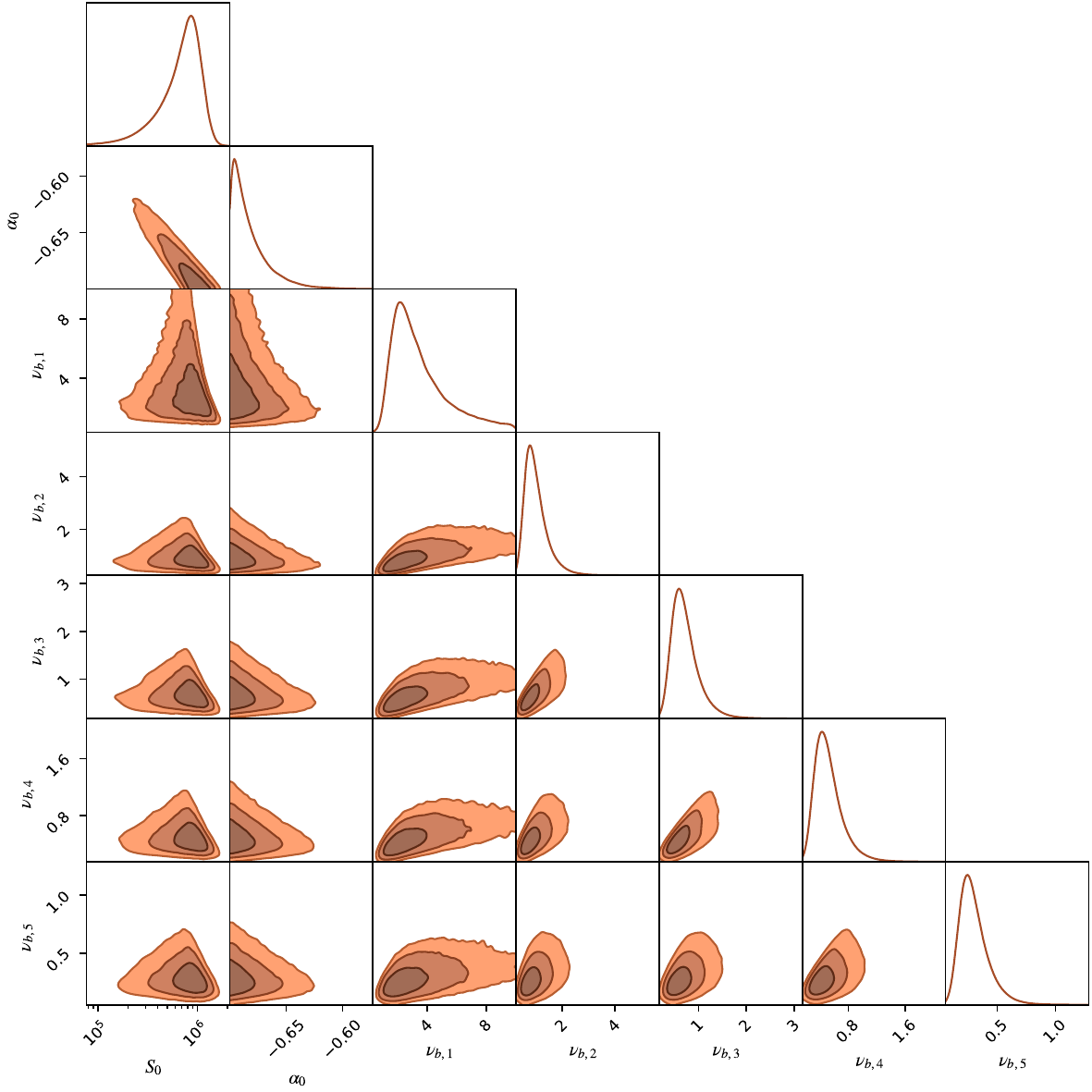}
\captionof{figure}{MCMC fitting results for the aging model along the tail.}
\label{fig:NGC4911_corner}
\end{minipage}}
\par\smallskip

\clearpage

\subsection{IC3913}

\par\smallskip\noindent
\centerline{\begin{minipage}{\textwidth}
\centering
\includegraphics[width = \textwidth]{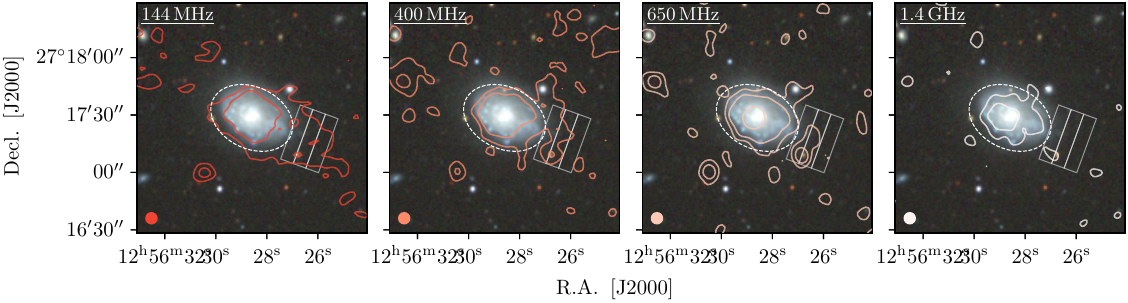}
\captionof{figure}{IC3913, see Fig.~\ref{fig:example_imgs_NGC4848} for details.}
\label{fig:example_imgs_IC3913}
\end{minipage}}
\par\smallskip

\par\smallskip\noindent
\centerline{\begin{minipage}{\textwidth}
\centering
\includegraphics[width = \textwidth]{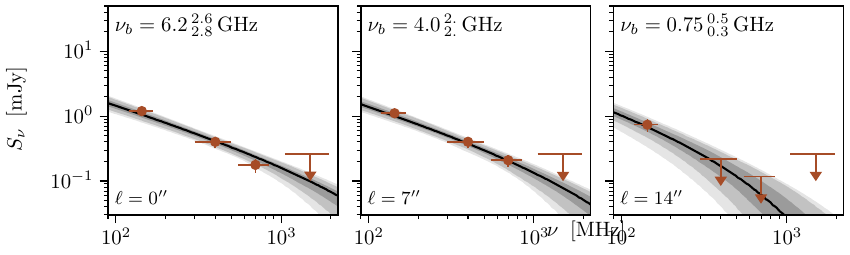}
\captionof{figure}{Tail spectral fitting results for IC3913.  See Fig.~\ref{fig:example_fit_result} for details.}
\label{fig:IC3913_spec}
\end{minipage}}
\par\smallskip

\par\smallskip\noindent
\centerline{\begin{minipage}{\textwidth}
\centering
\includegraphics[width = \textwidth]{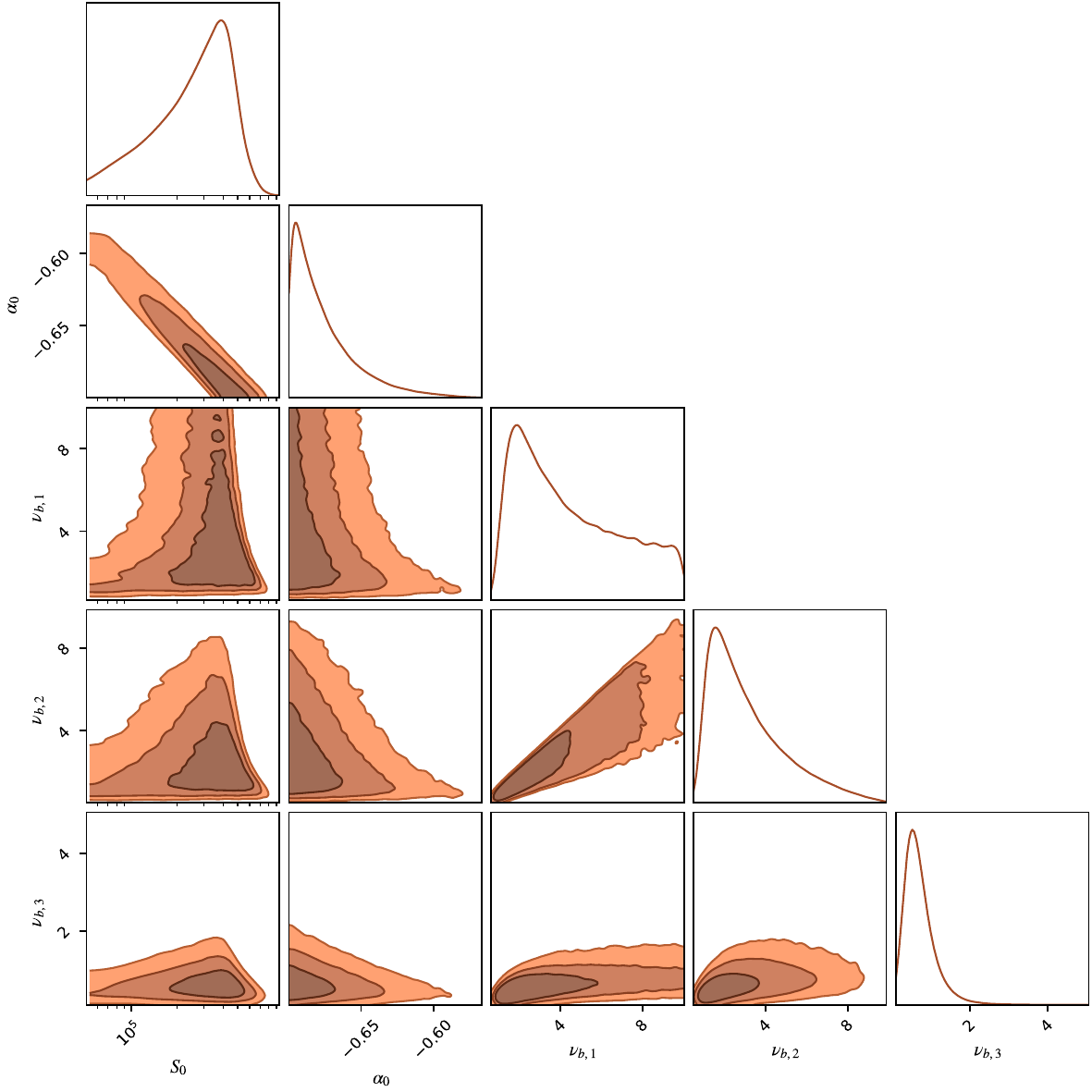}
\captionof{figure}{MCMC fitting results for the aging model along the tail.}
\label{fig:IC3913_corner}
\end{minipage}}
\par\smallskip

\clearpage

\subsection{IC3949}

\par\smallskip\noindent
\centerline{\begin{minipage}{\textwidth}
\centering
\includegraphics[width = \textwidth]{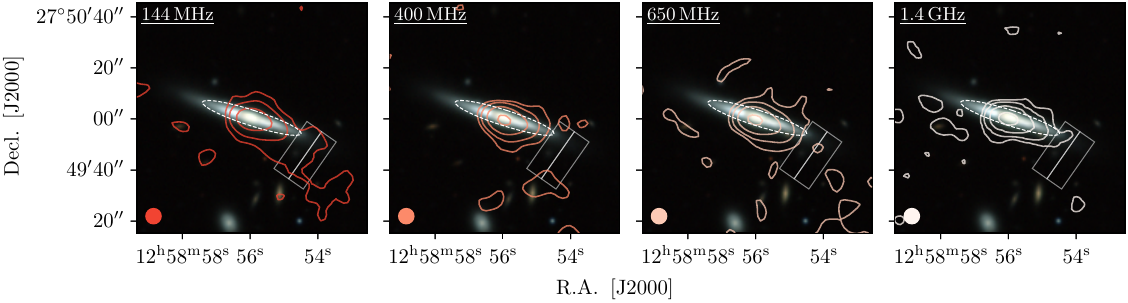}
\captionof{figure}{IC3949, see Fig.~\ref{fig:example_imgs_NGC4848} for details.}
\label{fig:example_imgs_IC3949}
\end{minipage}}
\par\smallskip

\clearpage

\subsection{IC4040}

\par\smallskip\noindent
\centerline{\begin{minipage}{\textwidth}
\centering
\includegraphics[width = \textwidth]{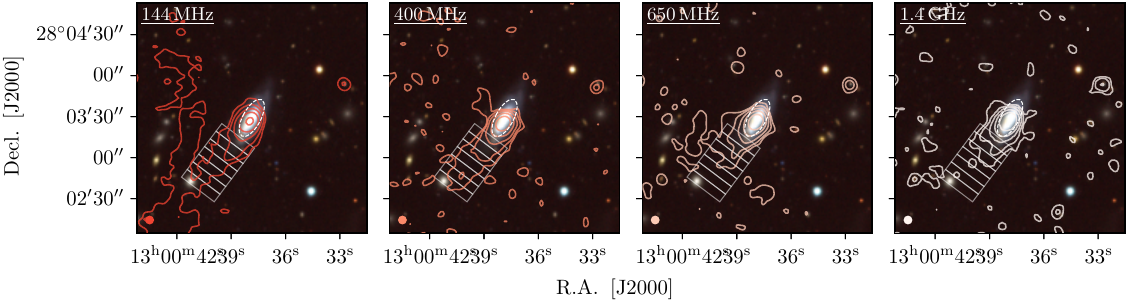}
\captionof{figure}{IC4040, see Fig.~\ref{fig:example_imgs_NGC4848} for details.}
\label{fig:example_imgs_IC4040}
\end{minipage}}
\par\smallskip

\par\smallskip\noindent
\centerline{\begin{minipage}{\textwidth}
\centering
\includegraphics[width = \textwidth]{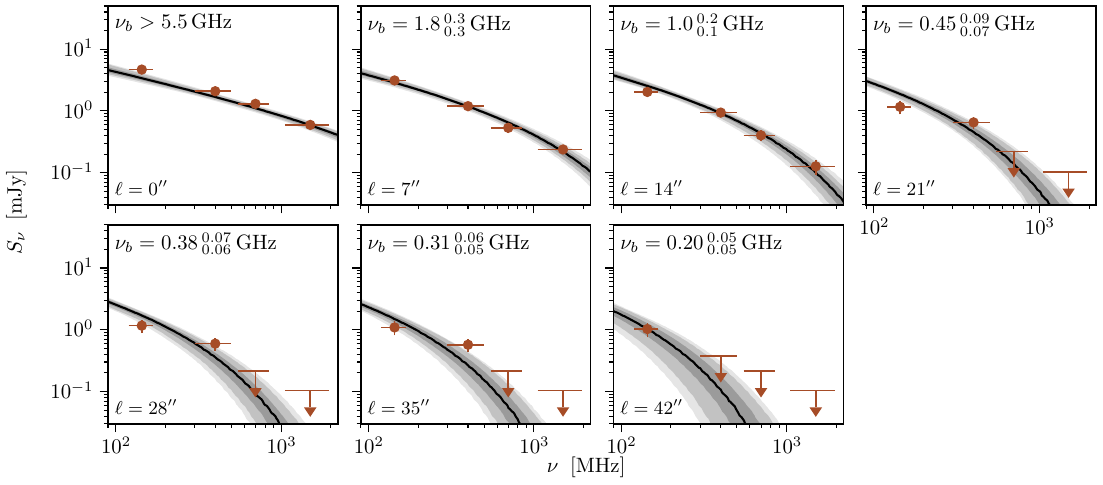}
\captionof{figure}{Tail spectral fitting results for IC4040.  See Fig.~\ref{fig:example_fit_result} for details.}
\label{fig:IC4040_spec}
\end{minipage}}
\par\smallskip

\par\smallskip\noindent
\centerline{\begin{minipage}{\textwidth}
\centering
\includegraphics[width = \textwidth]{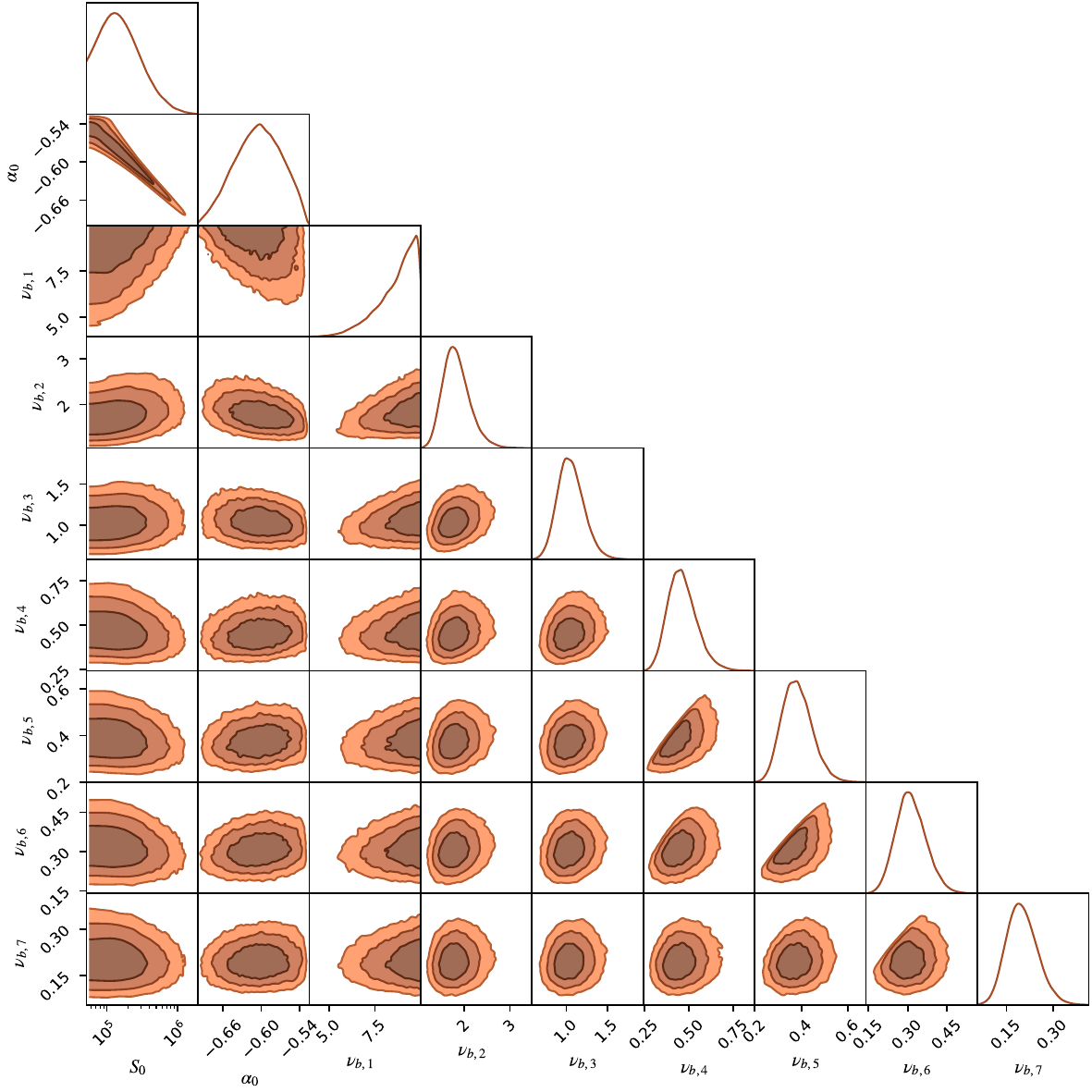}
\captionof{figure}{MCMC fitting results for the aging model along the tail.}
\label{fig:IC4040_corner}
\end{minipage}}
\par\smallskip

\clearpage

\subsection{KUG1250+276}

\par\smallskip\noindent
\centerline{\begin{minipage}{\textwidth}
\centering
\includegraphics[width = \textwidth]{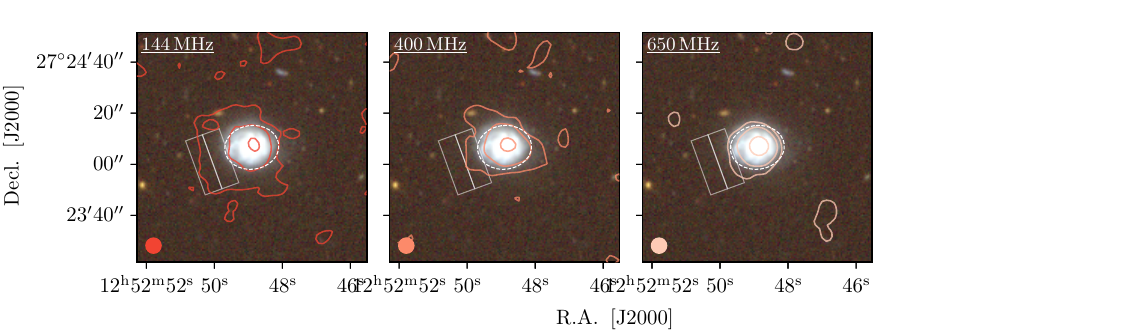}
\captionof{figure}{KUG1250+276, see Fig.~\ref{fig:example_imgs_NGC4848} for details.}
\label{fig:example_imgs_KUG1250+276}
\end{minipage}}
\par\smallskip

\clearpage

\subsection{KUG1255+275}

\par\smallskip\noindent
\centerline{\begin{minipage}{\textwidth}
\centering
\includegraphics[width = \textwidth]{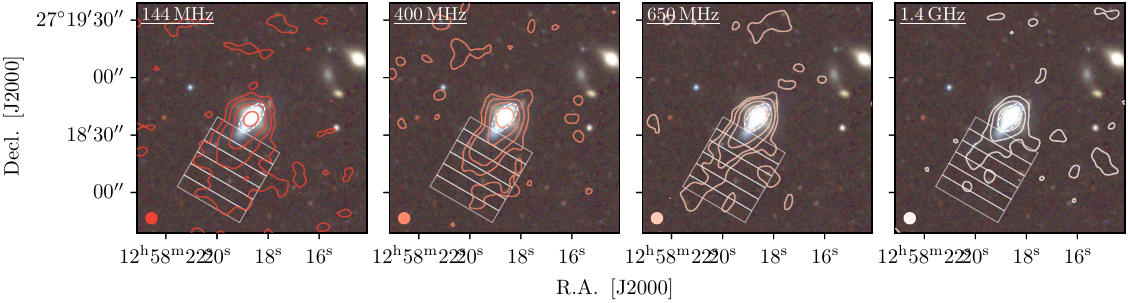}
\captionof{figure}{KUG1255+275, see Fig.~\ref{fig:example_imgs_NGC4848} for details.}
\label{fig:example_imgs_KUG1255+275}
\end{minipage}}
\par\smallskip

\par\smallskip\noindent
\centerline{\begin{minipage}{\textwidth}
\centering
\includegraphics[width = \textwidth]{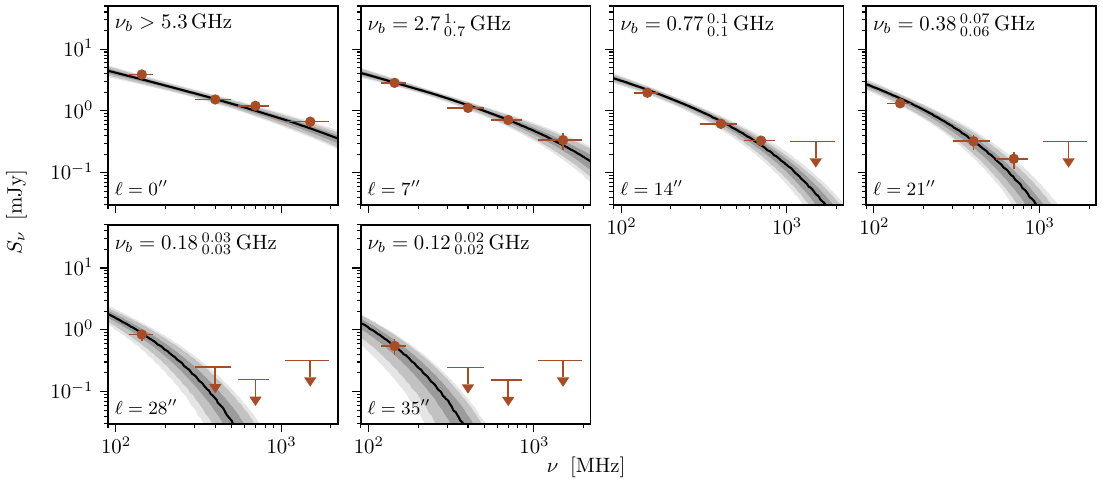}
\captionof{figure}{Tail spectral fitting results for KUG1255+275.  See Fig.~\ref{fig:example_fit_result} for details.}
\label{fig:KUG1255+275_spec}
\end{minipage}}
\par\smallskip

\par\smallskip\noindent
\centerline{\begin{minipage}{\textwidth}
\centering
\includegraphics[width = \textwidth]{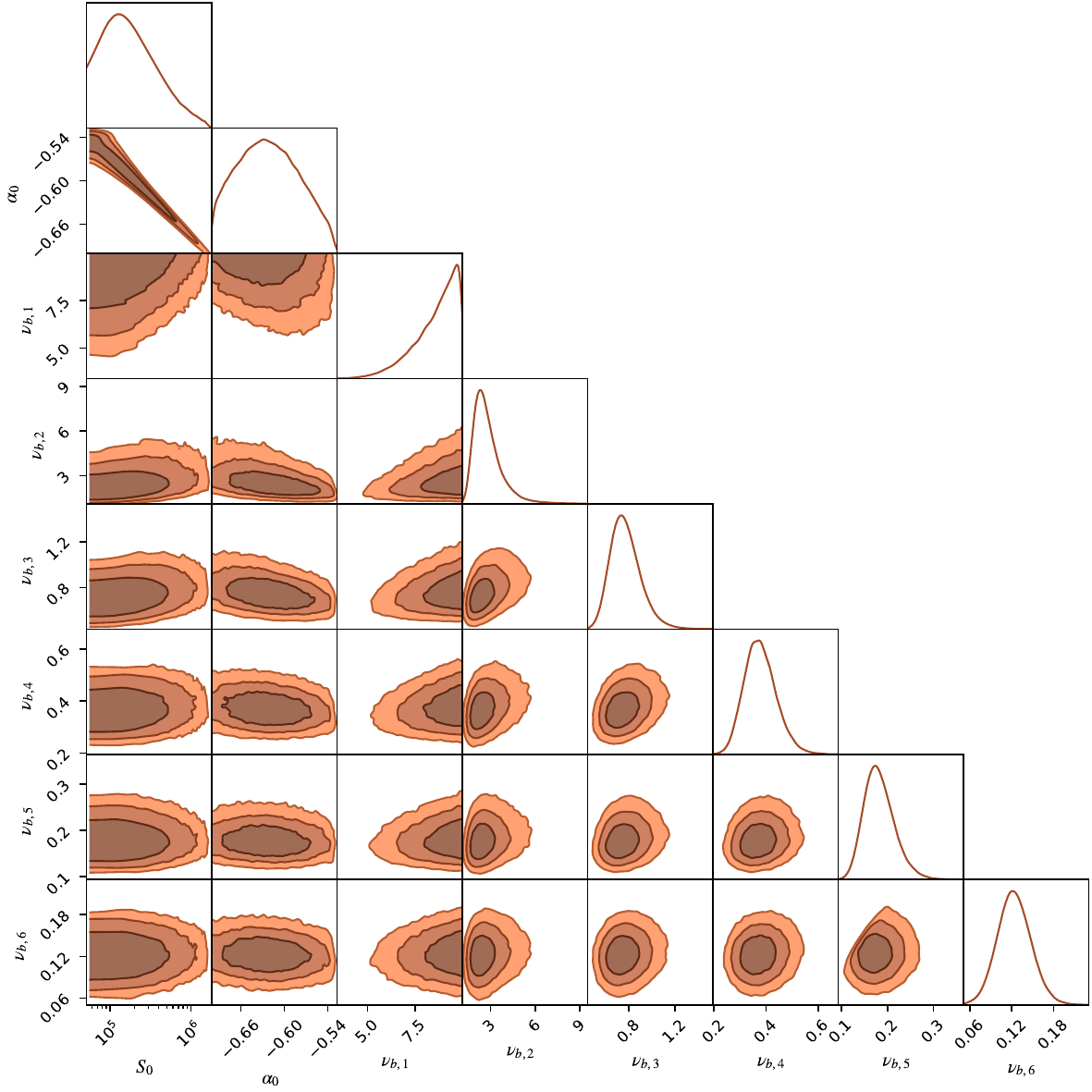}
\captionof{figure}{MCMC fitting results for the aging model along the tail.}
\label{fig:KUG1255+275_corner}
\end{minipage}}
\par\smallskip

\clearpage

\subsection{KUG1255+283}

\par\smallskip\noindent
\centerline{\begin{minipage}{\textwidth}
\centering
\includegraphics[width = \textwidth]{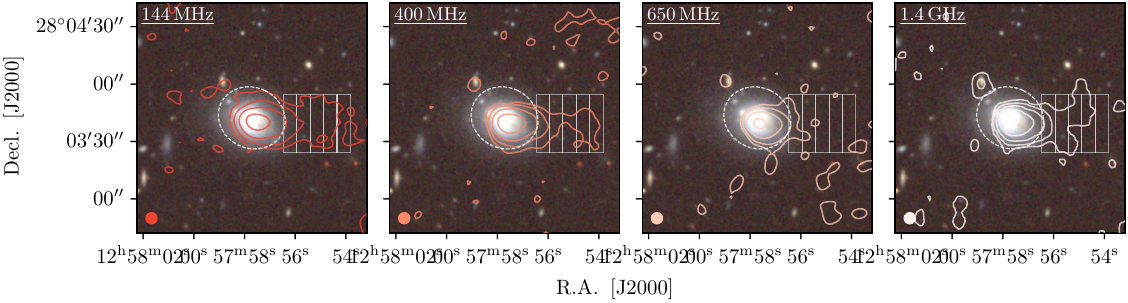}
\captionof{figure}{KUG1255+283, see Fig.~\ref{fig:example_imgs_NGC4848} for details.}
\label{fig:example_imgs_KUG1255+283}
\end{minipage}}
\par\smallskip

\par\smallskip\noindent
\centerline{\begin{minipage}{\textwidth}
\centering
\includegraphics[width = \textwidth]{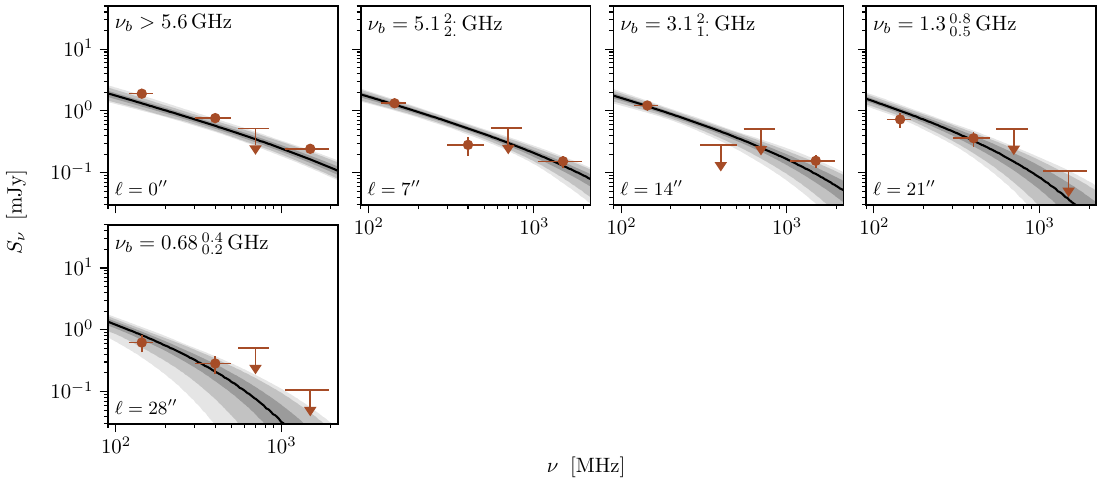}
\captionof{figure}{Tail spectral fitting results for KUG1255+283.  See Fig.~\ref{fig:example_fit_result} for details.}
\label{fig:KUG1255+283_spec}
\end{minipage}}
\par\smallskip

\par\smallskip\noindent
\centerline{\begin{minipage}{\textwidth}
\centering
\includegraphics[width = \textwidth]{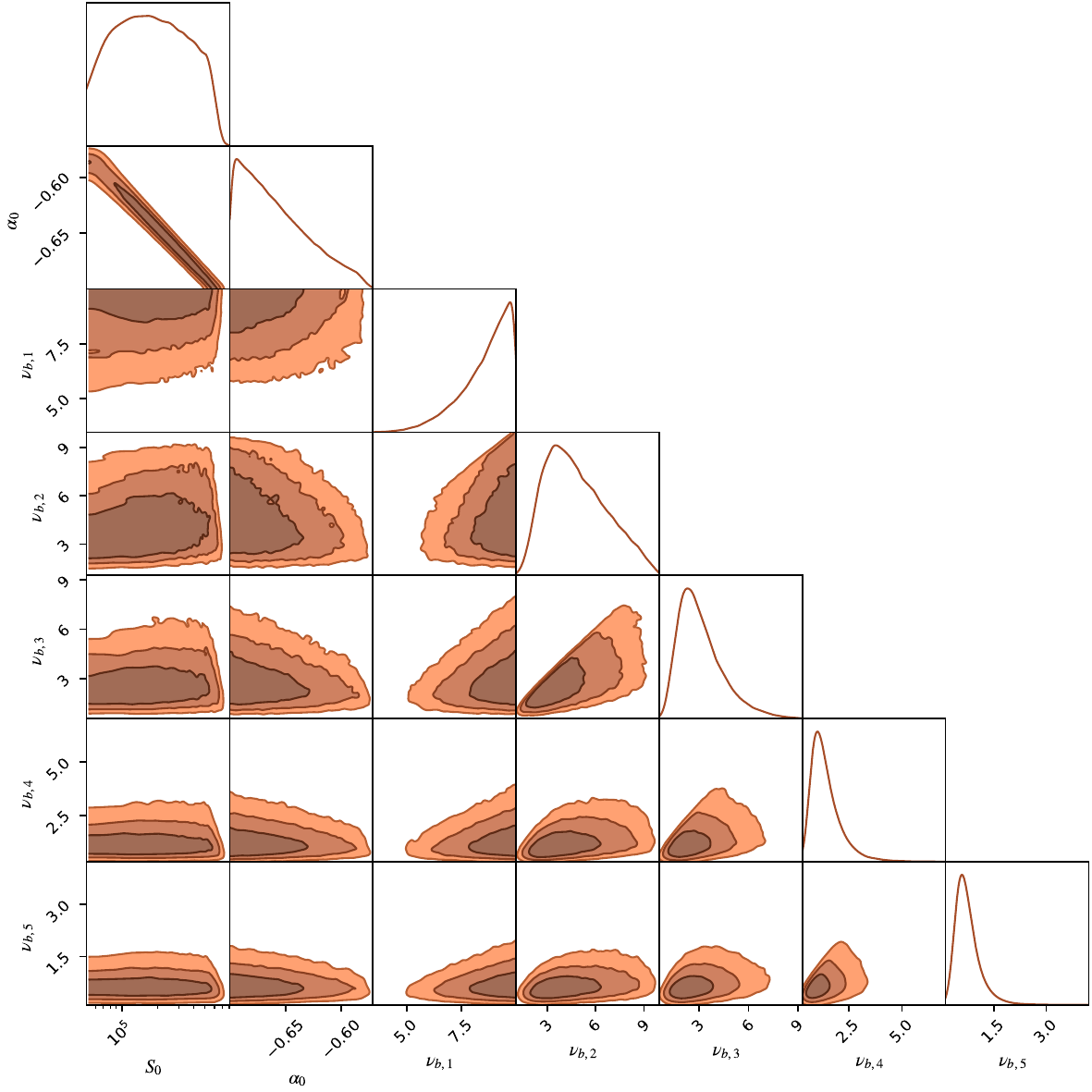}
\captionof{figure}{MCMC fitting results for the aging model along the tail.}
\label{fig:KUG1255+283_corner}
\end{minipage}}
\par\smallskip

\clearpage

\subsection{KUG1256+287B}

\par\smallskip\noindent
\centerline{\begin{minipage}{\textwidth}
\centering
\includegraphics[width = \textwidth]{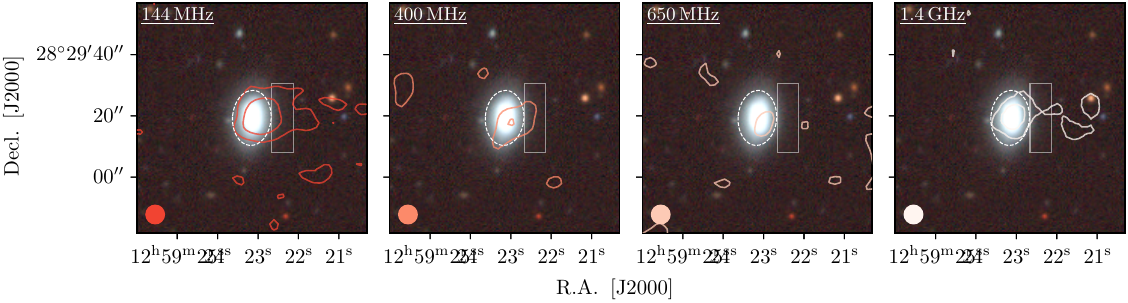}
\captionof{figure}{KUG1256+287B, see Fig.~\ref{fig:example_imgs_NGC4848} for details.}
\label{fig:example_imgs_KUG1256+287B}
\end{minipage}}
\par\smallskip

\clearpage

\subsection{KUG1257+288B}

\par\smallskip\noindent
\centerline{\begin{minipage}{\textwidth}
\centering
\includegraphics[width = \textwidth]{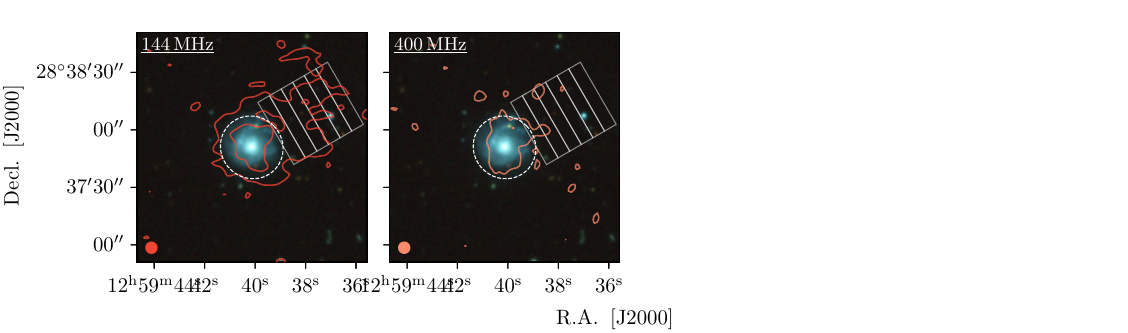}
\captionof{figure}{KUG1257+288B, see Fig.~\ref{fig:example_imgs_NGC4848} for details.}
\label{fig:example_imgs_KUG1257+288B}
\end{minipage}}
\par\smallskip

\par\smallskip\noindent
\centerline{\begin{minipage}{\textwidth}
\centering
\includegraphics[width = \textwidth]{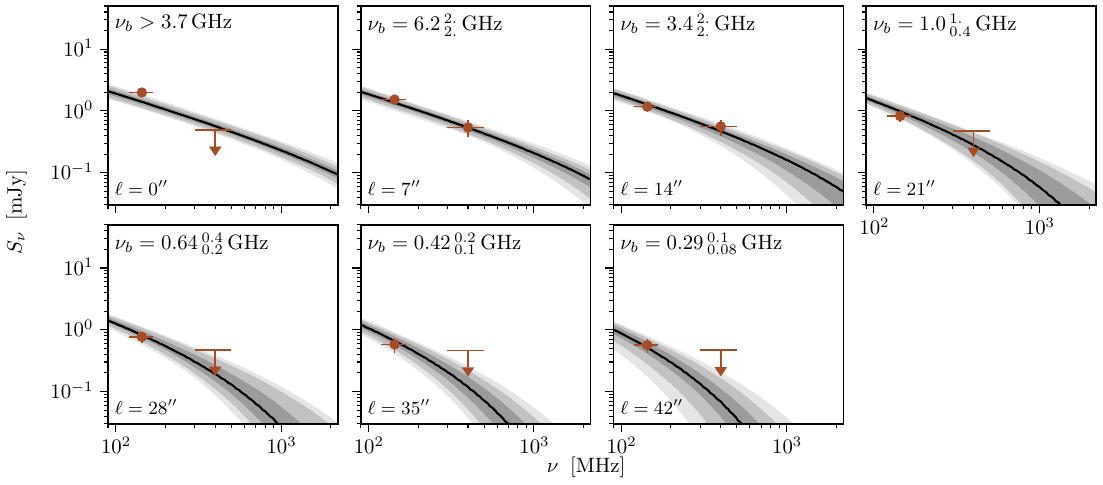}
\captionof{figure}{Tail spectral fitting results for KUG1257+288B.  See Fig.~\ref{fig:example_fit_result} for details.}
\label{fig:KUG1257+288B_spec}
\end{minipage}}
\par\smallskip

\par\smallskip\noindent
\centerline{\begin{minipage}{\textwidth}
\centering
\includegraphics[width = \textwidth]{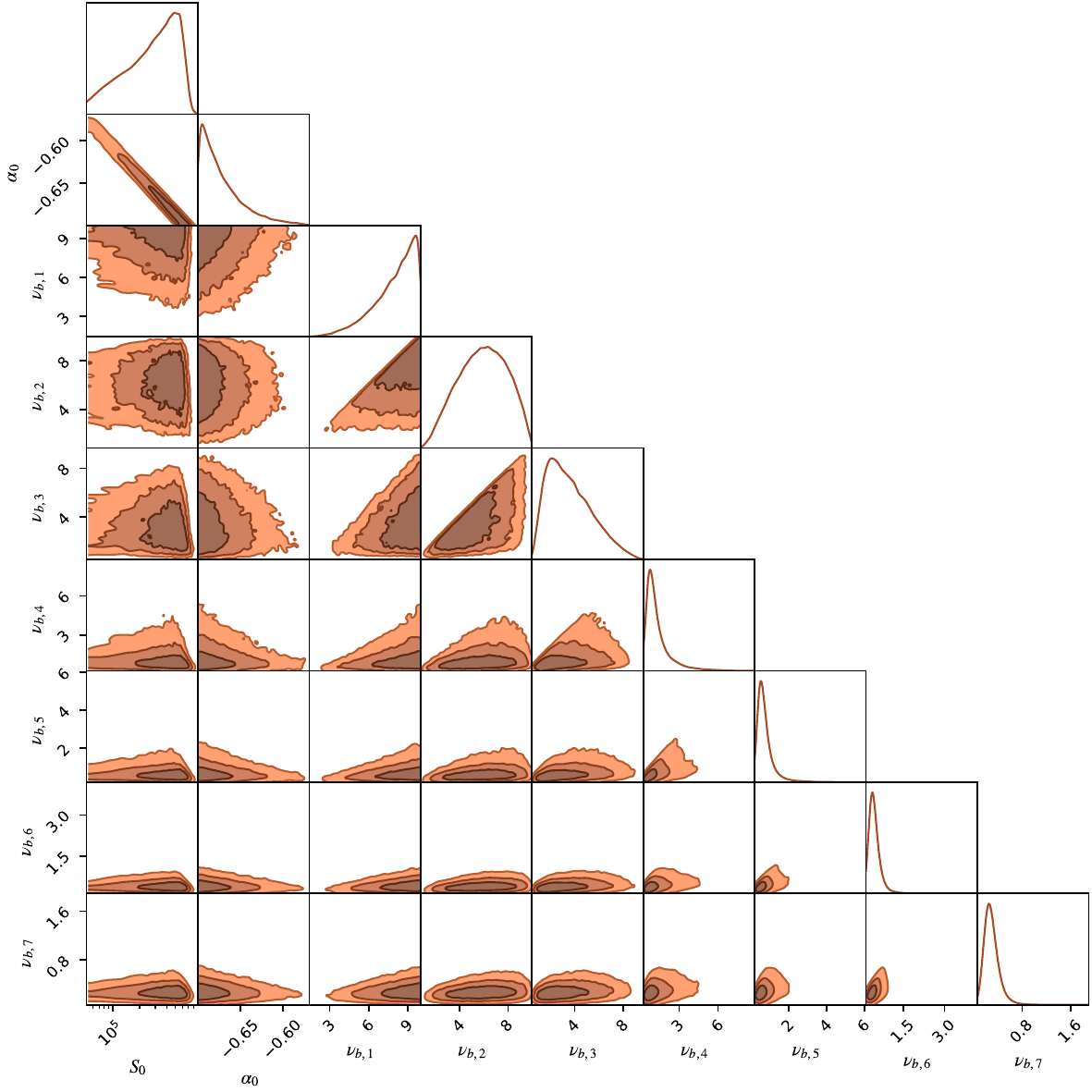}
\captionof{figure}{MCMC fitting results for the aging model along the tail.}
\label{fig:KUG1257+288B_corner}
\end{minipage}}
\par\smallskip

\clearpage

\subsection{KUG1258+279A}

\par\smallskip\noindent
\centerline{\begin{minipage}{\textwidth}
\centering
\includegraphics[width = \textwidth]{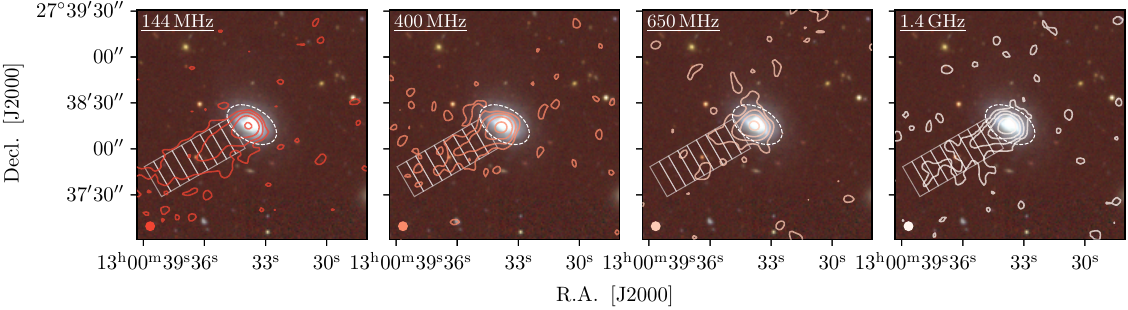}
\captionof{figure}{KUG1258+279A, see Fig.~\ref{fig:example_imgs_NGC4848} for details.}
\label{fig:example_imgs_KUG1258+279A}
\end{minipage}}
\par\smallskip

\par\smallskip\noindent
\centerline{\begin{minipage}{\textwidth}
\centering
\includegraphics[width = \textwidth]{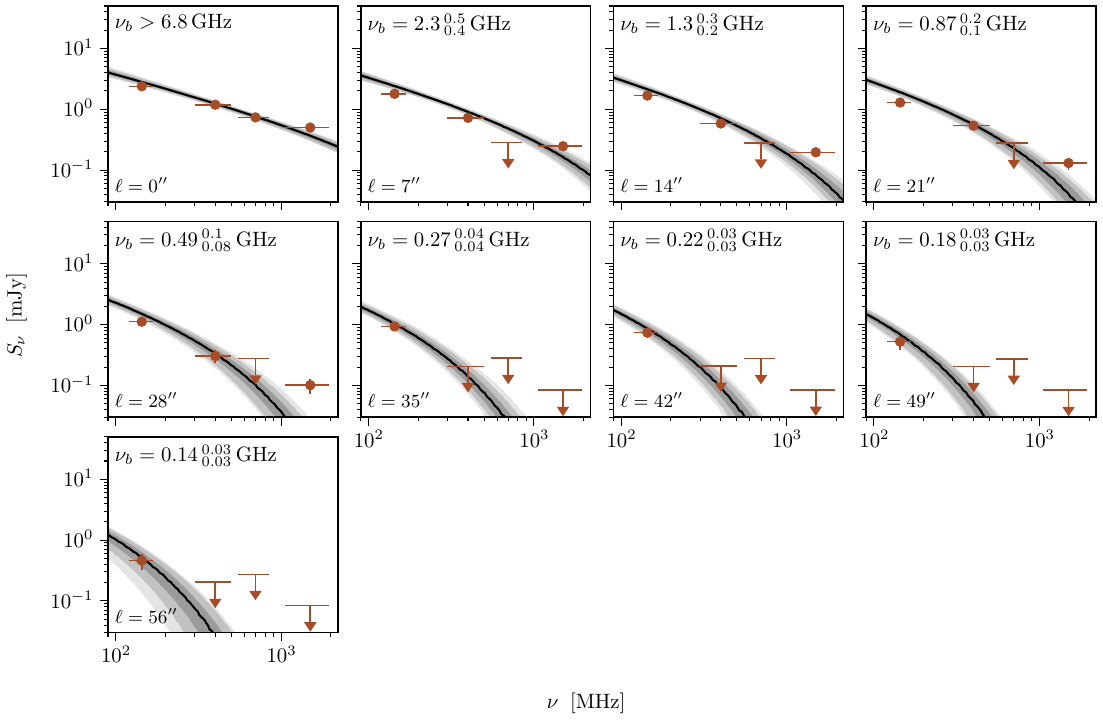}
\captionof{figure}{Tail spectral fitting results for KUG1258+279A.  See Fig.~\ref{fig:example_fit_result} for details.}
\label{fig:KUG1258+279A_spec}
\end{minipage}}
\par\smallskip

\par\smallskip\noindent
\centerline{\begin{minipage}{\textwidth}
\centering
\includegraphics[width = \textwidth]{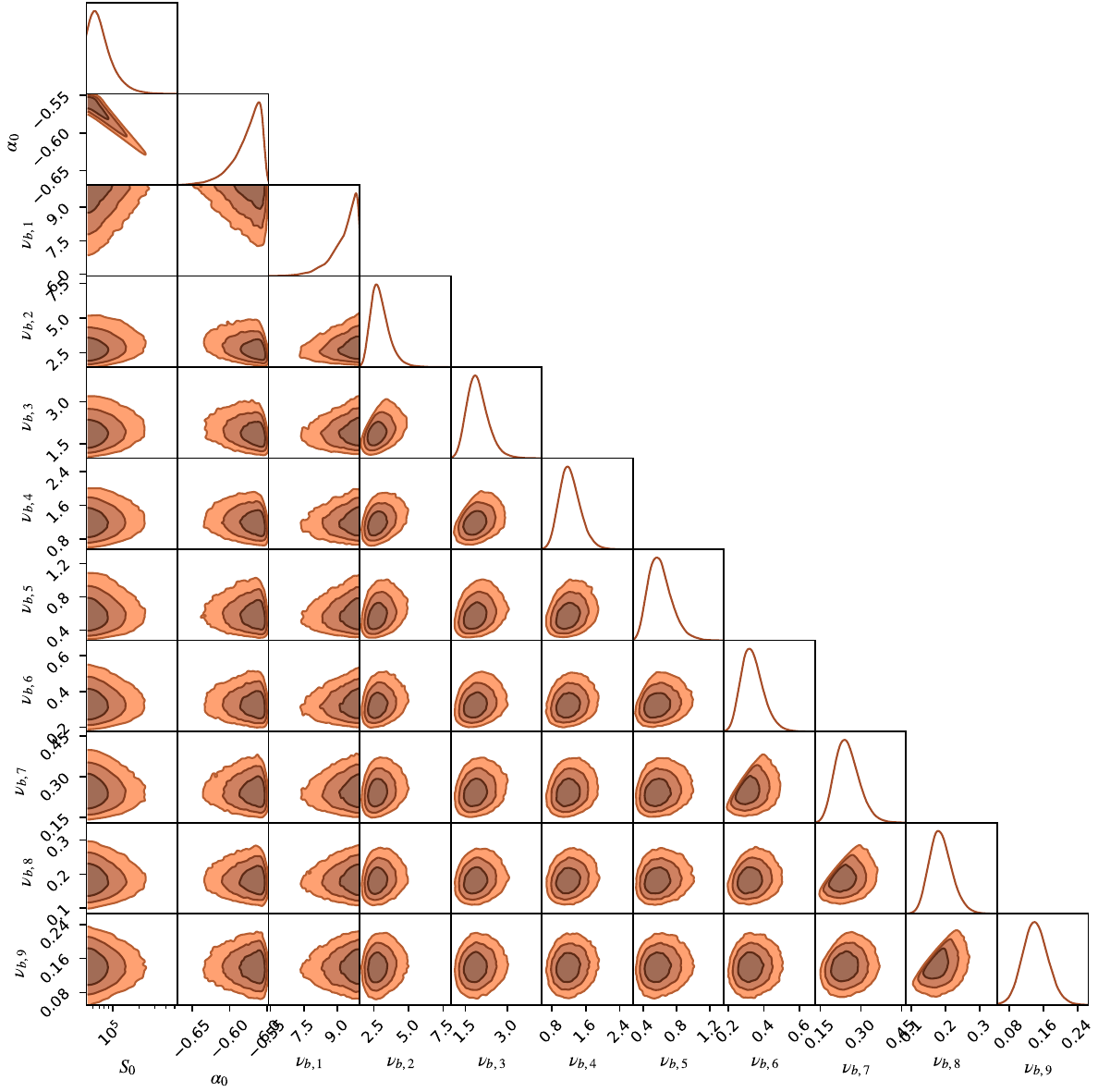}
\captionof{figure}{MCMC fitting results for the aging model along the tail.}
\label{fig:KUG1258+279A_corner}
\end{minipage}}
\par\smallskip

\clearpage

\subsection{KUG1258+287}

\par\smallskip\noindent
\centerline{\begin{minipage}{\textwidth}
\centering
\includegraphics[width = \textwidth]{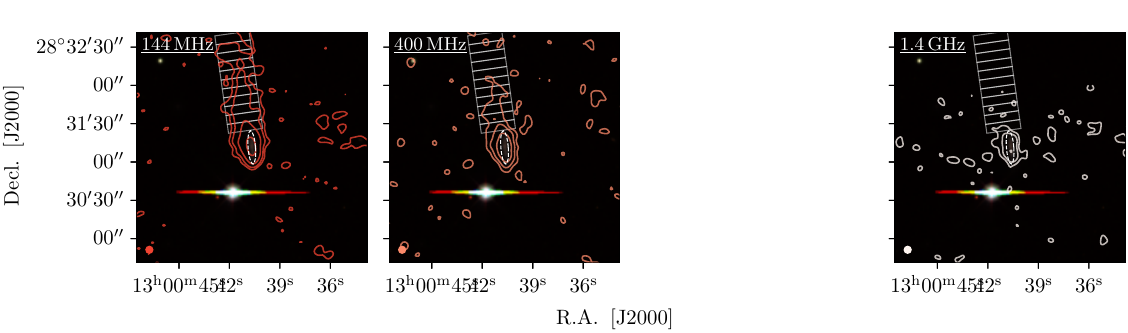}
\captionof{figure}{KUG1258+287, see Fig.~\ref{fig:example_imgs_NGC4848} for details.}
\label{fig:example_imgs_KUG1258+287}
\end{minipage}}
\par\smallskip

\par\smallskip\noindent
\centerline{\begin{minipage}{\textwidth}
\centering
\includegraphics[width = \textwidth]{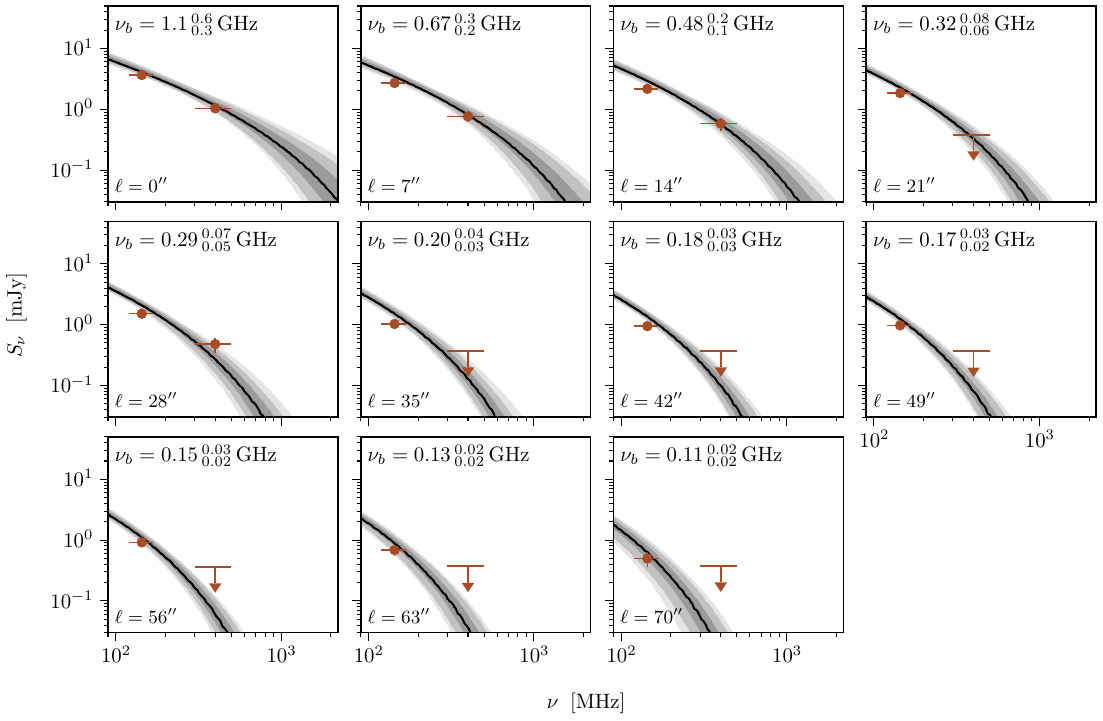}
\captionof{figure}{Tail spectral fitting results for KUG1258+287.  See Fig.~\ref{fig:example_fit_result} for details.}
\label{fig:KUG1258+287_spec}
\end{minipage}}
\par\smallskip

\par\smallskip\noindent
\centerline{\begin{minipage}{\textwidth}
\centering
\includegraphics[width = \textwidth]{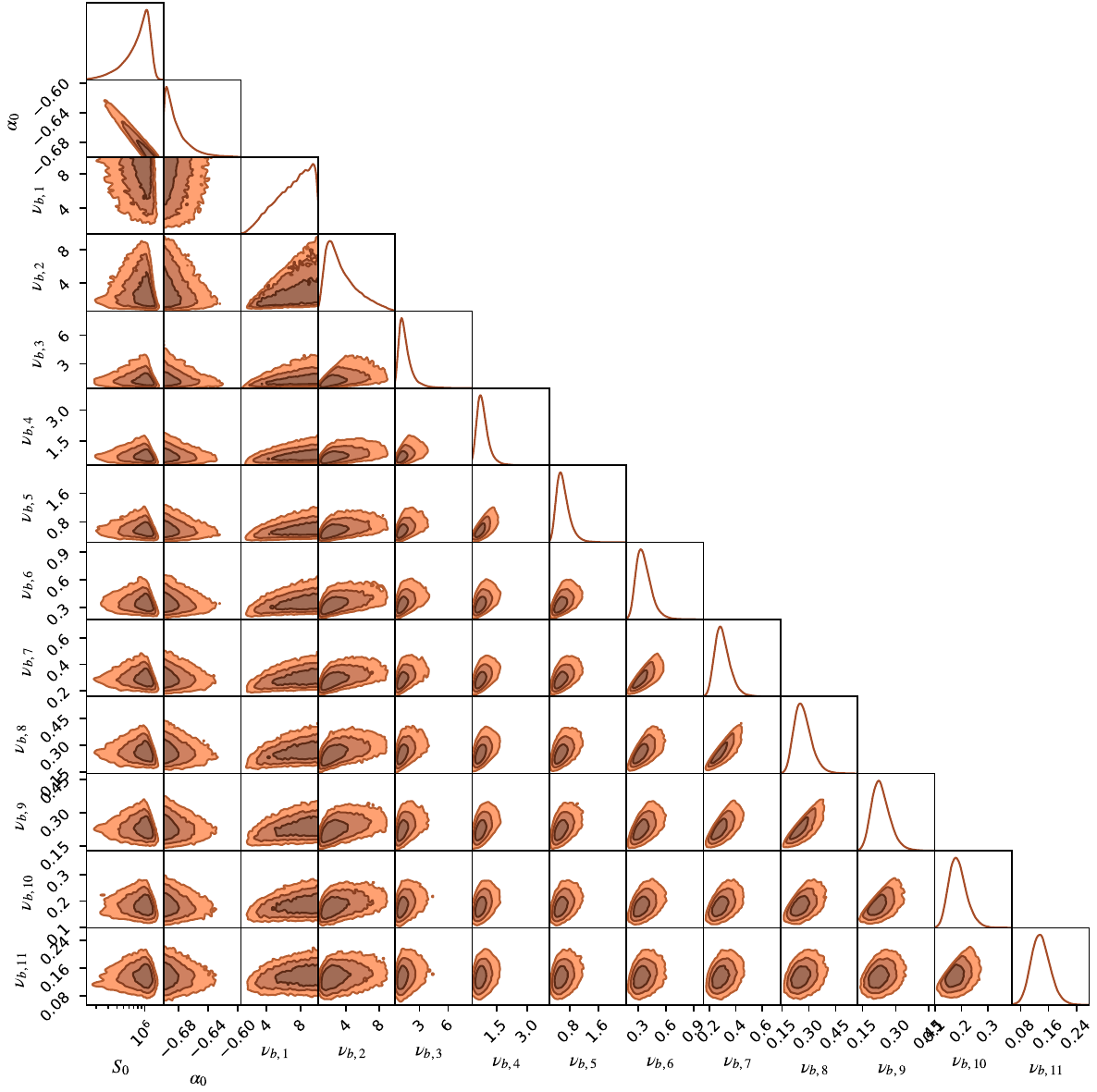}
\captionof{figure}{MCMC fitting results for the aging model along the tail.}
\label{fig:KUG1258+287_corner}
\end{minipage}}
\par\smallskip

\clearpage

\subsection{KUG1259+279}

\par\smallskip\noindent
\centerline{\begin{minipage}{\textwidth}
\centering
\includegraphics[width = \textwidth]{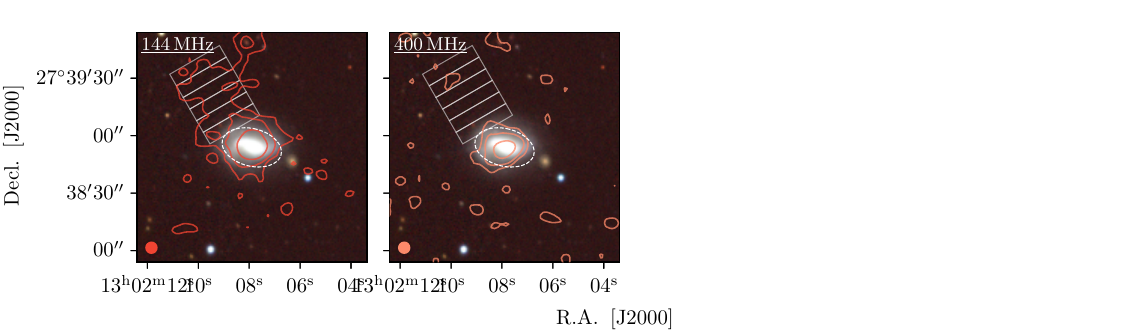}
\captionof{figure}{KUG1259+279, see Fig.~\ref{fig:example_imgs_NGC4848} for details.}
\label{fig:example_imgs_KUG1259+279}
\end{minipage}}
\par\smallskip

\par\smallskip\noindent
\centerline{\begin{minipage}{\textwidth}
\centering
\includegraphics[width = \textwidth]{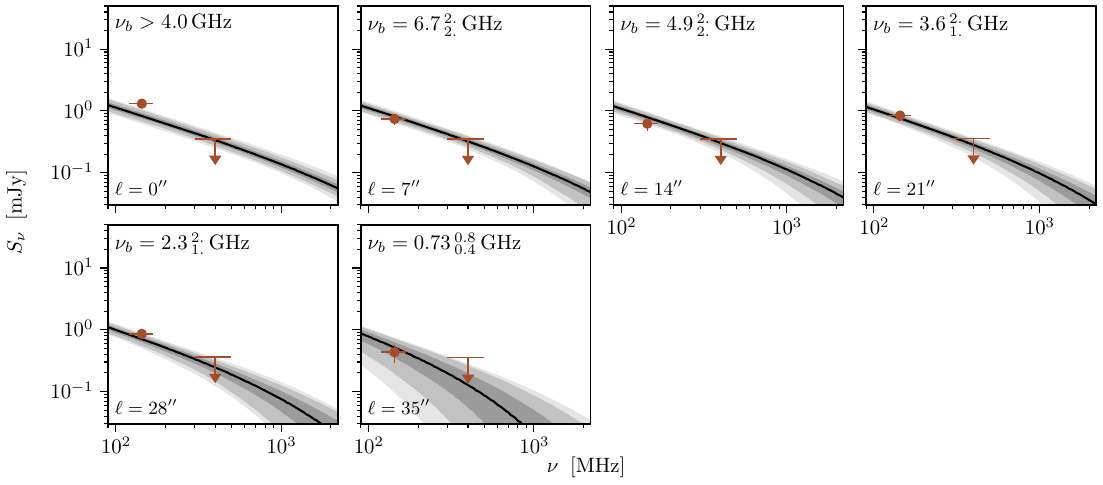}
\captionof{figure}{Tail spectral fitting results for KUG1259+279.  See Fig.~\ref{fig:example_fit_result} for details.}
\label{fig:KUG1259+279_spec}
\end{minipage}}
\par\smallskip

\par\smallskip\noindent
\centerline{\begin{minipage}{\textwidth}
\centering
\includegraphics[width = \textwidth]{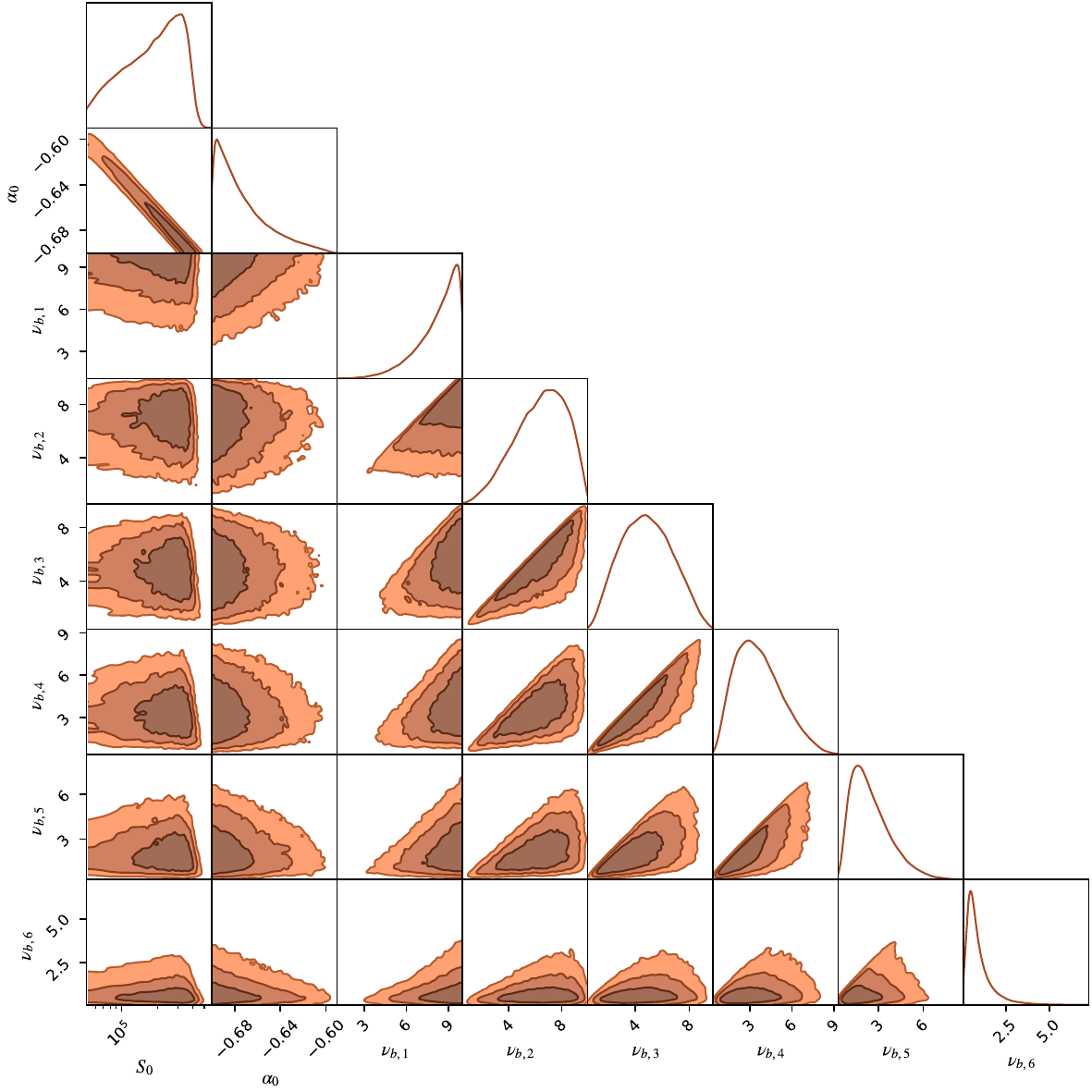}
\captionof{figure}{MCMC fitting results for the aging model along the tail.}
\label{fig:KUG1259+279_corner}
\end{minipage}}
\par\smallskip

\clearpage

\subsection{KUG1259+284}

\par\smallskip\noindent
\centerline{\begin{minipage}{\textwidth}
\centering
\includegraphics[width = \textwidth]{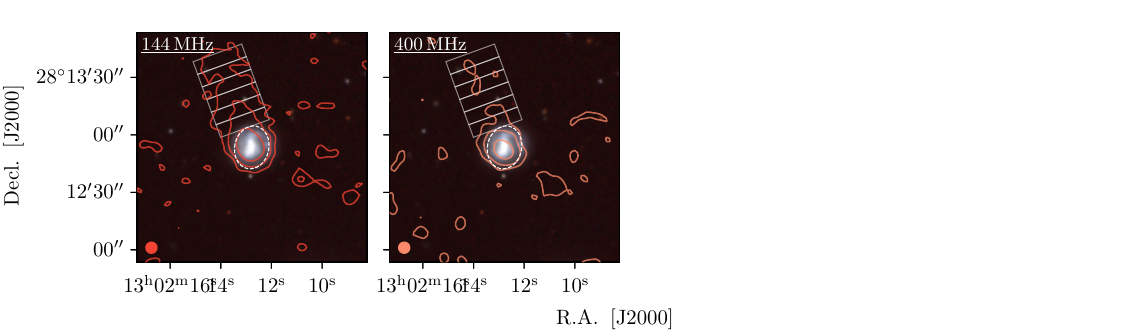}
\captionof{figure}{KUG1259+284, see Fig.~\ref{fig:example_imgs_NGC4848} for details.}
\label{fig:example_imgs_KUG1259+284}
\end{minipage}}
\par\smallskip

\par\smallskip\noindent
\centerline{\begin{minipage}{\textwidth}
\centering
\includegraphics[width = \textwidth]{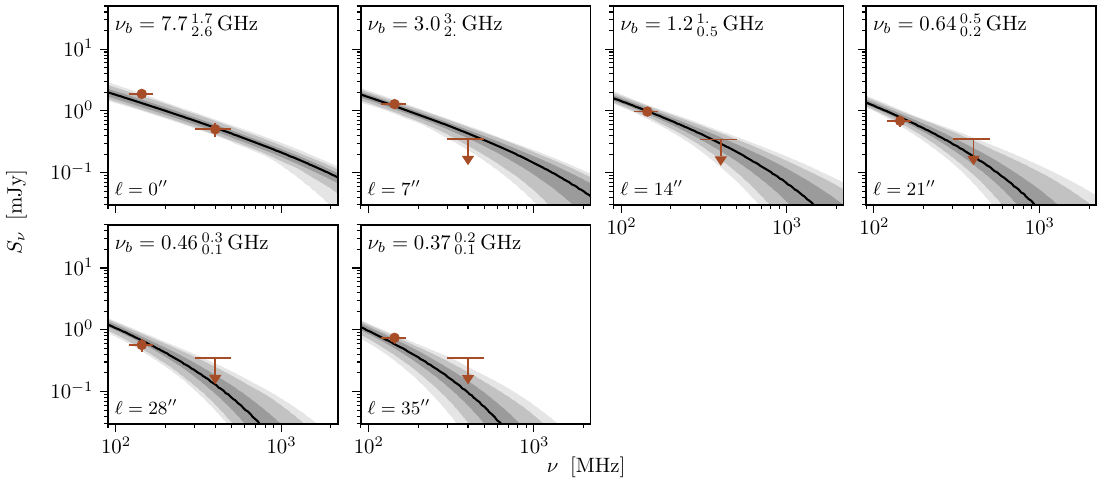}
\captionof{figure}{Tail spectral fitting results for KUG1259+284.  See Fig.~\ref{fig:example_fit_result} for details.}
\label{fig:KUG1259+284_spec}
\end{minipage}}
\par\smallskip

\par\smallskip\noindent
\centerline{\begin{minipage}{\textwidth}
\centering
\includegraphics[width = \textwidth]{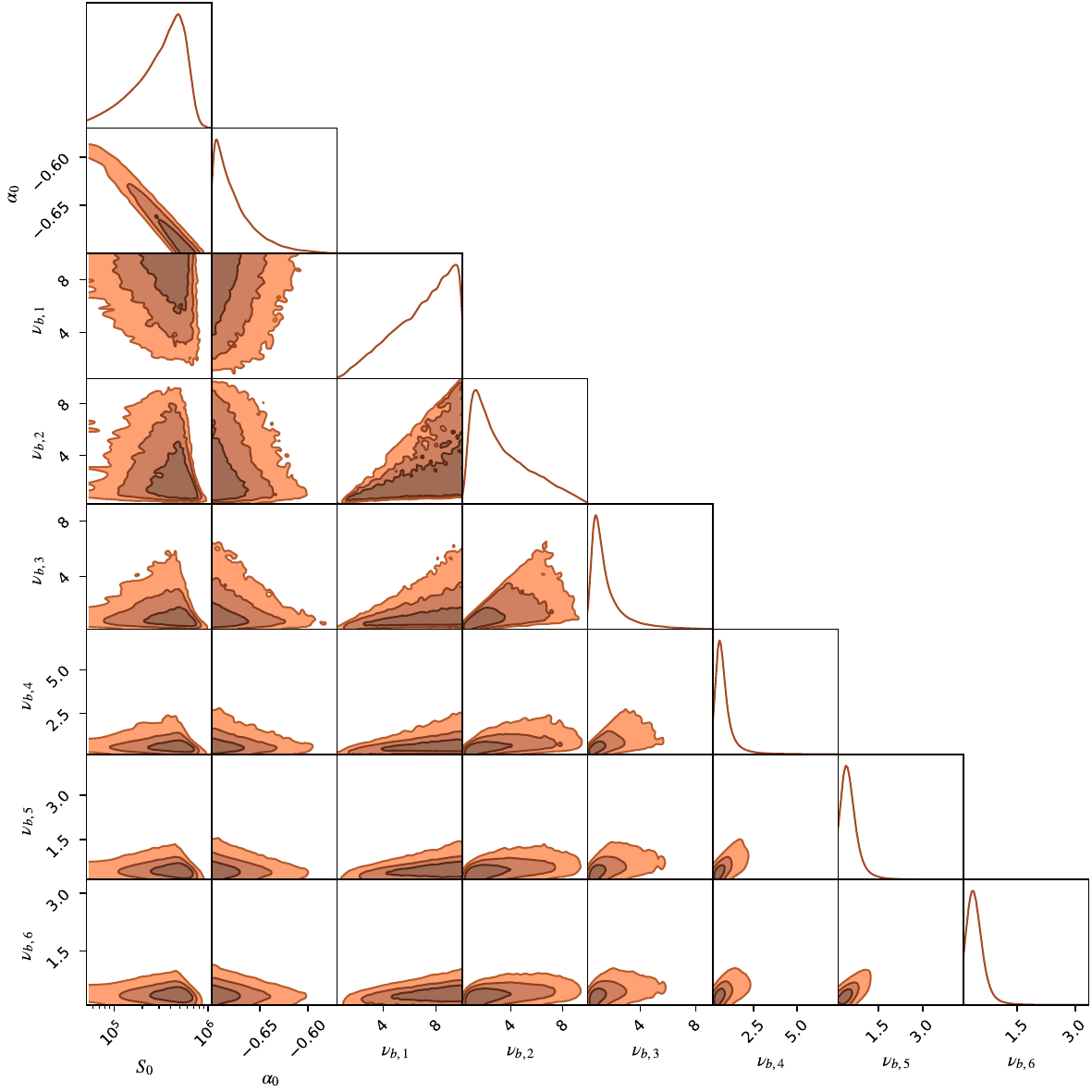}
\captionof{figure}{MCMC fitting results for the aging model along the tail.}
\label{fig:KUG1259+284_corner}
\end{minipage}}
\par\smallskip

\clearpage

\subsection{MRK0053}

\par\smallskip\noindent
\centerline{\begin{minipage}{\textwidth}
\centering
\includegraphics[width = \textwidth]{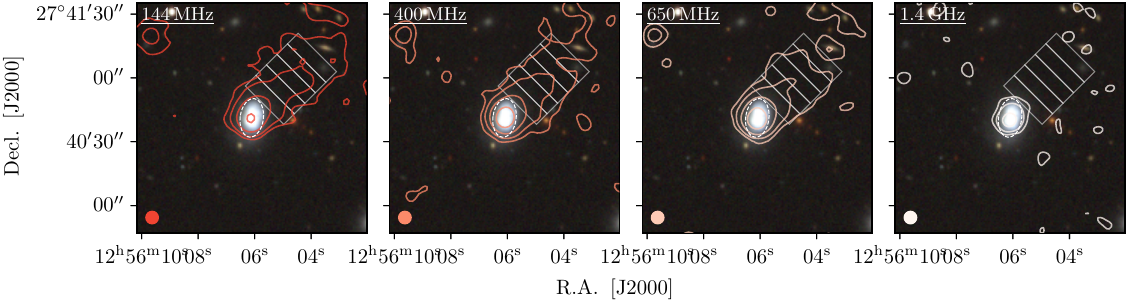}
\captionof{figure}{MRK0053, see Fig.~\ref{fig:example_imgs_NGC4848} for details.}
\label{fig:example_imgs_MRK0053}
\end{minipage}}
\par\smallskip

\par\smallskip\noindent
\centerline{\begin{minipage}{\textwidth}
\centering
\includegraphics[width = \textwidth]{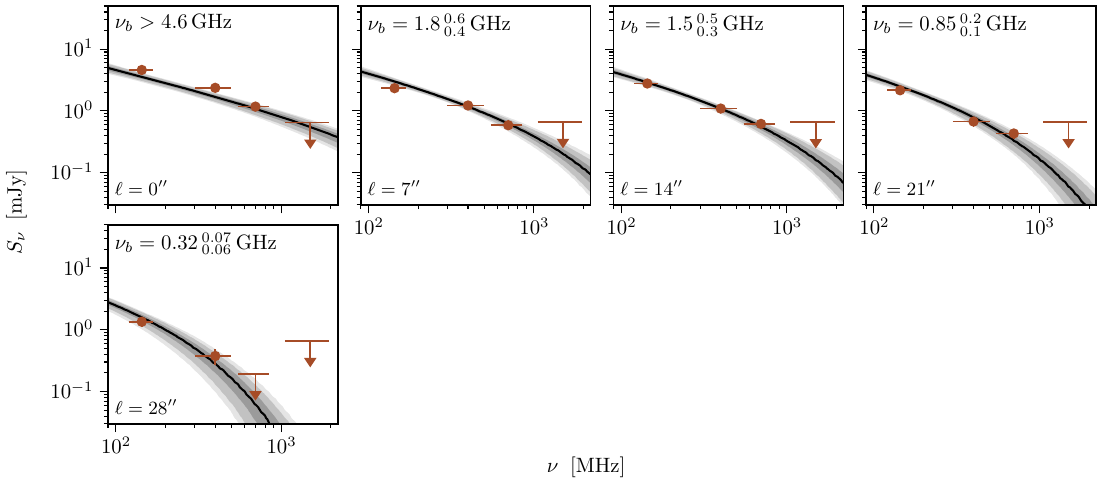}
\captionof{figure}{Tail spectral fitting results for MRK0053.  See Fig.~\ref{fig:example_fit_result} for details.}
\label{fig:MRK0053_spec}
\end{minipage}}
\par\smallskip

\par\smallskip\noindent
\centerline{\begin{minipage}{\textwidth}
\centering
\includegraphics[width = \textwidth]{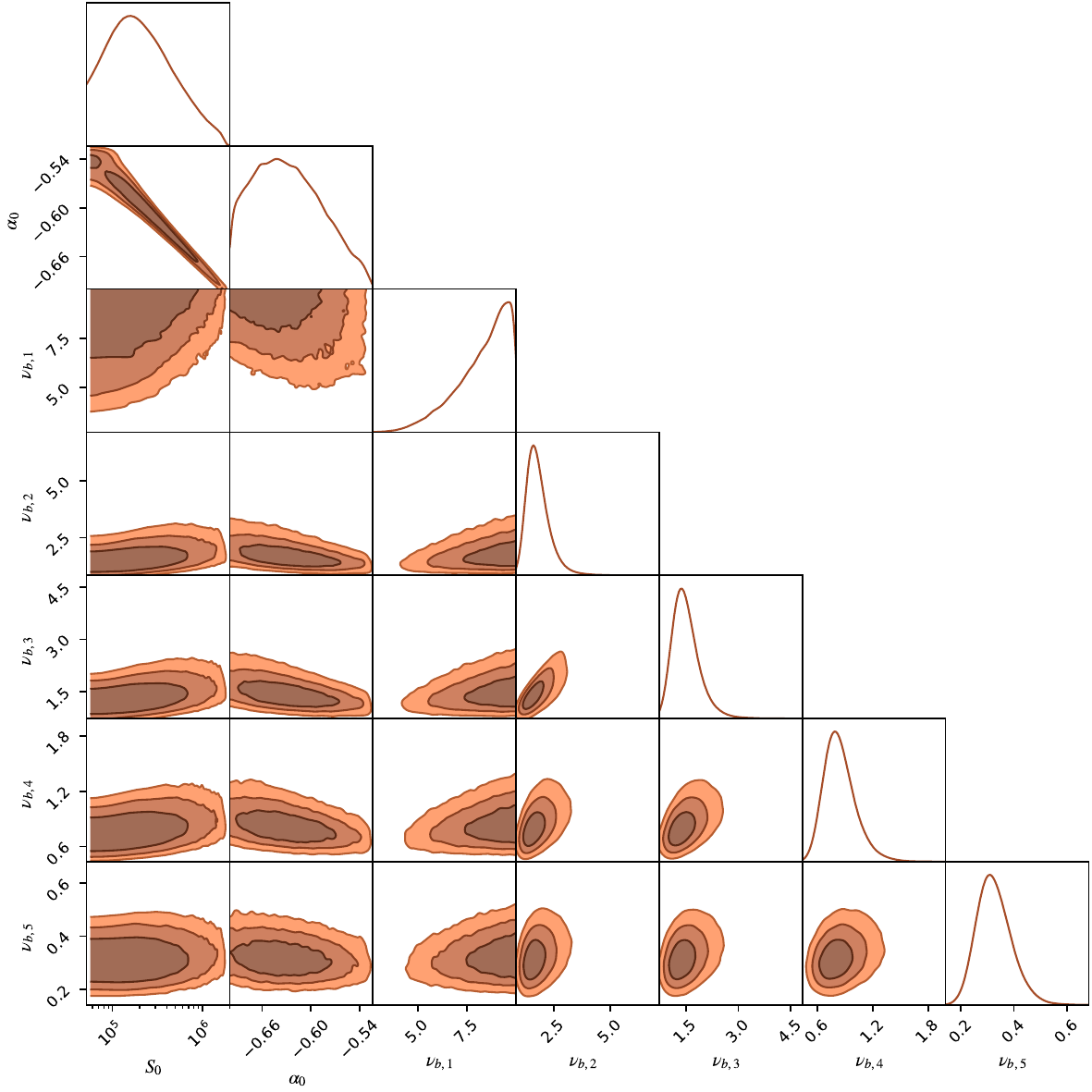}
\captionof{figure}{MCMC fitting results for the aging model along the tail.}
\label{fig:MRK0053_corner}
\end{minipage}}
\par\smallskip

\clearpage

\subsection{MRK0056}

\par\smallskip\noindent
\centerline{\begin{minipage}{\textwidth}
\centering
\includegraphics[width = \textwidth]{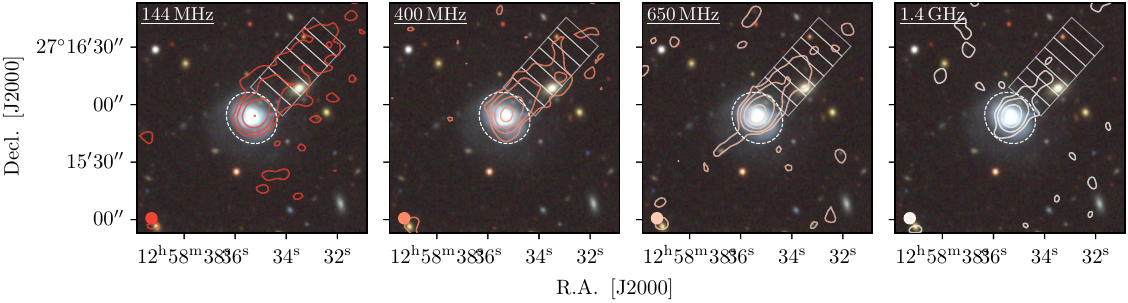}
\captionof{figure}{MRK0056, see Fig.~\ref{fig:example_imgs_NGC4848} for details.}
\label{fig:example_imgs_MRK0056}
\end{minipage}}
\par\smallskip

\par\smallskip\noindent
\centerline{\begin{minipage}{\textwidth}
\centering
\includegraphics[width = \textwidth]{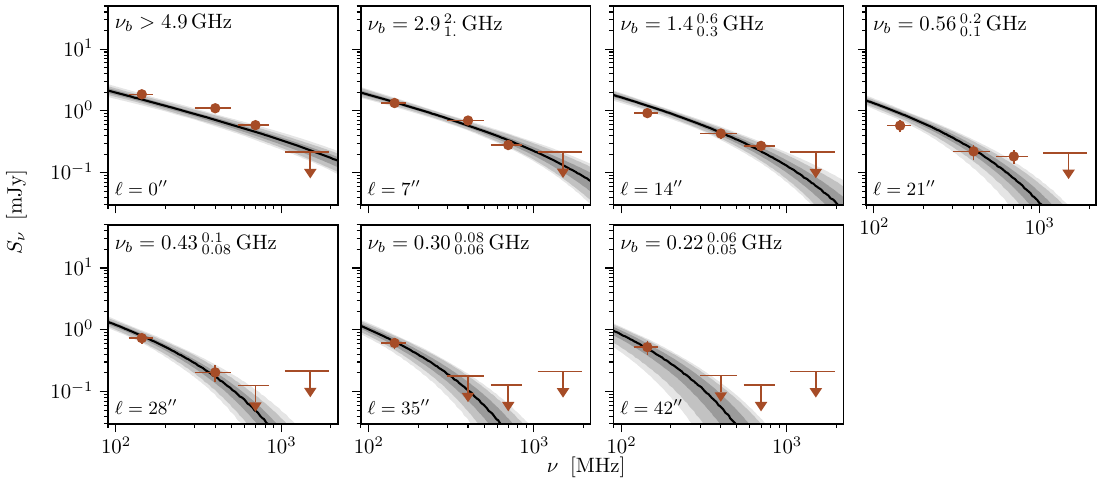}
\captionof{figure}{Tail spectral fitting results for MRK0056.  See Fig.~\ref{fig:example_fit_result} for details.}
\label{fig:MRK0056_spec}
\end{minipage}}
\par\smallskip

\par\smallskip\noindent
\centerline{\begin{minipage}{\textwidth}
\centering
\includegraphics[width = \textwidth]{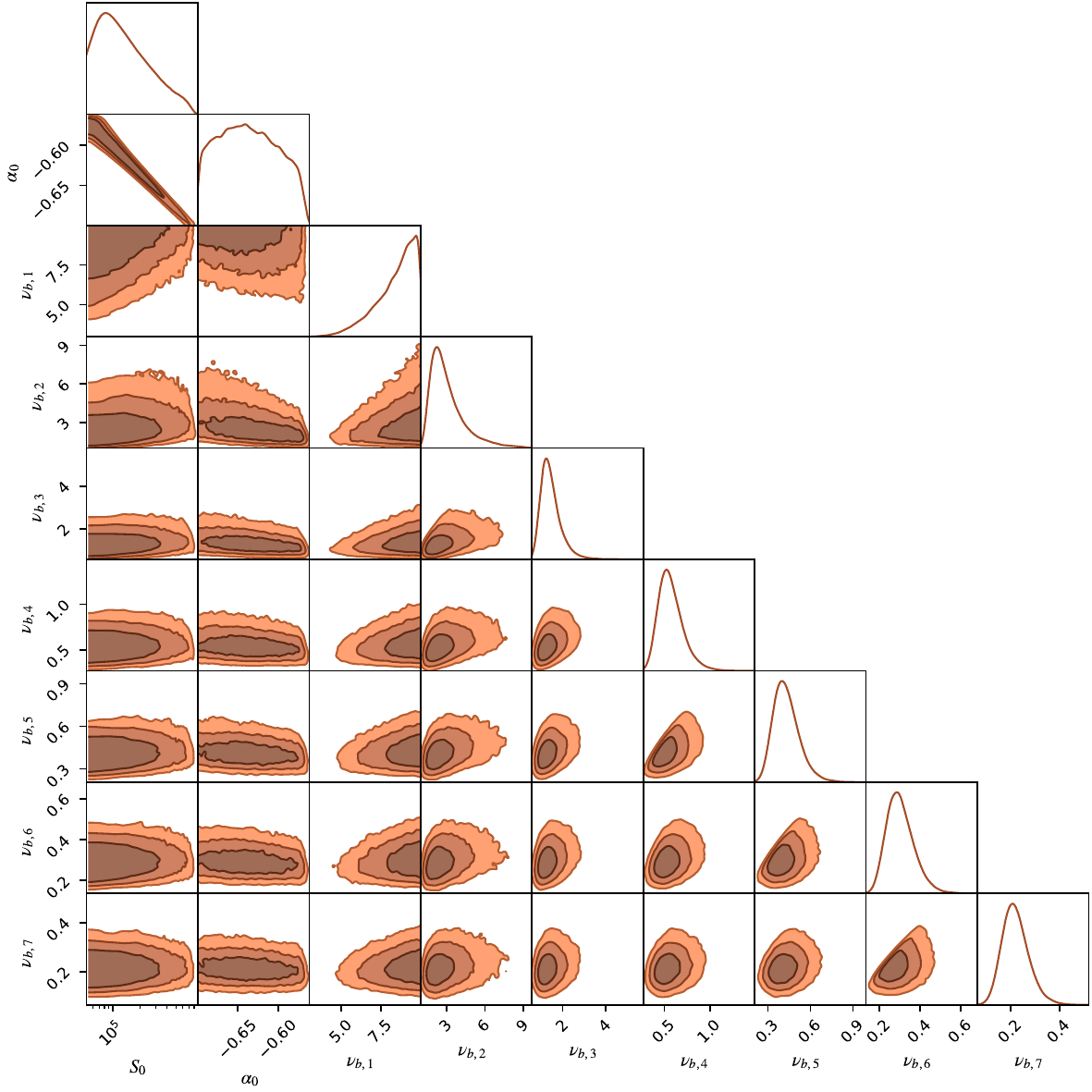}
\captionof{figure}{MCMC fitting results for the aging model along the tail.}
\label{fig:MRK0056_corner}
\end{minipage}}
\par\smallskip

\clearpage

\subsection{MRK0057}

\par\smallskip\noindent
\centerline{\begin{minipage}{\textwidth}
\centering
\includegraphics[width = \textwidth]{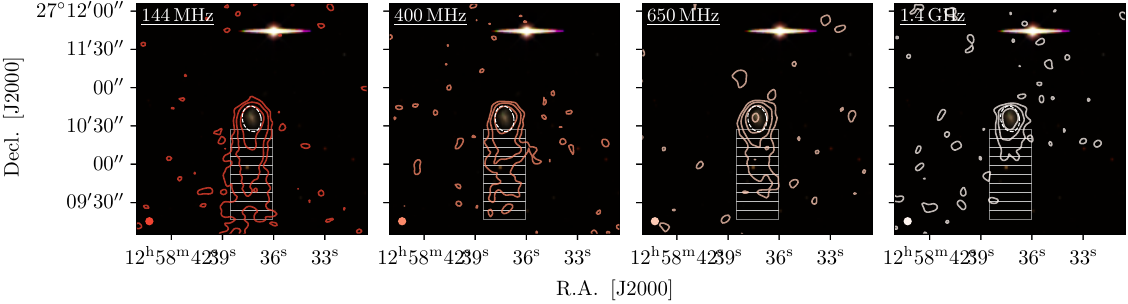}
\captionof{figure}{MRK0057, see Fig.~\ref{fig:example_imgs_NGC4848} for details.}
\label{fig:example_imgs_MRK0057}
\end{minipage}}
\par\smallskip

\par\smallskip\noindent
\centerline{\begin{minipage}{\textwidth}
\centering
\includegraphics[width = \textwidth]{figures/specs_MRK57.pdf}
\captionof{figure}{Tail spectral fitting results for MRK0057.  See Fig.~\ref{fig:example_fit_result} for details.}
\label{fig:MRK0057_spec}
\end{minipage}}
\par\smallskip

\par\smallskip\noindent
\centerline{\begin{minipage}{\textwidth}
\centering
\includegraphics[width = \textwidth]{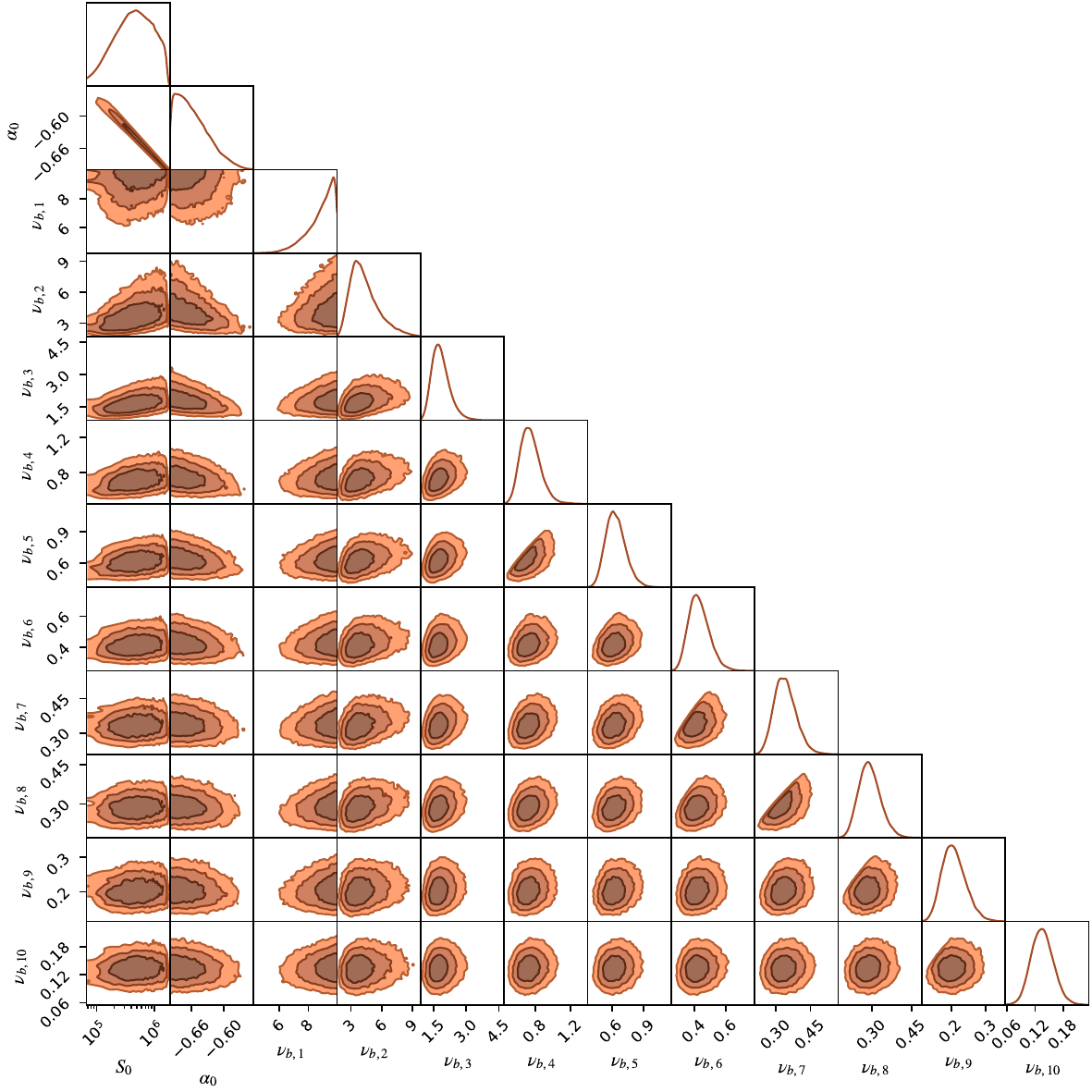}
\captionof{figure}{MCMC fitting results for the aging model along the tail.}
\label{fig:MRK0057_corner}
\end{minipage}}
\par\smallskip

\clearpage

\subsection{MRK0058}

\par\smallskip\noindent
\centerline{\begin{minipage}{\textwidth}
\centering
\includegraphics[width = \textwidth]{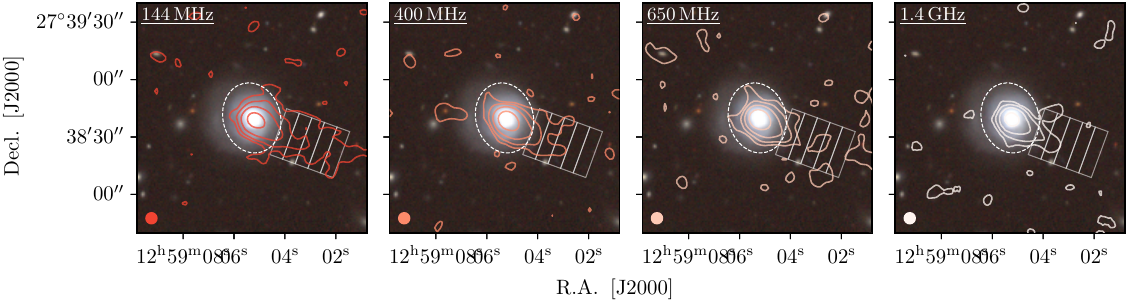}
\captionof{figure}{MRK0058, see Fig.~\ref{fig:example_imgs_NGC4848} for details.}
\label{fig:example_imgs_MRK0058}
\end{minipage}}
\par\smallskip

\par\smallskip\noindent
\centerline{\begin{minipage}{\textwidth}
\centering
\includegraphics[width = \textwidth]{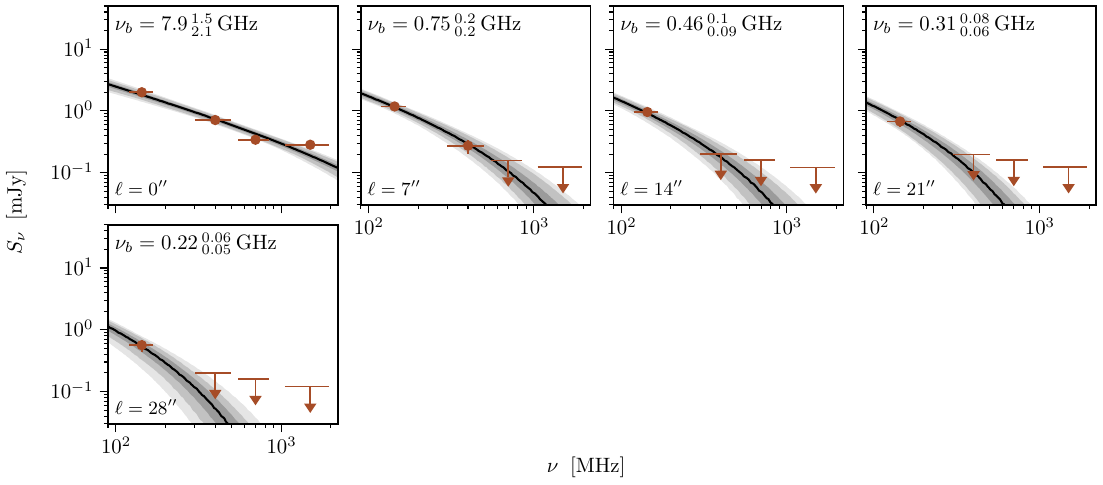}
\captionof{figure}{Tail spectral fitting results for MRK0058.  See Fig.~\ref{fig:example_fit_result} for details.}
\label{fig:MRK0058_spec}
\end{minipage}}
\par\smallskip

\par\smallskip\noindent
\centerline{\begin{minipage}{\textwidth}
\centering
\includegraphics[width = \textwidth]{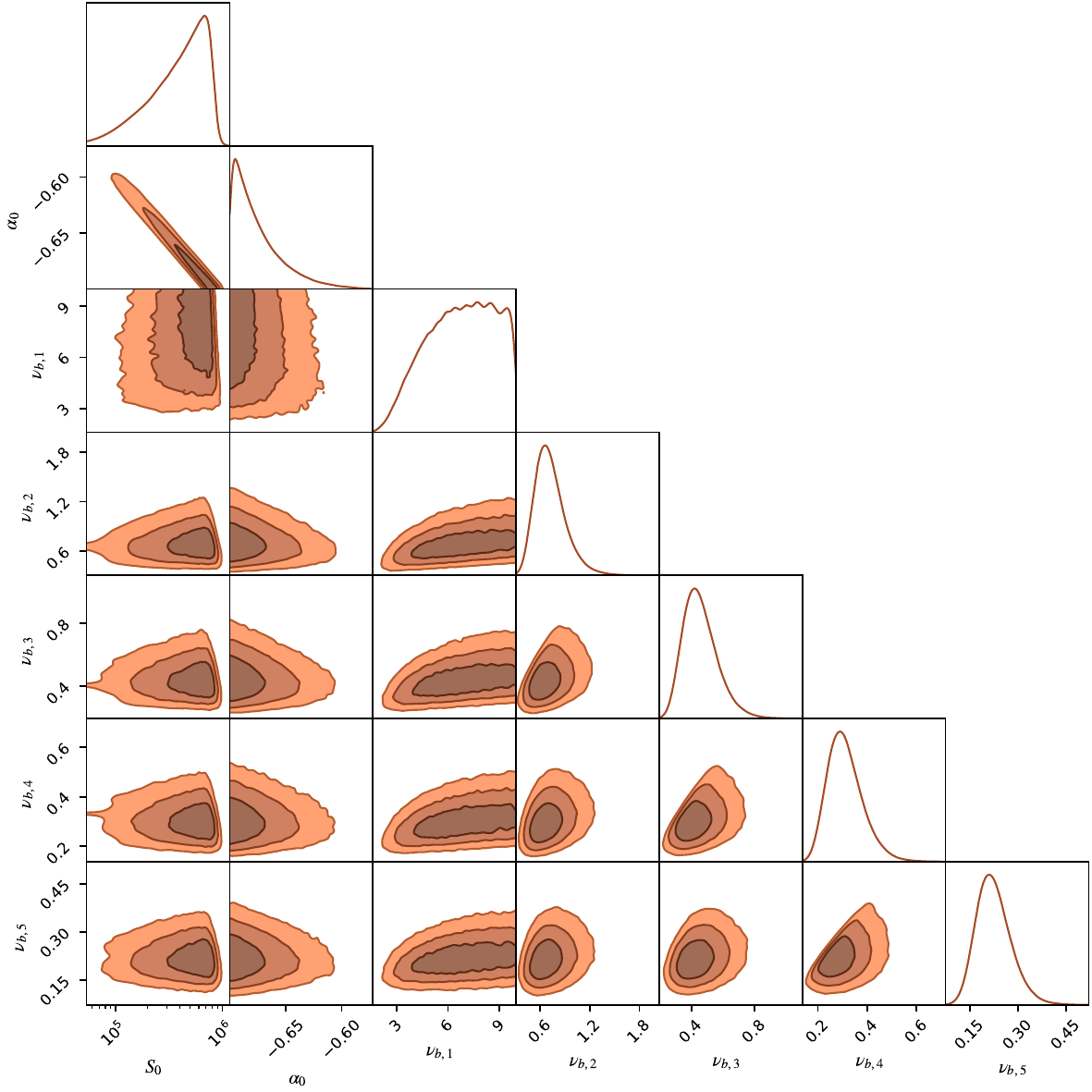}
\captionof{figure}{MCMC fitting results for the aging model along the tail.}
\label{fig:MRK0058_corner}
\end{minipage}}
\par\smallskip

\clearpage

\subsection{D100}

\par\smallskip\noindent
\centerline{\begin{minipage}{\textwidth}
\centering
\includegraphics[width = \textwidth]{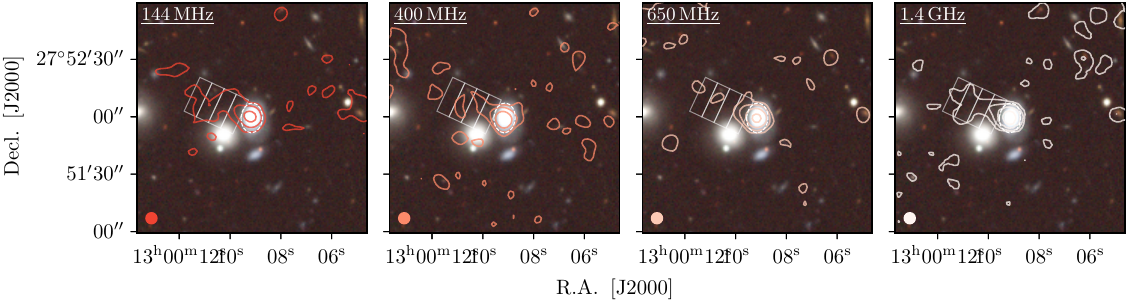}
\captionof{figure}{D100, see Fig.~\ref{fig:example_imgs_NGC4848} for details.}
\label{fig:example_imgs_MRK0060}
\end{minipage}}
\par\smallskip

\par\smallskip\noindent
\centerline{\begin{minipage}{\textwidth}
\centering
\includegraphics[width = \textwidth]{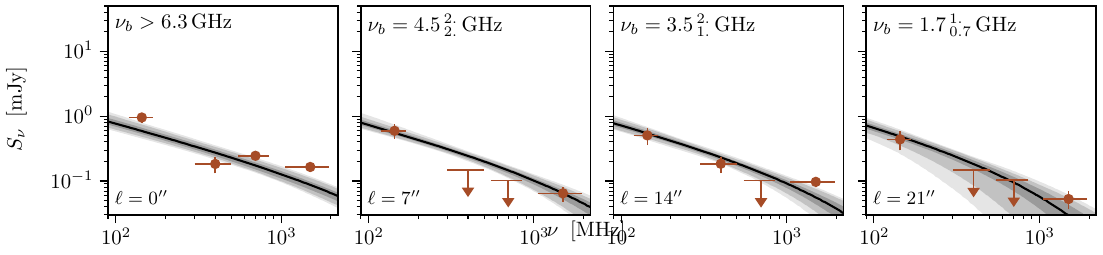}
\captionof{figure}{Tail spectral fitting results for D100.  See Fig.~\ref{fig:example_fit_result} for details.}
\label{fig:MRK0060_spec}
\end{minipage}}
\par\smallskip

\par\smallskip\noindent
\centerline{\begin{minipage}{\textwidth}
\centering
\includegraphics[width = \textwidth]{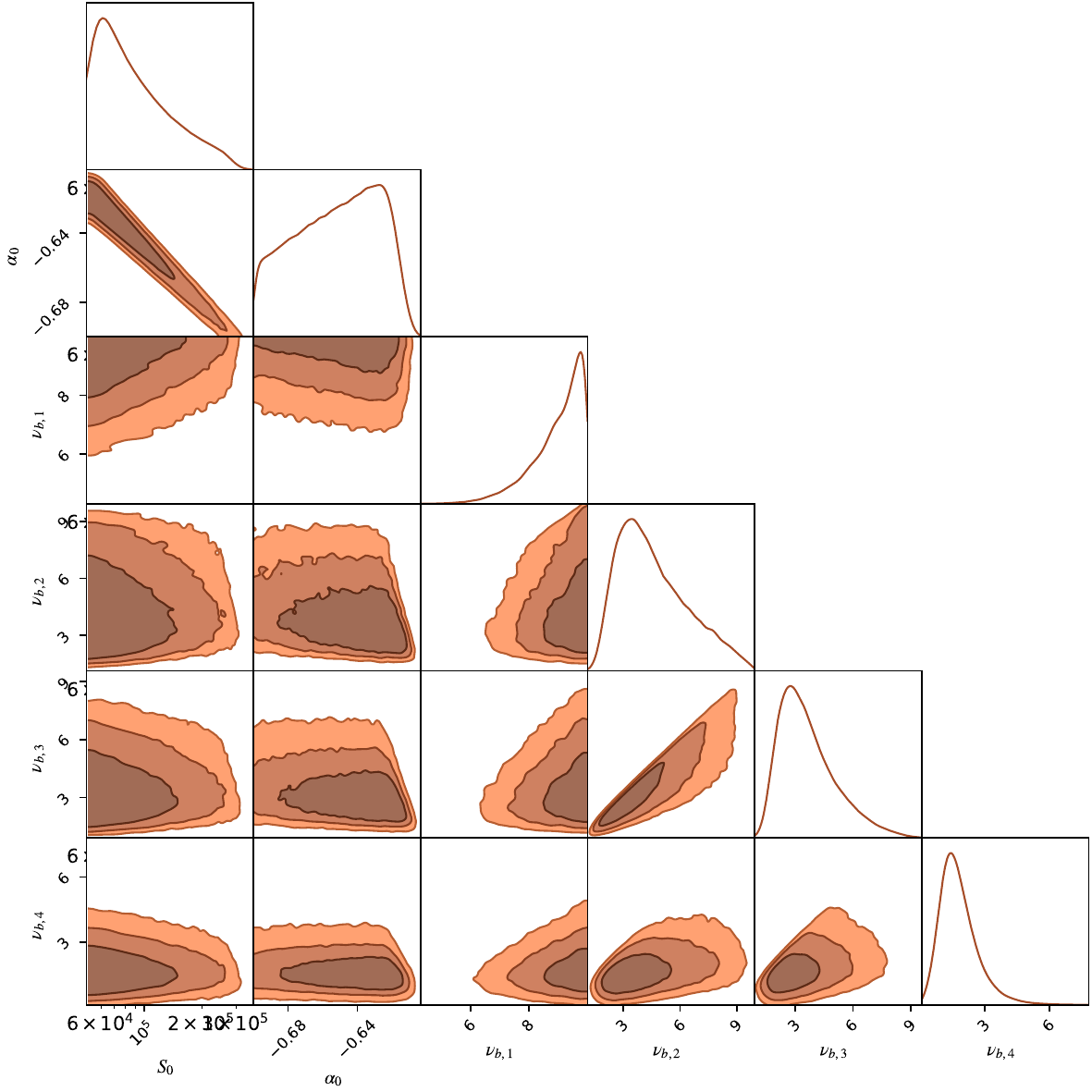}
\captionof{figure}{MCMC fitting results for the aging model along the tail.}
\label{fig:MRK0060_corner}
\end{minipage}}
\par\smallskip

\clearpage

\subsection{GMP2601}

\par\smallskip\noindent
\centerline{\begin{minipage}{\textwidth}
\centering
\includegraphics[width = \textwidth]{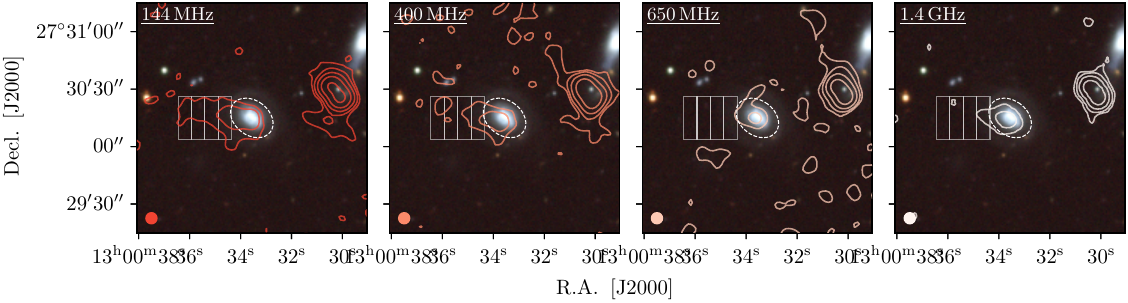}
\captionof{figure}{GMP2601, see Fig.~\ref{fig:example_imgs_NGC4848} for details.}
\label{fig:example_imgs_GMP2601}
\end{minipage}}
\par\smallskip

\par\smallskip\noindent
\centerline{\begin{minipage}{\textwidth}
\centering
\includegraphics[width = \textwidth]{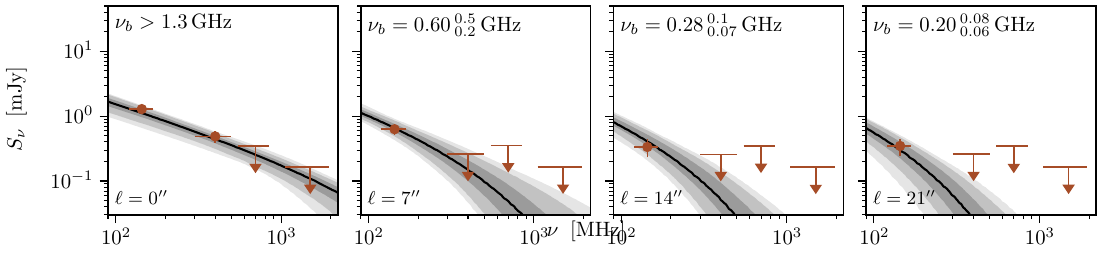}
\captionof{figure}{Tail spectral fitting results for GMP2601.  See Fig.~\ref{fig:example_fit_result} for details.}
\label{fig:GMP2601_spec}
\end{minipage}}
\par\smallskip

\par\smallskip\noindent
\centerline{\begin{minipage}{\textwidth}
\centering
\includegraphics[width = \textwidth]{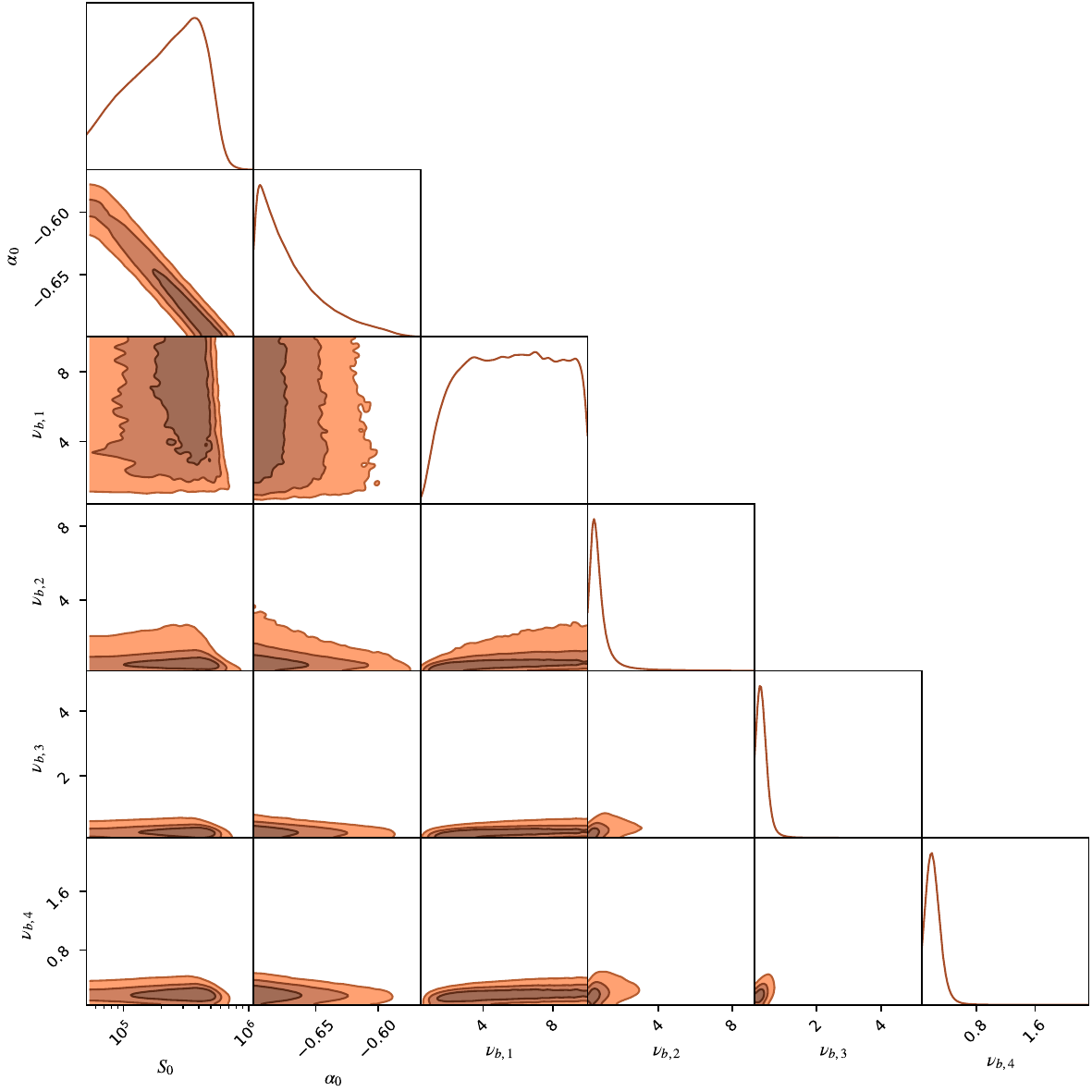}
\captionof{figure}{MCMC fitting results for the aging model along the tail.}
\label{fig:GMP2601_corner}
\end{minipage}}
\par\smallskip

\clearpage

\subsection{GMP3618}

\par\smallskip\noindent
\centerline{\begin{minipage}{\textwidth}
\centering
\includegraphics[width = \textwidth]{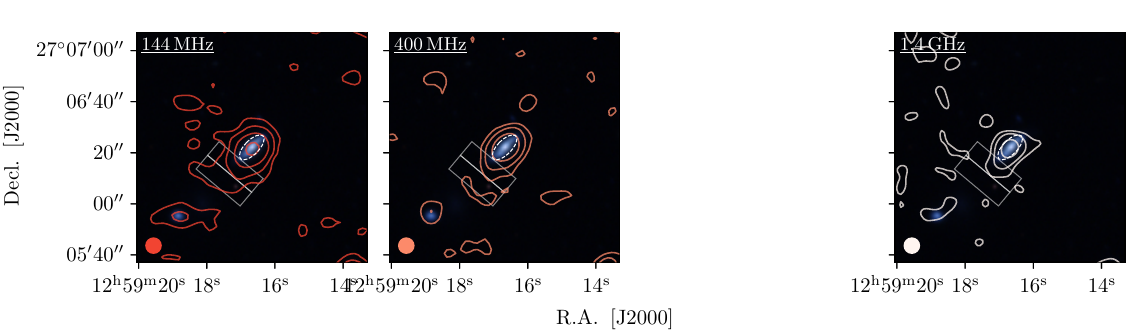}
\captionof{figure}{GMP3618, see Fig.~\ref{fig:example_imgs_NGC4848} for details.}
\label{fig:example_imgs_GMP3618}
\end{minipage}}
\par\smallskip

\clearpage

\subsection{GMP5226}

\par\smallskip\noindent
\centerline{\begin{minipage}{\textwidth}
\centering
\includegraphics[width = \textwidth]{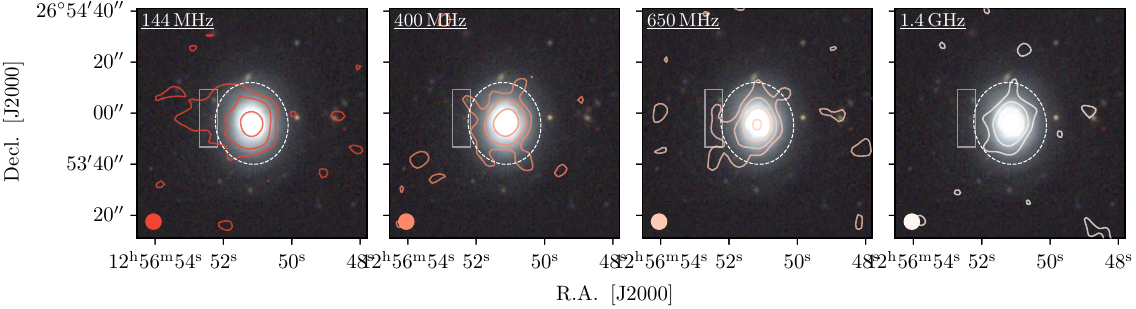}
\captionof{figure}{GMP5226, see Fig.~\ref{fig:example_imgs_NGC4848} for details.}
\label{fig:example_imgs_GMP5226}
\end{minipage}}
\par\smallskip

\end{document}